\documentclass{article}
\usepackage[margin=2.5cm,vmargin=2.5cm]{geometry}
\usepackage[T1]{fontenc}
\usepackage[utf8]{inputenc}
\usepackage[usenames,dvipsnames]{color}
\usepackage{pdflscape}
\usepackage{cite}
\usepackage[toc,page]{appendix}
\usepackage{amsmath}
\usepackage{amssymb}
\usepackage{mathtools}
\usepackage{dsfont}							
\usepackage{stmaryrd}							
\usepackage{cancel}
\usepackage{tikz}
\usetikzlibrary{matrix}
\usepackage{tabulary}
\usepackage{booktabs}
\usepackage{caption}
\usepackage{float}
\usepackage{multirow}
\usepackage[hidelinks]{hyperref}
\renewcommand*{\thefootnote}{\roman{footnote}}
\newcommand{\overbar}[1]{\mkern 1.5mu\overline{\mkern-1.5mu#1\mkern-1.5mu}\mkern 1.5mu}
\numberwithin{equation}{section}
\title{%
\vspace{-1cm}
\begin{flushright}
\normalsize{QMUL-PH-19-24}
\end{flushright}
$ $\\
\textbf{Reductions of Exceptional Field Theories}
}
\date{}
\author{\vspace{-1cm}}
\begin{document}
\maketitle
\begin{center}
\textsc{David S. Berman$^\dagger$}\footnote{\href{mailto:d.s.berman@qmul.ac.uk}{d.s.berman@qmul.ac.uk}},
\textsc{ and Ray Otsuki$^\dagger$}\footnote{\href{mailto:r.otsuki@qmul.ac.uk}{r.otsuki@qmul.ac.uk}}\\
\end{center}
\begin{center}
\vspace{1cm}
\textit{$^\dagger$Queen Mary University of London, Centre for Research in String Theory,\\
School of Physics and Astronomy, Mile End Road, London, England, E1 4NS}\\
\end{center}
\vspace{2cm}
\abstract{Double Field Theory (DFT) and Exceptional Field Theory (EFT), collectively called ExFTs, have proven to be a remarkably powerful new framework for string and M-theory. Exceptional field theories were constructed on a case by case basis as often each EFT has its own idiosyncrasies. Intuitively though, an $E_{n-1(n-1)}$ EFT must be contained in an $E_{n(n)}$ ExFT.  In this paper we propose a generalised Kaluza-Klein ansatz to relate different ExFTs. We then discuss in more detail the different aspects of the relationship \emph{between} various ExFTs including the coordinates, section condition and (pseudo)-Lagrangian densities. For the $E_{8(8)}$ EFT we describe a generalisation of the Mukhi -Papageorgakis mechanism to relate the $d=3$ topological term in the $E_{8(8)}$ EFT to a Yang-Mills action in the $E_{7(7)}$ EFT.}
\clearpage
\tableofcontents
\renewcommand*{\thefootnote}{\arabic{footnote}}
\setcounter{footnote}{0}
\section{Introduction}
Extended Field Theories (ExFT) are a class of theories that realise a manifest local $\operatorname{O}(D,D,\mathbb{R})$ symmetry or $E_{n(n)}(\mathbb{R})$ symmetry on an extended spacetime. The former is called Double Field Theory (DFT) whilst the latter is called Exceptional Field Theory (EFT). These symmetries are formed from the local symmetries of the supergravity fields (i.e.\ conventional diffeomorphisms and $p$-form gauge transformations) and are repackaged into a \emph{generalised diffeomorphism} that is generated, in analogy with GR, by a \emph{generalised Lie derivative} acting on this extended spacetime. Its action may be described as a conventional Lie derivative modified by a \emph{$Y$-tensor}, built from group invariants. This $Y$-tensor also plays a role in a consistency constraint on ExFTs called the \emph{section condition} that we shall describe in more detail later.\par
For example, the local symmetries of DFT are constructed from the gauge transformations of the metric and the Kalb-Ramond 2-form (namely diffeomorphisms and 1-form gauge transformations) which, together, generate an $\operatorname{O}(D,D, \mathbb{R})$ transformation\footnote{We have been explicit in specifying the continuous groups as they are not quite the discrete T- and U-duality groups. We shall henceforth drop the $\mathbb{R}$ and leave it implicit (see \cite{Berman:2014jsa} for a discussion of how these dualities appear in ExFTs).} that acts linearly on a doubled spacetime. The theory is then specified by a 2-derivative action that is constructed from terms invariant under generalised diffeomorphisms and whose on-shell degrees of freedom are parametrised by the supergravity fields. See \cite{Aldazabal:2013sca,Geissbuhler:2013uka,Berman:2013eva} for reviews on DFT.\par
EFT has an added complexity over DFT since we distinguish between an `internal' and `external' space that necessitates a tensor hierarchy of gauge fields and a covariantisation with respect to both internal (generalised) diffeomorphisms and external (conventional) diffeomorphisms. The details are rather more involved than DFT and we refer the reader to the original papers \cite{Hohm:2014fxa,Hohm:2013uia,Hohm:2013vpa,Abzalov:2015ega,Musaev:2015ces,Hohm:2015xna,Berman:2015rcc} for the details. We shall introduce the relevant details of each theory in the main text as needed. In particular, we shall restrict our discussion to the finite-dimensional cases\footnote{See also \cite{Bossard:2018utw,Bossard:2017aae} for progress on the $n=9$ case (the first instance where the extended spacetime is infinite-dimensional).}.\par

Thus far, the literature on relationship between ExFTs has remained rather sparse. Previous work in this area has generally focused on DFT-to-DFT reductions such as in gauged DFT (GDFT)\cite{Aldazabal:2011nj,Dibitetto:2012rk,Grana:2012rr,Berman:2012uy,Aldazabal:2013mya,Aldazabal:2013via}, which realised that one could produce DFTs with gauge deformations by a Scherk-Schwarz reduction of an ungauged DFT, or the Kaluza-Klein reduction of the DFT generalised metric. A key exception is \cite{Thompson:2011uw} which considered the reduction from the $\operatorname{SL}(5)$ extended field theory of \cite{Berman:2010is} to $\operatorname{O}(3,3)$ DFT\footnote{See also \cite{Sakatani:2017nfr} which studied the relation between the M-theory and Type IIB solutions of the \emph{same} EFT.}. This was not for the full exceptional field theory but just for the extended space and did not consider alternative reductions. In this paper, we extend these works by examining a host of EFT-to-EFT reductions. We mention that similar ideas have been leveraged in the infinite-dimensional EFTs. In particular, the embedding of $E_{8(8)}$ within $E_{9(9)}$ was used in \cite{Bossard:2018utw,Bossard:2017aae} and reductions within the context of the $E_{11}$ programme has been considered in \cite{Riccioni:2007au,Riccioni:2009xr,Berman:2011jh,West:2011mm,West:2012qz}. The work presented here is complementary to, but differs slightly from, previous work on the tensor hierarchy in that we work at the level of the action rather than the representations. In addition to this there was early seminal work in \cite{Obers:1999um} predating ExFT where the so called particle and string multiplets related to various U-duality groups were studied in detail. The coordinates and section constraints for the different EFTs then become related to these particle and string multiplets. Finally, the reduction considered here is similar to the decompactifiaction limit of curvature corrections that were first studied in \cite{Bossard:2016hgy} and later in the context of EFT in \cite{Bossard:2015foa}. \par

Whilst each EFT is constructed in the same way each theory nevertheless ends up with rather distinct features, necessitating that EFT-to-EFT reductions be treated on a case-by-case basis. This is in contrast to DFT where one simply needs to know the dimension, $D$ for the relevant $\operatorname{O}(D,D)$ group and everything else (the action, section condition, generalised Lie derivative etc.) is identical. 
For example, $E_{7(7)}$ EFT comes with a generalised Yang-Mills term in the action, a self-duality constraint on the generalised field strength and a symplectic structure $\Omega_{MN}$, none of which have an obvious origin in $E_{8(8)}$ EFT. Additionally, the $Y$-tensor in $E_{8(8)}$ EFT is not sufficient to close the generalised diffeomorphisms, requiring an extra gauge transformation to ensure closure and it is not obvious what happens to this extra gauge transformation if we reduce to $E_{7(7)}$ EFT.  Other examples of quirks of EFTs include the reducible coordinate representation of $\operatorname{SL}(2) \times \mathbb{R}^+$ EFT or the product group of $n=3$ EFT. \par

We begin in Section \ref{sec:Prelims} by laying out the relevant aspects of EFTs and setting up the notation that we use. In Section \ref{sec:ReductionGenMetric}, we introduce and give some motivation for the non-standard \textit{generalised Kaluza-Klein ansatz} by first reducing the DFT generalised metric in a particular manner before moving on to reducing the $\operatorname{SL}(5)$ EFT generalised metric. In this second example, we shall describe how both the $\operatorname{SL}(3) \times \operatorname{SL}(2)$ EFT and $\operatorname{O}(3,3)$ DFT generalised metric can both be obtained from the same reduction of the $\operatorname{SL}(5)$ generalised metric. We close that section by reducing a simplified $E_{8(8)}$ generalised metric by first reducing the extended coordinates. We use this example to highlight some of the difficulties one faces in reducing larger generalised metrics.\par

In Section \ref{sec:ReductionSection} we consider the reduction of the section condition for both $E_{8(8)}$ and $\operatorname{SL}(5)$ EFTs. We use the former to illustrate an explicit reduction of the $Y$-tensor and generalised Lie derivative. The reduction of the $\operatorname{SL}(5)$ theory is used to exhibit how there are alternative reductions to both $\operatorname{SL}(3) \times \operatorname{SL}(2)$ and $\operatorname{O}(3,3)$ theories which is, of course, equivalent to showing how the reductions are linked via the choices of the section condition in the higher dimensional EFT.\par
In Section \ref{sec:BF}, we shall discuss how the Yang-Mills term of $E_{7(7)}$ EFT can be obtained from the superficially different BF term of $E_{8(8)}$ EFT. This uses a generalised version of the mechanism described by Mukhi and Papageorgakis \cite{Mukhi:2008ux} in the context of Bagger-Lambert theory for converting scalar-Chern-Simons theory into a Yang-Mills theory while absorbing scalar degrees of freedom.

We end in Section \ref{sec:Discussion} with a discussion of the results obtained and possible further work in this area.

\section{Preliminaries and Notation}\label{sec:Prelims}
Throughout this paper, an EFT-to-EFT reduction should be understood as a spontaneous symmetry breaking from an  $E_{n(n)}$ EFT to an $E_{n-1(n-1)}$ EFT when there is a generalised isometry present. Note that the further we compactify down from eleven dimensions, the larger the exceptional group becomes and so the usual Kaluza-Klein reduction in supergravity yields an \emph{increase} in the dimension of the exceptional group which is made manifest in the reduced theory. Here, we are going in the opposite direction and spontaneously breaking the exceptional group to a subgroup. In the supergravity literature, this would be considered as an oxidation of the supergravity theory to one dimension higher (note the conflicting terminology; a reduction of the exceptional group corresponds to an oxidation of the spacetime dimension). A useful paper covering aspects of oxidation in supergravity before the ExFT programme is \cite{Keurentjes:2002xc}.\par

In this paper, wherever we need to differentiate between two EFTs, we shall refer to the larger ExFT (the theory with the larger group $\hat{G}$, though \emph{smaller} external space) as the `parent' theory and adorn all objects/indices in that ExFT with hats $(\widehat{\hphantom{G}})$ to distinguish them from the analogous structure in the `child' theory (whose associated group we denote as $G \subset \hat{G}$). However, when we speak in generality (as we shall for this section) we shall drop any hats to prevent cluttering the formulae. Hopefully, there should be no ambiguity in doing so.\par
We begin our description of EFTs by first splitting the 11 coordinates into $d$ `external coordinates' and $n =11-d$ `internal coordinates'. All of the local transformations of the supergravity fields of 11-dimensional supergravity, compactified on $T^n$, (namely diffeomorphisms and $p$-form gauge transformations) collectively generate a hidden local $G= E_{n(n)}$ action and we wish to construct a theory that makes this local symmetry manifest.\par
We linearly realise the action of $G$ by extending the internal coordinates $y^m$ to a set of extended coordinates $Y^M$ where $M =1, \ldots , \operatorname{dim} \mathcal{R}_1$ and $\mathcal{R}_1$ is a particular representation (which we call the \emph{coordinate representation}) of $G$. When these extra coordinates are taken to be toroidal, they can be interpreted as dual to the wrapping modes of the branes of M-theory. Upon this extended spacetime, we define a generalised Lie derivative which generates these generalised diffeomorphisms. Its action on a generalised vector $V^M$ (of weight $\lambda_V$) is taken to be of the form\footnote{This has to be slightly modified for $n=8$ which we shall discuss in Section \ref{sec:ReductionSection}.}
\begin{align}
\mathbb{L}_U V^M = U^N \partial_N V^M - V^N \partial_N U^M + Y^{MN}{}_{PQ} \partial_N U^P V^Q + (\lambda_V + \omega) \partial_N U^N V^M\,,
\end{align}
which can be loosely thought of as a modification of a Lie derivative (the first two terms) by a $Y$-tensor that is formed from $G$-invariants. Additionally, the weight term is modified by an extra \emph{universal weight} $\omega = - \frac{1}{9-n}$. Unlike the conventional Lie derivative, we require a closure constraint of the form
\begin{align}
{\left[ \mathbb{L}_U , \mathbb{L}_V \right]} = \mathbb{L}_{\llbracket U, V \rrbracket}\,,
\end{align}
where
\begin{align}
\llbracket U, V \rrbracket \coloneqq \frac{1}{2} \left( \mathbb{L}_{U} V - \mathbb{L}_V U \right)
\end{align}
is called the \emph{E-bracket} and is the antisymmetric part of the generalised Lie derivative (note that unlike the conventional Lie derivative, the generalised Lie derivative does not have a definite symmetry). This imposes constraints on the $Y$-tensor and, together with requiring $G$-invariance, fixes the form of the $Y$-tensor uniquely. These have been worked out for all the finite cases\footnote{See also \cite{Bossard:2017aae} for $n=9$.} $3 \leq n \leq 8$ in \cite{Berman:2012vc,Berman:2015rcc,Hohm:2015xna}. Additionally, closure requires we impose the following \emph{section constraint}
\begin{align}
Y^{MN}{}_{PQ} \partial_M \otimes \partial_N = 0\,,
\end{align}
where the notation above is understood to mean that the partial derivatives act either on the same field or different fields. There are only two inequivalent solutions (up to $G$-transformations) to the section condition, namely the M-theory solution and the Type IIB solution. Note that Type IIA naturally appears from the former upon assuming an extra isometry. The fields of the theory are
\begin{align}
\{\mathcal{M}_{MN}, g_{\mu \nu}, \mathcal{A}_\mu{}^M,\ldots\}\,.
\end{align}
$\mathcal{M}_{MN}$ is a generalised metric parametrising the coset $G/H$ (where $H$ is the maximal compact subgroup of $G$) and is usually given in terms of the supergravity fields in a Borel gauge. Additionally, $g_{\mu \nu}$ is a metric on the external space and $\mathcal{A}_{\mu}{}^M$ is the first level of a tensor hierarchy  that is to be modified by higher degree $p$-form fields (represented by the ellipsis) as required. The latter acts as a gauge field for the generalised Lie derivative. The dynamics of the EFT are constructed from terms that are invariant under both the (internal) generalised diffeomorphisms and external (conventional) diffeomorphisms and requiring reduction to the action of 11-dimensional supergravity upon imposing the section constraint.\par
DFT is comparatively simpler\footnote{Here we consider only the NS-NS sector, although R-R fields can be incorporated into the framework as well. See, for example, \cite{Hohm:2011dv, Jeon:2012kd}.}; the entire spacetime is doubled to fit a $\mathcal{R}_1 = \mathbf{2D}$ representation of $G = \operatorname{O}(D,D)$. There is no `external space' to speak of and so we require neither an external metric nor a tensor hierarchy of generalised gauge fields. Consequently, the only dynamical fields are a generalised metric $\mathcal{M}_{MN}$, parametrising the coset $\operatorname{O}(D,D)/(\operatorname{O}(D) \times \operatorname{O}(D))$ in terms of $g$ and $B_{(2)}$, and the DFT dilaton $d$ that is related to the supergravity dilaton $\varphi$ by a shift. The action is again constructed as the unique, 2-derivative, $\operatorname{O}(D,D)$-invariant action that reduces to the action of Type II supergravity upon solving the section condition. The generalised Lie derivative is also simpler; the $Y$-tensor is symmetric in both its upper and lower indices, given in terms of the $\operatorname{O}(D,D)$ structure $\eta$ as $Y^{MN}{}_{PQ} = \eta^{MN} \eta_{PQ}$, and there is no universal weight in DFT. Additionally, there is only one inequivalent solution to the DFT section condition.
\section{Reduction of the Generalised Metric}\label{sec:ReductionGenMetric}
The generalised metric of any ExFT is a representative of the coset $G/H$ and encodes the scalar degrees of freedom of the theory. The form of the generalised metric depends on the theory but the generalised metric for each EFT (in a Borel gauge) has been known for a while and can be found in \cite{Lee:2016qwn,Berman:2011jh}. Note that one could also work with the Vielbein and examine a reduction of the Vielbein (from which one of course could work out the generalised metric reduction). This approach has been considered for the case of $E7$ to $E6$ in the paper \cite{Bossard:2014lra}. \par
Here, we shall consider two cases; a reduction of the DFT generalised metric and the reduction of the $\operatorname{SL}(5)$ generalised metric. The first will be useful as an illustrative example of what a reduction may look like in ExFT and agrees with previous results in the area. We will then explicitly reduce the $\operatorname{SL}(5)$ generalised metric and show how both the $\operatorname{O}(3,3)$ generalised metric and the $\operatorname{SL}(3) \times \operatorname{SL}(2)$ generalised metric can both be obtained from the same circle reduction, just with different reduction ansatzes. The reduction to the $\operatorname{O}(3,3)$ DFT matches that described in \cite{Thompson:2011uw,Blair:2018lbh,Berman:2019izh}.
\subsection{\texorpdfstring{$\operatorname{O}(D,D)$}{O(D,D)} DFT to \texorpdfstring{$\operatorname{O}(d,d) \times \operatorname{O}(n,n)$}{O(d,d)xO(n,n)} DFT}
Our starting point is the $\hat{G} = \operatorname{O}(D,D)$ DFT generalised metric
\begin{align}
{\hat{\mathcal{M}}}_{\hat{M} \hat{N}} & = \begin{pmatrix}
{\hat{g}}_{\hat{m} \hat{n}} - {\hat{B}}_{\hat{m} \hat{p}} {\hat{g}}^{\hat{p} \hat{q}} {\hat{B}}_{\hat{q} \hat{n}} & {\hat{B}}_{\hat{m} \hat{p}} {\hat{g}}^{\hat{p} \hat{n}}\\
- {\hat{g}}^{\hat{m} \hat{p}} {\hat{B}}_{\hat{p} \hat{n}} & {\hat{g}}^{\hat{m} \hat{n}}
\end{pmatrix}\,,
\end{align}
where $\hat{M},\hat{N} = ({}^{\hat{m}}, {}_{\hat{m}}) = 1, \ldots,2D$ and $\hat{m}, \hat{n} = 1, \ldots, D$. We now consider a Kaluza-Klein (KK) decomposition of the underlying fields according to
\begin{align}
{\hat{g}}_{\hat{m} \hat{n}} & = \begin{pmatrix}
e^{2\alpha \phi} g_{\mu \nu} + e^{2\beta \phi} A_\mu{}^m g_{mn} A^n{}_\nu & e^{2\beta \phi} A_\mu{}^m g_{mn}\\
e^{2 \beta \phi} g_{mn} A^n{}_\nu & e^{2\beta \phi} g_{mn}\\
\end{pmatrix}\label{eq:KKMetric}\,,\\
{\hat{g}}^{\hat{m} \hat{n}} & = \begin{pmatrix}
e^{-2\alpha \phi} g^{\mu \nu} & - e^{-2 \alpha \phi} g^{\mu \nu} A_\nu{}^n\\
- e^{-2 \alpha \phi} A^m{}_\mu g^{\mu \nu} & e^{-2\alpha \phi} A^m{}_\mu g^{\mu \nu} A_\nu{}^n + e^{-2\beta \phi} g^{mn}
\end{pmatrix}\label{eq:KKInverse}\,,\\
{\hat{B}}_{\hat{m} \hat{n}} & = \begin{pmatrix}
B_{\mu \nu} + B_{\mu n} A^n{}_\nu + A_\mu{}^m B_{m \nu} + A_\mu{}^m B_{mn} A^n{}_\nu & B_{\mu n} + A_\mu{}^m B_{mn}\\
B_{m \nu} + B_{mn} A^n{}_\nu & B_{mn}
\end{pmatrix}\,,
\end{align}
where we have now decomposed the index $\hat{m}$ to $\hat{m} = (\mu, m)$ with ranges $\mu = 1, \ldots, d$ and $m = 1, \ldots, n$ such that $d+n = D$. One finds that this can be reorganised into a generalised Kaluza-Klein ansatz for ${\hat{\mathcal{M}}}_{\hat{M} \hat{N}}$ according to
\begin{align}\label{eq:GeneralisedKK}
{\hat{\mathcal{M}}}_{\hat{M} \hat{N}} & = \begin{pmatrix}
\mathcal{M}_{MN} + \mathcal{A}_M{}^A \mathcal{G}_{AB} \mathcal{A}^B{}_N &  {\mathcal{A}}_M{}^A \mathcal{G}_{AB} + \mathcal{M}_{MN} \mathcal{B}^N{}_B\\
\mathcal{G}_{AB} \mathcal{A}^B{}_N + \mathcal{B}_A{}^M \mathcal{M}_{MN} & \mathcal{G}_{AB} + \mathcal{B}_A{}^M \mathcal{M}_{MN} \mathcal{B}^N{}_B
\end{pmatrix}\,,
\end{align}
where
\begin{align}
\mathcal{M}_{MN} & = \begin{pmatrix}
e^{2\alpha \phi} g_{\mu \nu} - e^{-2\alpha \phi} (B_{\mu \sigma} + A_\mu{}^m B_{m \sigma}) g^{\sigma \rho} ( B_{\rho \nu} + B_{\rho n} A^n{}_\nu) & e^{-2\alpha \phi} (B_{\mu \sigma} + A_\mu{}^m B_{m \sigma}) g^{\sigma \nu}\\
- e^{-2\alpha \phi} g^{\mu \sigma} ( B_{\sigma \nu} + B_{\sigma n} A^n{}_\nu) & e^{-2\alpha \phi} g^{\mu \nu}\\
\end{pmatrix}\\
\mathcal{G}_{AB} & = \begin{pmatrix}
e^{2 \beta \phi} g_{mn} - e^{-2 \beta \phi} B_{mp} g^{pq} B_{qn} & e^{-2 \beta \phi} B_{mp} g^{pn}\\
- e^{-2 \beta \phi} g^{mp} B_{pn} & e^{-2\beta \phi} g^{mn}
\end{pmatrix}\\
{\mathcal{A}}_M{}^A & = \begin{pmatrix} A_\mu{}^m & B_{\mu m} \\ 0 & 0 \end{pmatrix}, \qquad
{\mathcal{A}}^A{}_M = \begin{pmatrix} A^m{}_\mu & 0 \\ - B_{m \mu} & 0 \end{pmatrix} = {\left( {\mathcal{A}}_M{}^A \right)}^T\\
{\mathcal{B}}^M{}_A & = \begin{pmatrix} 0 & 0 \\ - B_{\mu m} & - A_\mu{}^m \end{pmatrix}, \qquad
{\mathcal{B}}_A{}^M = \begin{pmatrix} 0 & B_{m \mu} \\ 0 & - A^m{}_\mu \end{pmatrix} = {\left( {\mathcal{B}}^M{}_A \right)}^T\,.
\end{align}
The new indices are DFT-type doubled indices given by ${}_M = ({}_\mu, {}^\mu)$, ${}_A = ({}_m, {}^m)$. In effect, we have split the parent DFT coordinates into two sets of doubled coordinates ${\hat{Y}}^{\hat{M}} = (Y^M, Y^A)$ for which we recognise $\mathcal{M}_{MN}$ as a generalised metric on the $Y^M$ space (constructed from a metric $e^{2\alpha \phi} g_{\mu \nu}$ and 2-form $B_{\mu \nu} + A_\mu{}^m B_{m \nu}$) and $\mathcal{G}_{AB}$ as a generalised metric on the $Y^A$ space (itself constructed from a metric $e^{2\beta \phi} g_{mn}$ and 2-form $B_{mn}$). Finally, $\mathcal{A}$ and $\mathcal{B}$ take on the roles of generalised Kaluza-Klein vectors, though not in the standard form \eqref{eq:KKMetric} but rather in a manner that treats the internal DFT and external DFT on an equal footing \eqref{eq:GeneralisedKK} by introducing extra $\mathcal{B}$-twisted terms. As such, we shall refer to this type of ansatz as a \emph{generalised Kaluza-Klein ansatz}. We shall comment on this again when we consider the reduction of the $\operatorname{SL}(5)$ the generalised metric next.\par
Note that if we split the indices of the $\operatorname{O}(D,D)$ structure in the same way, we obtain
\begin{align}
{\hat{\eta}}_{\hat{M} \hat{N}} & = \begin{pmatrix}
\eta_{MN} & 0\\
0 & \eta_{AB}	
\end{pmatrix}
\end{align}
with
\begin{align}
\eta_{MN} & = \begin{pmatrix} 0 & \delta_\mu^\nu\\ \delta^\mu_\nu & 0\end{pmatrix}\,, \qquad \eta_{AB} = \begin{pmatrix} 0 & \delta_m^n\\ \delta^m_n & 0 \end{pmatrix}\,.
\end{align}
Then the generalised Kaluza-Klein vectors can be related to each other as
\begin{align}
\eta_{AB} \mathcal{A}^B{}_N \eta^{NM} = - \mathcal{B}_A{}^M\,,
\end{align}
and it is then tempting to write this in term of the Y-tensor as follows:
\begin{align}
{\hat{Y}}^{MN}{}_{BA} {\mathcal{A}}^B{}_N = - {\mathcal{B}}_A{}^M\,.
\end{align}
Then this gives us the general KK ansatz for any ExFT and the usual KK ansatz can then be understood as a particular case since the $Y$-tensor vanishes for $\hat{G} = \operatorname{GL}(d)$ in GR (when viewed an ExFT with no external space at all), causing all of the extra $\mathcal{B}$-twisted terms to drop out. The fact that $\mathcal{A}$ and $\mathcal{B}$ are related is also consistent with the counting of degrees of freedom: ${\hat{\mathcal{M}}}_{\hat{M} \hat{N}}$ parametrises the coset $\operatorname{O}(D,D) / (\operatorname{O}(D) \times \operatorname{O}(D))$ and thus has $D^2 = {(n+ d)}^2 = n^2 + d^2 + 2 nd$ components. The $d^2$ and $n^2$ components enter into $\mathcal{M}_{MN} \in \operatorname{O}(d,d) / (\operatorname{O}(d) \times \operatorname{O}(d) )$ and $\mathcal{G}_{AB} \in \operatorname{O}(n,n) / (\operatorname{O}(n) \times \operatorname{O}(n))$ respectively and the $2nd$ components entered into $\mathcal{A}_M{}^A \sim \mathcal{B}^M{}_A \sim (A_\mu{}^m, B_{\mu m})$.\par

In order to invert this generalised KK ansatz, we note that we may diagonalise the matrix ${\hat{\mathcal{M}}}_{\hat{M} \hat{N}}$ as follows:
\begin{align}
{\hat{\mathcal{M}}}_{\hat{P} \hat{Q}} = {\hat{\mathcal{E}}}_{\hat{P}}{}^{\hat{M}} {\tilde{\mathcal{M}}}_{\hat{M} \hat{N}} {\hat{\mathcal{E}}}^{\hat{N}}{}_{\hat{Q}}
\end{align}
where
\begin{gather}
{\tilde{\mathcal{M}}}_{\hat{M} \hat{N}} = \begin{pmatrix}
\mathcal{M}_{MN} & 0 \\
0 & \mathcal{G}_{AB}
\end{pmatrix}\,,\\
{\hat{\mathcal{E}}}_{\hat{P}}{}^{\hat{M}} = \begin{pmatrix}
\delta^M_P & \mathcal{A}_P{}^A\\
\mathcal{B}_C{}^M & \delta^A_C
\end{pmatrix}, \qquad 
{\hat{\mathcal{E}}}^{\hat{N}}{}_{\hat{Q}} = \begin{pmatrix}
\delta^N_Q & \mathcal{B}^N{}_D\\
\mathcal{A}^B{}_Q & \delta^B_D
\end{pmatrix}\,.
\end{gather}
We then proceed by noting that the combinations $\mathcal{A}^A{}_M \mathcal{B}^M{}_B$ and $\mathcal{B}_A{}^M \mathcal{A}_M{}^B$ both vanish such that the inverses of ${\hat{\mathcal{A}}}_{\hat{M}}{}^{\hat{P}}$ and ${\hat{\mathcal{A}}}^{\hat{N}}{}_{\hat{Q}}$ are given by
\begin{align}
{\left({\hat{\mathcal{E}}}^{-1} \right)}_{\hat{P}}{}^{\hat{N}} & = \begin{pmatrix}
\delta_P^N + \mathcal{A}_P{}^B \mathcal{B}_B{}^N & - \mathcal{A}_P{}^B\\
- \mathcal{B}_C{}^N & \delta^B_C \end{pmatrix}\,, \qquad
{\left( {\hat{\mathcal{E}}}^{-1} \right)}^{\hat{Q}}{}_{\hat{R}} = \begin{pmatrix}
\delta^Q_R + \mathcal{B}^Q{}_C \mathcal{A}^C{}_R & - \mathcal{B}^Q{}_E\\
- \mathcal{A}^D{}_R & \delta^D_E
\end{pmatrix}\,.
\end{align}
The inverse of the generalised KK ansatz is then given by
\begingroup
\renewcommand*{\arraystretch}{4}
\begin{align}
{\hat{\mathcal{M}}}^{\hat{M} \hat{N}} & \coloneqq {\left( {\hat{\mathcal{E}}}^{-1} \right)}^{\hat{M}}{}_{\hat{P}} {\left( {\tilde{\mathcal{M}}}^{-1} \right)}^{\hat{P} \hat{Q}}  {\left( {\hat{\mathcal{E}}}^{-1} \right)}_{\hat{Q}}{}^{\hat{N}}\\
& = \begin{pmatrix} \label{eq:InverseGeneralisedKK}
\begingroup
\renewcommand*{\arraystretch}{1.5}
\begin{matrix}
\mathcal{M}^{MN} +  \mathcal{B}^M{}_E \mathcal{A}^E{}_P {\mathcal{M}}^{PQ} \mathcal{A}_Q{}^F \mathcal{B}_F{}^N\\
\mathcal{B}^M{}_E \mathcal{A}^E{}_P \mathcal{M}^{PN} + {\mathcal{M}}^{MQ} \mathcal{A}_Q{}^F \mathcal{B}_F{}^N\\
+ \mathcal{B}^M{}_C \mathcal{G}^{CD} \mathcal{B}_D{}^N \end{matrix}
\endgroup
&
\begingroup
\renewcommand*{\arraystretch}{1.5}
\begin{matrix} -  \left( \delta^M_P + \mathcal{B}^M{}_E \mathcal{A}^E{}_P \right) \mathcal{M}^{PQ} \mathcal{A}_Q{}^B\\
- \mathcal{B}^M{}_C \mathcal{G}^{CB}
\end{matrix}
\endgroup
\\
\begingroup
\renewcommand*{\arraystretch}{1.5}
\begin{matrix}
- \mathcal{A}^A{}_P \mathcal{M}^{PQ} \left( \delta_Q^N + \mathcal{A}_Q{}^F \mathcal{B}_F{}^N \right)\\
- \mathcal{G}^{AD} \mathcal{B}_D{}^N
\end{matrix}
\endgroup
&
\mathcal{A}^A{}_P \mathcal{M}^{PQ} \mathcal{A}_Q{}^B + \mathcal{G}^{AB}
\end{pmatrix}\,.
\end{align}
\endgroup
The parametrisation that we obtain here matches the one obtained in \cite{Hohm:2013nja}, in which they constructed a tensor hierarchy for DFT (essentially enhancing it to a full EFT with group $\operatorname{O}(D,D)$) which, in turn, was found to be consistent with results from the heterotic theory. We have also constructed the second generalised metric explicitly. So as to compare our results with those in the literature we write their generalised vector as, $A_\mu{}^A[\text{HS}]$ in terms of the fields here as
\begin{align}
A_\mu{}^A[\text{HS}] = \begin{pmatrix}
A_\mu{}^m [\text{here}]\\
- B_{\mu m} [\text{here}]
\end{pmatrix}
\end{align}
and their generalised metric $\mathcal{M}_{MN}[\text{HS}]$ is should be thought of as our $\mathcal{G}_{AB}$, parametrised in terms of the internal components $g_{mn}[\text{here}]$ and $B_{mn} [\text{here}]$. However, the 2-form that they introduce as part of the tensor hierarchy is related to ours by field redefinitions: we have
\begin{align}
C_{\mu \nu}[\text{HS}] = B_{\mu \nu}[\text{HS}] + \frac{1}{2} A_\mu{}^A [\text{HS}] A_{\nu A}[\text{HS}] = B_{\mu \nu} [\text{here}] +  A_\mu{}^m [\text{here}] B_{m \nu}[\text{here}]
\end{align}
and so $B_{\mu \nu}[\text{HS}] = B_{\mu \nu}[\text{here}] + A_{[\mu|}{}^m [\text{here}]B_{m| \nu]} [\text{here}]$. The crucial difference between these two constructions is the location of the isometries. In the canonical Kaluza-Klein set-up, one must take the internal space (with coordinates $Y^m$) to be an isometry such that all the fields transform covariantly under the symmetries of the reduced theory. When lifting this ansatz to the doubled spacetime, one must presumably require the entire doubled space $Y^A = (Y^m, Y_m)$ to be an isometry such that any DFT frame will have the same number of isometries. However, this differs from the coordinate dependence of \cite{Hohm:2013nja} where the fields are allowed to depend on (translating into our notation) $Y^m$ and $Y_m$. 
\subsection{Reduction of the \texorpdfstring{$\operatorname{SL}(5)$}{SL(5)} Generalised Metric}
We now turn to the reduction of the $\hat{G} = \operatorname{SL}(5)$ generalised metric which is given by
\begin{align}
{\hat{\mathcal{M}}}_{\hat{M}\hat{N}} = \begin{pmatrix}
{\hat{g}}_{\hat{m} \hat{n}} + \frac{1}{2} {\hat{C}}_{\hat{m} \hat{k} \hat{l}} {\hat{g}}^{\hat{k} \hat{l}, \hat{p} \hat{q}} {\hat{C}}_{\hat{p} \hat{q} \hat{n}} & \frac{1}{\sqrt{2}} {\hat{C}}_{\hat{m} \hat{k} \hat{l}} {\hat{g}}^{\hat{k} \hat{l}, {\hat{n}_1 {\hat{n}}_2}}\\
\frac{1}{\sqrt{2}} {\hat{g}}^{{\hat{m}}_1 {\hat{m}}_2, \hat{p} \hat{q}} {\hat{C}}_{\hat{p} \hat{q} \hat{n}} & {\hat{g}}^{{\hat{m}}_1 {\hat{m}}_2, {\hat{n}}_1 {\hat{n}}_2}
\end{pmatrix}\,,
\end{align}
where ${\hat{g}}^{\hat{m}\hat{n}, \hat{p} \hat{q}} = \frac{1}{2} ( {\hat{g}}^{\hat{m} \hat{p}} {\hat{g}}^{\hat{q} \hat{n}} - {\hat{g}}^{\hat{m} \hat{q}} {\hat{g}}^{\hat{p} \hat{n}})$ and $\hat{m}, \hat{n}= 1, \ldots, 4$ such that $\hat{M} = 1, \ldots, 10$. We consider the reduction of this metric to the generalised metric of both $\operatorname{O}(3,3)$ DFT and $\operatorname{SL}(3) \times \operatorname{SL}(2)$ EFT. The first of these was explored in \cite{Thompson:2011uw}. Here we give an equivalent description that facilitates the comparison with the reduction to the $\operatorname{SL}(3) \times \operatorname{SL}(2)$ generalised metric.\par
We split the index $\hat{m} = (m, z)$ with $m = 1,2,3$ for the KK ansatz
\begin{align}
{\hat{g}}_{\hat{m} \hat{n}} & = \begin{pmatrix}
e^{2\alpha \phi} g_{mn} + e^{2\beta \phi} A_m A_n & e^{2\beta \phi} A_m\\
e^{2 \beta \phi} A_n & e^{2\beta \phi}
\end{pmatrix}\label{eq:CircleReduction}\,,\\
{\hat{g}}^{\hat{m} \hat{n}} & = \begin{pmatrix}
e^{-2\alpha \phi} g^{mn} & -  e^{-2 \alpha \phi} A^m\\
- e^{- 2 \alpha \phi} A^n & e^{-2 \alpha \phi} A^m A_m + e^{- 2 \beta \phi}
\end{pmatrix}\label{eq:InverseCircleReduction}\,,\\
{\hat{C}}_{\hat{m} \hat{n} \hat{p}} & = \begin{pmatrix}
C_{m n p} + 3 B_{[m n} A_{p]}\\
3B_{mn}
\end{pmatrix}\,,
\end{align}
which induces the reduction of the $\operatorname{SL}(5)$ generalised metric
\begin{align}
{\hat{\mathcal{M}}}_{\hat{M}\hat{N}} = \begin{pmatrix}
{\hat{\mathcal{M}}}_{m n} & {\hat{\mathcal{M}}}_{m z} & {\hat{\mathcal{M}}}_{m}{}^{n_1 n_2} & {\hat{\mathcal{M}}}_{m}{}^{n_1 z}\\
{\hat{\mathcal{M}}}_{z n} & {\hat{\mathcal{M}}}_{z z} & {\hat{\mathcal{M}}}_{z}{}^{n_1 n_2} & {\hat{\mathcal{M}}}_{z}{}^{n_1 z}\\
{\hat{\mathcal{M}}}^{m_1 m_2}{}_n & {\hat{\mathcal{M}}}^{m_1 m_2}{}_z & {\hat{\mathcal{M}}}^{m_1 m_2, n_1 n_2} & {\hat{\mathcal{M}}}^{m_1 m_2, n_1 z}\\
{\hat{\mathcal{M}}}^{m_1 z}{}_n & {\hat{\mathcal{M}}}^{m_1 z}{}_z & {\hat{\mathcal{M}}}^{m_1 z, n_1 n_2} &{\hat{\mathcal{M}}}^{m_1 z, n_1 z}\\
\end{pmatrix}\,,
\end{align}
where
\begin{align}
{\hat{\mathcal{M}}}_{mn} & = \begin{array}[t]{l}
e^{2\alpha \phi} g_{mn} + e^{2\beta \phi} A_m A_n - 9 e^{-2(\alpha + \beta) \phi} B_{mp} g^{pq} B_{qn}\\
\qquad + \frac{1}{2} e^{-4 \alpha \phi} \left( C_{m k l} - 3 B_{[m k} A_{l]} \right) g^{k l, p q} \left( C_{p q n} - 3 A_{[p} B_{q n]} \right)
\end{array}\\
{\hat{\mathcal{M}}}_{mz} & = \frac{3}{2} e^{-4 \alpha \phi} (C_{m k l} - 3 B_{[m k} A_{l]} )g^{k l, p q} B_{p q} + e^{2 \beta \phi} A_m\\
{\hat{\mathcal{M}}}_m{}^{n_1 n_2} & = \frac{1}{\sqrt{2}} e^{-4 \alpha \phi} (C_{m k l} - 3 B_{[m k} A_{l]} ) g^{k l, n_1 n_2}\\
{\hat{\mathcal{M}}}_m{}^{n_1 z} & = \frac{1}{\sqrt{2}} e^{-4 \alpha \phi} (C_{m k l} - 3 B_{[m k} A_{l]}) g^{k l, p n_1} A_{p} + \frac{3}{\sqrt{2}} e^{-2(\alpha + \beta)} B_{m p} g^{p n_1}\\
{\hat{\mathcal{M}}}_{zn} & = \frac{3}{2} e^{-4 \alpha \phi} B_{k l} g^{k l, p q} (C_{p q n} - 3 B_{[p q} A_{n]}) + e^{2\beta \phi} A_n\\
{\hat{\mathcal{M}}}_{zz} & = e^{2\beta \phi} + \frac{9}{2} e^{-4\alpha \phi} B_{k l} g^{k l, p q} B_{p q}\\
{\hat{\mathcal{M}}}_z{}^{n_1 n_2} & = \frac{3}{\sqrt{2}} e^{-4 \alpha \phi} B_{p q} g^{p q, n_1 n_2}\\
{\hat{\mathcal{M}}}_z{}^{n_1 z} & = \frac{3}{\sqrt{2}} e^{-4\alpha \phi} B_{k l} A_p g^{k l, p n_1}\\
{\hat{\mathcal{M}}}^{m_1 m_2}{}_n & = \frac{1}{\sqrt{2}} e^{-4\alpha \phi}  g^{m_1 m_2, p q} (C_{p q n} - 3 A_{[p} B_{q n]})\\
{\hat{\mathcal{M}}}^{m_1 m_2}{}_z & = \frac{3}{\sqrt{2}} e^{-4 \alpha \phi} g^{m_1 m_2, k l} B_{k l}\\
{\hat{\mathcal{M}}}^{m_1 m_2, n_1 n_2} & = e^{-4\alpha \phi} g^{m_1 m_2, n_1 n_2}\\
{\hat{\mathcal{M}}}^{m_1 m_2, n_1 z} & = e^{-4 \alpha \phi} g^{m_1 m_2, p n_1} A_p\\
{\hat{\mathcal{M}}}^{m_1 z}{}_n & = - \frac{1}{\sqrt{2}} e^{-4 \alpha \phi} g^{m_1 k, p q} A_k( C_{p q n} - 3B_{[p q} A_{n]}) -\frac{3}{\sqrt{2}} e^{-2 (\alpha + \beta) \phi} g^{m_1 p} B_{p n}\\
{\hat{\mathcal{M}}}^{m_1 z}{}_z & = - \frac{3}{\sqrt{2}} e^{-4\alpha} g^{m_1 k, p q} A_k B_{p q}\\
{\hat{\mathcal{M}}}^{m_1 z, n_1 n_2} & =  - e^{-4 \alpha \phi} A_k g^{m_1 k, n_1 n_2}\\
{\hat{\mathcal{M}}}^{m_1 z, n_1 z} & = \frac{1}{2} e^{-2(\alpha  + \beta) \phi} g^{m_1 n_1} - \frac{1}{2} e^{-4\alpha} A_k g^{m_1 k, p n_1} A_p
\end{align}
We now have two possible reductions, depending on which components we choose to form the reduced generalised metric from\footnote{In principle, one should also be able to do the same for the DFT case considered above.}. In the following, the pieces that are identified with the components that lie wholly in the lower-dimensional section ${\hat{\mathcal{M}}}_{MN}$ and the components that lie wholly outside of the lower-dimensional section ${\hat{\mathcal{M}}}_{AB}$ are boxed in red and green respectively. For $\operatorname{SL}(5) \rightarrow \operatorname{SL}(3) \times \operatorname{SL}(2)$ we choose
\begin{equation}\label{eq:SL3SL2Reduction}
{\hat{\mathcal{M}}}_{\hat{M} \hat{N}} =
\begin{tikzpicture}[baseline=(current bounding box.center)]
  \matrix (m)[
    matrix of math nodes,
    nodes in empty cells,
    left delimiter={(},
    right delimiter={)},
    minimum width=2.3cm,
    minimum height=0.6cm,
  ] {
{\hat{\mathcal{M}}}_{m n} &
{\hat{\mathcal{M}}}_{m z} &
{\hat{\mathcal{M}}}_{m}{}^{n_1 n_2} &
{\hat{\mathcal{M}}}_{m}{}^{n_1 z}\\
{\hat{\mathcal{M}}}_{z n} &
{\hat{\mathcal{M}}}_{z z} &
{\hat{\mathcal{M}}}_{z}{}^{n_1 n_2} &
{\hat{\mathcal{M}}}_{z}{}^{n_1 z}\\
{\hat{\mathcal{M}}}^{m_1 m_2}{}_n &
{\hat{\mathcal{M}}}^{m_1 m_2}{}_z &
{\hat{\mathcal{M}}}^{m_1 m_2, n_1 n_2} &
{\hat{\mathcal{M}}}^{m_1 m_2, n_1 z}\\
{\hat{\mathcal{M}}}^{m_1 z}{}_n &
{\hat{\mathcal{M}}}^{m_1 z}{}_z &
{\hat{\mathcal{M}}}^{m_1 z, n_1 n_2} &
{\hat{\mathcal{M}}}^{m_1 z, n_1 z}\\
  } ;
  \draw[draw=red] (m-1-1.south west) rectangle (m-1-1.north east);
  \draw[draw=red] (m-3-1.south west) rectangle (m-3-1.north east);
  \draw[draw=red] (m-1-3.south west) rectangle (m-1-3.north east);
  \draw[draw=red] (m-3-3.south west) rectangle (m-3-3.north east);
  \draw[draw=green] (m-2-2.south west) rectangle (m-2-2.north east);
  \draw[draw=green] (m-4-2.south west) rectangle (m-4-2.north east);
  \draw[draw=green] (m-2-4.south west) rectangle (m-2-4.north east);
  \draw[draw=green] (m-4-4.south west) rectangle (m-4-4.north east);
\end{tikzpicture}\,,
\end{equation}
whilst for $\operatorname{SL}(5) \rightarrow \operatorname{O}(3,3)$, we choose
\begin{equation}\label{eq:O33Reduction}
{\hat{\mathcal{M}}}_{\hat{M} \hat{N}} =
\begin{tikzpicture}[baseline=(current bounding box.center)]
  \matrix (m)[
    matrix of math nodes,
    nodes in empty cells,
    left delimiter={(},
    right delimiter={)},
    minimum width=2.3cm
  ] {
{\hat{\mathcal{M}}}_{m n} &
{\hat{\mathcal{M}}}_{m z} &
{\hat{\mathcal{M}}}_{m}{}^{n_1 n_2} &
{\hat{\mathcal{M}}}_{m}{}^{n_1 z}\\
{\hat{\mathcal{M}}}_{z n} &
{\hat{\mathcal{M}}}_{z z} &
{\hat{\mathcal{M}}}_{z}{}^{n_1 n_2} &
{\hat{\mathcal{M}}}_{z}{}^{n_1 z}\\
{\hat{\mathcal{M}}}^{m_1 m_2}{}_n &
{\hat{\mathcal{M}}}^{m_1 m_2}{}_z &
{\hat{\mathcal{M}}}^{m_1 m_2, n_1 n_2} &
{\hat{\mathcal{M}}}^{m_1 m_2, n_1 z}\\
{\hat{\mathcal{M}}}^{m_1 z}{}_n &
{\hat{\mathcal{M}}}^{m_1 z}{}_z &
{\hat{\mathcal{M}}}^{m_1 z, n_1 n_2} &
{\hat{\mathcal{M}}}^{m_1 z, n_1 z}\\
  } ;
  \draw[draw=red] (m-1-1.north west) rectangle (m-1-1.south east);
  \draw[draw=red] (m-2-3.north east) rectangle (m-1-4.north east);
  \draw[draw=red] (m-4-1.south west) rectangle (m-4-1.north east);
  \draw[draw=red] (m-3-3.south east) rectangle (m-4-4.south east);
  \draw[draw=green] (m-1-1.south east) rectangle (m-3-3.south east);
  \draw[draw=green] (m-1-3.south west) -- (m-3-3.south west);
  \draw[draw=green] (m-2-2.south west) -- (m-2-3.south east);
\end{tikzpicture}\,.
\end{equation}
In both cases we have suppressed the components that enter into the off-diagonal pieces but they should hopefully be clear from the above: each piece takes one component from each quadrant of ${\hat{\mathcal{M}}}_{\hat{M} \hat{N}}$. We shall use the indices $M, N$ to index the coordinate representation of the reduced theory and $A,B$ to index the remaining coordinates. For the $\operatorname{SL}(3) \times \operatorname{SL}(2)$ reduction, we have $Y^M = (Y^m, Y_{m_1 m_2})$ whilst for the $\operatorname{O}(3,3)$ reduction we have $Y^M = (Y^m ,Y_{m_1 z})$.\par
Note that we have introduced a slight abuse of terminology; the direction $z$ plays different roles in the two reductions since it corresponds to the M-theory circle in the $\operatorname{O}(3,3)$ reduction but to the decompactification of one of the directions of the M-theory 4-torus in oxidising $d=7$ to $d=8$. It is thus perhaps better to think of $z$ as just some `distinguished' direction rather than a compactification circle. The shift in perspective is indicated by the change in identification of the dilaton below since the fixing of the dilaton determines the particular embedding of the $\operatorname{SL}(3) \subset \operatorname{SL}(5)$.
\subsubsection{\texorpdfstring{$\operatorname{SL}(5)$}{SL(5)} EFT to \texorpdfstring{$\operatorname{SL}(3) \times \operatorname{SL}(2)$}{SL(3)xSL(2)} EFT Reduction}\label{sec:SL3xSL2GenMetric}
We begin by rescaling the whole $\operatorname{SL}(5)$ generalised metric ${\hat{\mathcal{M}}}_{\hat{M} \hat{N}}$ by ${\hat{g}}^{\frac{1}{5}} = g^{\frac{1}{5}} e^{\frac{6\alpha + 2 \beta}{5}\phi}$ to obtain a determinant 1 generalised metric. From \eqref{eq:SL3SL2Reduction}, we rearrange the components of the $\operatorname{SL}(5)$ generalised metric into blocks that expose the underlying structure that will become apparent in a moment:
\begin{align}
\begin{split}
{\hat{\mathcal{M}}}_{MN} & = g^{\frac{1}{5}} e^{\frac{-14\alpha + 2\beta}{5} \phi}
\begin{pmatrix}
e^{6\alpha \phi} g_{m n} + \frac{1}{2} {\tilde{C}}_{m k l} g^{k l, p q}  {\tilde{C}}_{p q n} & \frac{1}{\sqrt{2}} {\tilde{C}}_{m k l}  g^{k l, n_1 n_2}\\
\frac{1}{\sqrt{2}}g^{m_1 m_2, p q} {\tilde{C}}_{p q n} & g^{m_1 m_2, n_1 n_2}
\end{pmatrix}\\
& \qquad + g^{\frac{1}{5}} e^{\frac{-4 \alpha -8 \beta}{5} \phi}
\begin{pmatrix}
e^{(2 \alpha + 4 \beta) \phi} A_m A_n - \frac{1}{2} {\tilde{B}}_{m p} g^{p q} {\tilde{B}}_{q n} & 0\\
0 & 0\\
\end{pmatrix}\,,
\end{split}\\
\begin{split}
{\hat{\mathcal{M}}}_{MB} & = g^{\frac{1}{5}}e^{\frac{-14\alpha + 2 \beta}{5} \phi}
\begin{pmatrix}
\frac{1}{2\sqrt{2}} {\tilde{C}}_{m k l} g^{k l, p q} {\tilde{B}}_{p q} &  \frac{1}{\sqrt{2}} {\tilde{C}}_{m k l} g^{k l, p n_1} A_p\\
\frac{1}{2} g^{m_1 m_2, p q} {\tilde{B}}_{p q} & g^{m_1 m_2, p n_1} A_p
\end{pmatrix}\\
& \qquad +g^{\frac{1}{5}} e^{\frac{-4\alpha - 8 \beta}{5}}
\begin{pmatrix}
e^{(2 \alpha + 4\beta) \phi} A_m & \frac{1}{2} {\tilde{B}}_{m p} g^{p n_1}\\
0 & 0
\end{pmatrix}\,,
\end{split}\\
\begin{split}
{\hat{\mathcal{M}}}_{AN} & = g^{\frac{1}{5}} e^{\frac{-14\alpha +2 \beta}{5}\phi}
\begin{pmatrix}
\frac{1}{2 \sqrt{2}} {\tilde{B}}_{k l} g^{k l, p q} {\tilde{C}}_{p q n} & \frac{1}{2} {\tilde{B}}_{k l} g^{k l, n_1 n_2}\\
- \frac{1}{\sqrt{2}} g^{m_1 l, p q} A_l {\tilde{C}}_{p q n} & - A_l g^{m_1 l, n_1 n_2}
\end{pmatrix}\\
& \qquad + g^{\frac{1}{5}} e^{\frac{-4\alpha - 8 \beta}{5} \phi}
\begin{pmatrix}
e^{(2 \alpha + 4 \beta) \phi} A_n & 0\\
- \frac{1}{2} g^{m k} {\tilde{B}}_{k n} & 0\\
\end{pmatrix}\,,
\end{split}\\
\begin{split}
{\hat{\mathcal{M}}}_{AB} & =g^{\frac{1}{5}} e^{\frac{-4\alpha - 8\beta}{5}}
\begin{pmatrix}
e^{(2\alpha + 4 \beta) \phi} & 0\\
0 & \frac{1}{2}  g^{m_1 n_1}
\end{pmatrix}\\
& \qquad + g^{\frac{1}{5}}e^{\frac{-14\alpha + 2 \beta}{5} \phi}
\begin{pmatrix}
\frac{1}{4} {\tilde{B}}_{k l} g^{k l, p q} {\tilde{B}}_{p q} & \frac{1}{2} {\tilde{B}}_{k l} A_p g^{k l, p n_1}\\
- \frac{1}{2} g^{m_1 l, p q} A_l {\tilde{B}}_{p q} & - A_k g^{m_1 k, p n_1} A_p
\end{pmatrix}\,,
\end{split}
\end{align}
where we have defined
\begin{align}
{\tilde{C}}_{m n p} & \coloneqq C_{m n p} - 3 B_{[m n} A_{p]}\,,\qquad {\tilde{B}}_{m n} \coloneqq 3 \sqrt{2} B_{m n}\,.
\end{align}
We can make the group structure more explicit by defining dual coordinates
\begin{align}
Y^{\bar{m}} \coloneqq \varepsilon^{m n_1 n_2} Y_{n_1 n_2}\,,
\end{align}
where $\bar{m} =1, 2,3$ indexes a distinct $\mathbf{3}$ of $\operatorname{SL}(3)$ to the first one that we indexed by $m = 1, 2,3$ (and we have thus adorned with an overbar to distinguish the two) but that is raised and lowered with the \emph{same} 3-dimensional metric such that $g_{\bar{m} \bar{n}} = g_{m n} = g_{m \bar{n}} = g_{\bar{m} n}$, in the same way that $g^{m_1 m_2, n_1 n_2}$ contains the same metric degrees of freedom as $g_{mn}$. In terms of these coordinates, we have
\begin{align}
\begin{split}
{\hat{\mathcal{M}}}_{MN} & = g^{\frac{1}{5}} e^{\frac{-14\alpha + 2\beta}{5} \phi}
\begin{pmatrix}
e^{6\alpha \phi} g_{m n} + {\tilde{C}}^2 g_{m n} & \sqrt{\frac{2}{g}} \tilde{C} g_{m \bar{n}}\\
\sqrt{\frac{2}{g}} \tilde{C} g_{\bar{m}n} & \frac{2}{g} g_{\bar{m} \bar{n}} 
\end{pmatrix}\\
& \qquad + g^{\frac{1}{5}} e^{\frac{-4 \alpha -8 \beta}{5} \phi}
\begin{pmatrix}
e^{(2 \alpha + 4 \beta) \phi} A_m A_n - \frac{1}{2} {\tilde{B}}_{m k} g^{l q} {\tilde{B}}_{q n} & 0\\
0 & 0\\
\end{pmatrix}\,,
\end{split}\\
\begin{split}
{\hat{\mathcal{M}}}_{MB} & = g^{\frac{1}{5}}e^{\frac{-14\alpha + 2 \beta}{5} \phi}
\begin{pmatrix}
\frac{1}{2\sqrt{2}} {\tilde{C}}_{m k l} g^{k l, p q} {\tilde{B}}_{p q} &  \frac{1}{\sqrt{2}} {\tilde{C}}_{m k l} g^{k l, q n_1} A_q\\
\frac{1}{2\sqrt{2}} g^{-\frac{1}{2}} g_{m k} \epsilon^{k p q} {\tilde{B}}_{p q} & g^{-\frac{1}{2}} g_{m k} \epsilon^{k q n_1} A_q
\end{pmatrix}\\
& \qquad +g^{\frac{1}{5}} e^{\frac{-4\alpha - 8 \beta}{5}}
\begin{pmatrix}
e^{(2 \alpha + 4\beta) \phi} A_m & \frac{1}{2} {\tilde{B}}_{m k} g^{k n_1}\\
0 & 0
\end{pmatrix}\,,
\end{split}\\
\begin{split}
{\hat{\mathcal{M}}}_{AN} & = g^{\frac{1}{5}} e^{\frac{-14\alpha +2 \beta}{5}\phi}
\begin{pmatrix}
\frac{1}{2\sqrt{2}} {\tilde{B}}_{k l} g^{k l, p q} {\tilde{C}}_{p q n} & \frac{1}{2} g^{-\frac{1}{2}} {\tilde{B}}_{k l} \epsilon^{k l q} g_{q n}\\
- \frac{1}{\sqrt{2}} g^{m_1 l, p q} A_l {\tilde{C}}_{p q n} & - A_k \epsilon^{m_1 k l} g_{l n}
\end{pmatrix}\\
& \qquad + g^{\frac{1}{5}} e^{\frac{-4\alpha - 8 \beta}{5} \phi}
\begin{pmatrix}
e^{(2 \alpha + 4 \beta) \phi} A_n & 0\\
- \frac{1}{2} g^{m q} {\tilde{B}}_{q n} & 0\\
\end{pmatrix}\,,
\end{split}\\
\begin{split}
{\hat{\mathcal{M}}}_{AB} & =g^{\frac{1}{5}} e^{\frac{-4\alpha - 8\beta}{5}}
\begin{pmatrix}
e^{(2\alpha + 4 \beta) \phi} & 0\\
0 & \frac{1}{2}  g^{m_1 n_1}
\end{pmatrix}\\
& \qquad + g^{\frac{1}{5}}e^{\frac{-14\alpha + 2 \beta}{5} \phi}
\begin{pmatrix}
\frac{1}{4} {\tilde{B}}_{k l} g^{k l, p q} {\tilde{B}}_{p q} & \frac{1}{2} {\tilde{B}}_{k l} A_q g^{k l, q n_1}\\
- \frac{1}{2} g^{m_1 k, p q} A_\lambda {\tilde{B}}_{p q} & - A_k g^{m_1 k, q n_1} A_q
\end{pmatrix}\,.
\end{split}
\end{align}
Note that we have used the fact that ${\tilde{C}}_{m n p} \coloneqq C_{m n p} - 3 B_{[m n} A_{p]}$ is a top form such that ${\tilde{C}}_{m n p} \propto \epsilon_{m n p}$. In particular, we find
\begin{align}
\epsilon^{m k l} {\tilde{C}}_{k l n} = 2{\tilde{C}} \delta^m_n\,, \qquad \tilde{C} \coloneqq \frac{1}{3!} \epsilon^{m n p} C_{m n p}
\end{align}
by taking the trace, giving ${\tilde{C}}_{m n p} = {\tilde{C}} \epsilon_{m n p}$. Under the above splitting, we can rewrite the $\operatorname{SL}(5)$ generalised metric in the same generalised KK ansatz as the DFT case \eqref{eq:GeneralisedKK}
\begin{align}\label{eq:GeneralisedKKAnsatz}
{\hat{\mathcal{M}}}_{\hat{M} \hat{N}} & = \begin{pmatrix}
e^{2A\phi} \mathcal{M}_{MN} + e^{2B\phi} {\mathcal{A}}_M{}^A {\mathcal{G}}_{AB} {\mathcal{A}}^B{}_N & e^{2B\phi} {\mathcal{A}}_M{}^A {\mathcal{G}}_{AB} + e^{2A\phi} {\mathcal{M}}_{MN} {\mathcal{B}}^N{}_B\\
e^{2B\phi} \mathcal{G}_{AB} {\mathcal{A}}^B{}_N + e^{2A\phi} {\mathcal{B}}_A{}^M {\mathcal{M}}_{MN} & e^{2B\phi} \mathcal{G}_{AB} +  e^{2A\phi} {\mathcal{B}}_A{}^M {\mathcal{M}}_{MN} {\mathcal{B}}^N{}_B
\end{pmatrix}\,,
\end{align}
where
\begin{align}
{\mathcal{M}}_{MN} & =  \begin{pmatrix}
e^{6\alpha \phi} g_{m n} + {\tilde{C}}^2 g_{m n} & \sqrt{\frac{2}{g}} \tilde{C} g_{m \bar{n}}\\
\sqrt{\frac{2}{g}} \tilde{C} g_{\bar{m} n} & \frac{2}{g} g_{\bar{m} \bar{n}} 
\end{pmatrix}\\
{\mathcal{A}}_M{}^A & = \begin{pmatrix}
A_m & {\tilde{B}}_{m p_1}\\
0 & 0
\end{pmatrix}, \qquad {\mathcal{A}}^B{}_N = \begin{pmatrix}
A_n & 0\\
- {\tilde{B}}_{q_1 n} & 0\\
\end{pmatrix} = {\left( {\mathcal{A}}^T \right)}^B{}_N\\
\mathcal{B}^N{}_B & = \begin{pmatrix}
0 & 0\\
\frac{1}{4} \varepsilon^{\bar{m} k l} {\tilde{B}}_{k l} &  \frac{1}{2} \varepsilon^{\bar{n} p n_1} A_p
\end{pmatrix}\,, \qquad \mathcal{B}_A{}^M = \begin{pmatrix}
0 & \frac{1}{4} \varepsilon^{\bar{m} p q} {\tilde{B}}_{p q}\\
0 &  - \frac{1}{2} A_q \varepsilon^{m_1 q \bar{m}} 
\end{pmatrix} = {\left( {\mathcal{B}}^T \right)}_A{}^M \label{eq:SL5B}\\
{\mathcal{G}}_{AB} & = \begin{pmatrix}
e^{(2\alpha + 4 \beta) \phi} & 0\\
0 & \frac{1}{2}  g^{m_1 n_1}
\end{pmatrix} \\
e^{2A\phi} & = g^{\frac{1}{5}}e^{\frac{-14\alpha + 2 \beta}{5} \phi}\\
e^{2B\phi} & = g^{\frac{1}{5}}e^{\frac{-4\alpha - 8 \beta}{5} \phi}\,.
\end{align}
Note that both $\mathcal{A}$ and $\mathcal{B}$ do not contain metric degrees of freedom as required. In particular $\mathcal{B}$ is defined with the alternating \emph{symbol} $\varepsilon$ without reference to the metric determinant. As in the DFT case, we have that ${\mathcal{A}}^A{}_M {\mathcal{B}}^M{}_B = {\mathcal{B}}_A{}^M {\mathcal{A}}_M{}^B = \mathbf{0}$ and so the inverse reduction ansatz is given by \eqref{eq:InverseGeneralisedKK} except with $\mathcal{M}_{MN} \rightarrow e^{2A\phi} \mathcal{M}_{MN}$ and $\mathcal{G}_{AB} \rightarrow e^{2B\phi} \mathcal{G}_{AB}$.\par
To demonstrate that $\mathcal{M}_{MN}$ is indeed the $\operatorname{SL}(3) \times \operatorname{SL}(2)$ generalised metric, we define objects upon which the $\operatorname{SL}(2)$ action is manifest:
\begin{align}
C_{(0)} & = \sqrt{\frac{g}{2}} \tilde{C} \,, \qquad e^{\Phi} = e^{-3\alpha \phi} \sqrt{\frac{2}{g}}\,.
\end{align}
Note that, like $\mathcal{B}$ (though unlike $\tilde{C}$), the scalar $C_{(0)} = \frac{1}{\sqrt{2} \cdot 3!} \varepsilon^{m n p} C_{m n p}$ is also defined with the alternating symbol and so does not include the metric degree of freedom that $\tilde{C}$ included; it is an independent degree of freedom from the metric determinant, as required.
Then,
\begin{align}
\mathcal{M}_{MN} = \sqrt{\frac{2}{g}} e^{3\alpha \phi} g_{m n} \otimes \frac{1}{e^{-\Phi}} \begin{pmatrix}
e^{-2\Phi} + C_{(0)}^2 &  C_{(0)}\\
C_{(0)} & 1
\end{pmatrix}\,.
\end{align}
In this form, it is clear that the generalised metric can be factorised into an $\operatorname{SL}(3)$ component and an $\operatorname{SL}(2)$ component as $\mathcal{M}_{MN} = \mathcal{M}_{m n} \otimes \mathcal{M}_{\alpha \beta}$. As in the usual KK ansatz, we have the freedom to fix $\alpha$ to a convenient value. One way to fix it would be to require that we reduce to a generalised metric that also has determinant 1. For the $\operatorname{SL}(3)$ `generalised metric' on the coordinate representation $\mathcal{R}_1 = \mathbf{3}$ (which is really just the usual 3-dimensional metric), this occurs for $\mathcal{M}_{m n} = g^{-\frac{1}{3}} g_{m n}$ from which
\begin{align}\label{eq:AlphaFix}
\sqrt{\frac{2}{g}} e^{3\alpha \phi} = g^{-\frac{1}{3}}\,,\qquad \Rightarrow \qquad e^{2\alpha \phi} = {\left( \frac{g}{2^3} \right)}^{\frac{1}{9}}\,.
\end{align}
The other constant $\beta$ can be fixed in the same way as the conventional Kaluza-Klein theory, namely by a choice of frame (by which we mean choice of Weyl scaling to give the string or Einstein frame of the theory). Although we shall not conduct the full reduction of the potential, we illustrate what we mean by singling out one of the terms that appears in the reduction. To simplify the following analysis, we take $C_{m n p} = B_{m n} = 0$ which implies
\begin{align}
{\mathcal{A}}_M{}^A & = \begin{pmatrix}
A_m & 0\\
0 & 0\\
\end{pmatrix}\,,\\
{\mathcal{B}}^M{}_A & = \begin{pmatrix}
0 & 0\\
0 & \frac{1}{2} \varepsilon^{\bar{m} q m_1} A_q
\end{pmatrix}\,,\\
{\mathcal{B}}^M{}_A {\mathcal{A}}^A{}_N & = \mathbf{0}\,,\\
{\hat{\mathcal{M}}}^{\hat{M} \hat{N}} & = \begin{pmatrix}
e^{-2A\phi} {\mathcal{M}}^{MN} + e^{-2B\phi} {\mathcal{B}}^M{}_A {\mathcal{G}}^{AB} {\mathcal{B}}_B{}^N & - e^{-2A\phi} {\mathcal{M}}^{MQ} {\mathcal{A}}_Q{}^B - e^{-2B\phi} {\mathcal{B}}^M{}_A {\mathcal{G}}^{AB}\\
- e^{-2A\phi} {\mathcal{A}}^A{}_P {\mathcal{M}}^{PN} - e^{-2B\phi} {\mathcal{G}}^{AB} {\mathcal{B}}_B{}^N & e^{-2A\phi} {\mathcal{A}}^A{}_P {\mathcal{M}}^{PQ} {\mathcal{A}}_Q{}^B + e^{-2B\phi} {\mathcal{G}}^{AB}
\end{pmatrix}\,.
\end{align}
One of the terms in the $\operatorname{SL}(5)$ potential is
\begin{align}
\frac{1}{2} {\hat{\mathcal{M}}}^{\hat{M} \hat{N}} \partial_{\hat{M}} {\hat{\mathcal{M}}}^{\hat{K} \hat{L}} \partial_{\hat{K}} {\hat{\mathcal{M}}}_{\hat{N} \hat{L}} = - \frac{1}{4} g e^{-2B\phi} \partial_{\bar{m}} A_q \partial^{\bar{m}} A^q  + \ldots
\end{align}
In particular, the first term contributes to an additional Maxwell term that appears in addition to the $\operatorname{SL}(3) \times \operatorname{SL}(2)$ potential. Taking into account the fact that the $\operatorname{SL}(3) \times \operatorname{SL}(2)$ potential comes with the scaling
\begin{align}
-\frac{1}{12} {\hat{\mathcal{M}}}^{\hat{M} \hat{N}} \partial_{\hat{M}} {\hat{\mathcal{M}}}^{\hat{K} \hat{L}} \partial_{\hat{N}} {\hat{\mathcal{M}}}_{\hat{K} \hat{L}} = - \frac{1}{12} e^{-2A\phi} {\mathcal{M}}^{MN}\partial_M {\mathcal{M}}^{KL} \partial_N {\mathcal{M}}_{KL} + \ldots\,,
\end{align}
one sees that the potential and new Maxwell terms have a relative scaling (up to constant factors) of $g e^{2(A-B)\phi} = g e^{-2(\alpha - \beta) \phi}$ which can be fixed to land on any frame that one may wish by an appropriate choice of $\beta$.

\subsubsection{\texorpdfstring{$\operatorname{SL}(5)$}{SL(5)} EFT to \texorpdfstring{$\operatorname{O}(3,3)$}{O(3,3)} DFT Reduction}
As in \cite{Thompson:2011uw} the reduction to the $\operatorname{O}(3,3)$ DFT generalised metric,  in accordance with \eqref{eq:O33Reduction}, is:
\begin{align}
\begin{split}
{\hat{\mathcal{M}}}_{MN} & = e^{\frac{16 \alpha + 2\beta}{5} \phi} g^{\frac{1}{5}}
\begin{pmatrix}
g_{m n} - {\tilde{B}}_{m p} g^{p q} {\tilde{B}}_{q n} & {\tilde{B}}_{m p} g^{p n_1}\\
- g^{m q} {\tilde{B}}_{q n_1} & g^{m_1 n_1}
\end{pmatrix}\\
& \qquad + g^{\frac{1}{5}} e^{\frac{6 \alpha + 12 \beta}{5}\phi}
\begin{pmatrix}
{\tilde{C}}_{m k l} g^{k l, p q} {\tilde{C}}_{p q n} + A_m A_n & \sqrt{2} {\tilde{C}}_{m k l} g^{k l, p n_1} A_p\\
-\sqrt{2} A_k g^{m_1 k, p q} {\tilde{C}}_{p q n} & -2 A_k g^{m_1 k, q n_1} A_q
\end{pmatrix}\,,
\end{split}\\
{\hat{\mathcal{M}}}_{MB} & = e^{\frac{6\alpha + 12 \beta}{5} \phi} g^{\frac{1}{5}}
\begin{pmatrix}
A_m + \frac{1}{\sqrt{2}} {\tilde{C}}_{m k l} g^{k l, p q} {\tilde{B}}_{p q} & \sqrt{2} {\tilde{C}}_{m k l} g^{k l, n_1 n_2}\\
- g^{m k, p q} A_k {\tilde{B}}_{p q} & - 2 A_k g^{m_1 k, n_1 n_2} 
\end{pmatrix}\,,\\
{\hat{\mathcal{M}}}_{AN} & = e^{\frac{6\alpha + 12 \beta}{5}\phi} g^{\frac{1}{5}}
\begin{pmatrix}
A_\nu + \frac{1}{\sqrt{2}} {\tilde{B}}_{k l} g^{k l, p q} {\tilde{C}}_{p q n} & {\tilde{B}}_{k l} A_{p} g^{k l, p n_1}\\
\sqrt{2} g^{m_1 m_2, p q} {\tilde{C}}_{p q n} & 2 g^{m_1 m_2, k n_1} A_k
\end{pmatrix}\,,\\
{\hat{\mathcal{M}}}_{AB} & = e^{\frac{6\alpha + 12 \beta}{5} \phi} g^{\frac{1}{5}}
\begin{pmatrix}
1 + \frac{1}{2} {\tilde{B}}_{k l} g^{k l, p q} {\tilde{B}}_{p q} & {\tilde{B}}_{k l} g^{k l, n_1 n_2}\\
g^{m_1 m_2, p q} {\tilde{B}}_{p q} & 2 g^{m_1 m_2, n_1 n_2}
\end{pmatrix}\,,
\end{align}
where we have chosen different values of $\alpha$ and $\beta$ from the $\operatorname{SL}(3) \times \operatorname{SL}(2)$ reduction, instead taking
\begin{align}
e^{-(4\alpha + 2 \beta) \phi} = 2
\end{align}
that will enable us to land on the canonical form of the DFT metric. Then, subject to the following identifications

\begin{align}
{\mathcal{M}}_{MN} & = \begin{pmatrix}
g_{mn} - {\tilde{B}}_{mp} g^{pq} {\tilde{B}}_{qn} & {\tilde{B}}_{mp} g^{pn_1}\\
- g^{mp} {\tilde{B}}_{pn_1} & g^{m_1 n_1}
\end{pmatrix}\,,\\
\mathcal{A}_M{}^A & = \begin{pmatrix}
A_m & \frac{1}{\sqrt{2}} {\tilde{C}}_{m p_1 p_2} - \frac{1}{2} A_{[m} {\tilde{B}}_{p_1 p_2]}\\
0 & - \delta^m_{[p_1} A_{p_2]}
\end{pmatrix}\,, \quad \mathcal{A}^B{}_N & = \begin{pmatrix}
A_n & 0\\
\frac{1}{\sqrt{2}}  {\tilde{C}}_{q_1 q_2 n} - \frac{1}{2} {\tilde{B}}_{[q_1 q_2} A_{n]} & A_{[q_1} \delta_{q_2]}^n
\end{pmatrix}\,,\\
{\mathcal{G}}_{AB} & = \begin{pmatrix}
1 + \frac{1}{2} {\tilde{B}}_{kl} g^{kl, pq} {\tilde{B}}_{pq} & {\tilde{B}}_{pq} g^{pq, n_1 n_2}\\
g^{m_1 m_2, pq} {\tilde{B}}_{pq} & 2 g^{m_1 m_2, n_1 n_2}
\end{pmatrix}\,,\\
e^{2A\phi} & = g^{\frac{1}{5}} e^{\frac{16\alpha + 2\beta}{5}}\,,\\
e^{2B\phi} & = g^{\frac{1}{5}} e^{\frac{6\alpha + 12\beta}{5}}\,,
\end{align}
we can rewrite the $\operatorname{SL}(5)$ generalised metric as
\begin{align}\label{eq:KK}
{\hat{\mathcal{M}}}_{MN} & = \begin{pmatrix}
e^{2A \phi} \mathcal{M}_{MN} + e^{2B\phi} {\mathcal{A}}_M{}^A {\mathcal{G}}_{AB} {\mathcal{A}}^B{}_N & e^{2B\phi} {\mathcal{A}}_M{}^A {\mathcal{G}}_{AB}\\
e^{2B\phi} {\mathcal{G}}_{AB} {\mathcal{A}}^B{}_N & e^{2B\phi} {\mathcal{G}}_{AB}
\end{pmatrix}
\end{align}
which is the \emph{doubled KK ansatz} \cite{Thompson:2011uw}. Like the conventional KK ansatz, this can be understood as a particular case of the generalised KK ansatz given by (\ref{eq:GeneralisedKKAnsatz}). One may verify that the pieces of the $\operatorname{SL}(5)$ $Y$-tensor that would have entered into the $\mathcal{B}$-twisted terms under this reduction happen to vanish and so the reduction of the generalised KK ansatz to this doubled KK anstz in this case is non-trivial.\par
We note that the appearance of the $Y$-tensor in the generalised KK ansatz may be justified as follows: in the Kaluza-Klein reduction ansatz, the reduced fields are required to transform under the symmetries of the lower dimensional theory. In ExFTs, these must include the lower-dimensional (generalised) diffeomorphisms and so any appearance of the $Y$-tensor in the reduction ansatz could come about as a compensatory term to ensure the fields transform correctly.\par
\subsection{Some notes on the Reduction of Larger Generalised Metrics}\label{sec:ReductionLargerGenMetric}
Larger generalised metrics are much more difficult to reduce in full; for $E_{6(6)}$ and upwards it also contains the 6-form that couples electrically to the M5 whilst for $E_{8(8)}$, it further contains the dual graviton as propagating degrees of freedom. It is evident that reductions from $E_{6(6)}$ to $\operatorname{SO}(5,5)$ and $E_{8(8)}$ to $E_{7(7)}$ must somehow exclude the 6-form and dual graviton respectively from the reduced generalised metric.\par
Additionally, the generalised metric grows with the coordinate representation. In particular this means that the number of blocks appearing in the generalised metric also grows; for $E_{7(7)}$, the $\mathcal{R}_1 = \mathbf{56}$ decomposed under $\operatorname{GL}(7)$ to $\mathbf{7} \oplus \mathbf{21} \oplus \overbar{\mathbf{21}} \oplus \overbar{\mathbf{7}}$ produces $4 \times 4$ block matrices and so it is not clear whether even the generalised KK ansatz \eqref{eq:GeneralisedKK} is sufficient. The case for $E_{8(8)}$ is even worse with $\mathcal{R}_1 = \mathbf{248}$. Setting all internal potentials to zero for simplicity, the generalised metric for $E_{8(8)}$ (which we have rescaled to give determinant 1) when decomposed under $\operatorname{GL}(8)$ takes the form
\begin{align}\label{eq:E8GenMetric}
\begin{split}
{\hat{\mathcal{M}}}_{\hat{M} \hat{N}} = \operatorname{diag} \biggl[ & \hat{g} {\hat{g}}_{\hat{m} \hat{n}}, \hat{g} {\hat{g}}^{\hat{m}_1 \hat{m}_2, \hat{n}_1 \hat{n}_2}, {\hat{g}}_{\hat{m}_1 \hat{m}_2 \hat{m}_3, \hat{n}_1 \hat{n}_2 \hat{n}_3}, {\hat{g}}^{\hat{m}_1 \hat{n}_1} {\hat{g}}_{\hat{m}_2 \hat{n}_2} - \frac{1}{8} \delta^{\hat{m}_1}_{\hat{m}_2} \delta^{\hat{n}_1}_{\hat{n}_2}, 1,\\
& {\hat{g}}^{\hat{m}_1 \hat{m}_2 \hat{m}_3, \hat{n}_1 \hat{n}_2, \hat{n}_3}, {\hat{g}}^{-1} {\hat{g}}_{\hat{m}_1 \hat{m}_2, \hat{n}_1 \hat{n}_2}, {\hat{g}}^{-1} {\hat{g}}^{\hat{m} \hat{n}} \biggr]\,,
\end{split}
\end{align}
where $m_i, n_i = 1, \ldots, 8$ and we have chosen the conventions
\begin{align}
{\hat{g}}_{{\hat{m}}_1 {\hat{m}}_2, {\hat{n}}_1 {\hat{n}}_2} & \coloneqq {\hat{g}}_{\hat{{m}}_1[{\hat{n}}_1|} {\hat{g}}_{{\hat{m}}_2|{\hat{n}}_2]}\label{eq:G2}\,,\\
{\hat{g}}^{{\hat{m}}_1 {\hat{m}}_2, {\hat{n}}_1 {\hat{n}}_2} & \coloneqq {\hat{g}}^{\hat{{m}}_1[{\hat{n}}_1|} {\hat{g}}^{{\hat{m}}_2|{\hat{n}}_2]} \label{eq:InverseG2}\,,\\
{\hat{g}}_{{\hat{m}}_1 {\hat{m}}_2 {\hat{m}}_3, {\hat{n}}_1 {\hat{n}}_2 {\hat{n}}_3} & \coloneqq {\hat{g}}_{{\hat{m}}_1[{\hat{n}}_1|} {\hat{g}}_{{\hat{m}}_2 |{\hat{n}}_2|} {\hat{g}}_{{\hat{m}}_3 |{\hat{n}}_3]} \label{eq:G3}\,,\\
{\hat{g}}^{{\hat{m}}_1 {\hat{m}}_2 {\hat{m}}_3, {\hat{n}}_1 {\hat{n}}_2 {\hat{n}}_3} & \coloneqq {\hat{g}}^{{\hat{m}}_1[{\hat{n}}_1|} {\hat{g}}^{{\hat{m}}_2 |{\hat{n}}_2|} {\hat{g}}^{{\hat{m}}_3 |{\hat{n}}_3]}\label{eq:InverseG3}\,,
\end{align}
for the metrics on the antisymmetric representations. In principle, one could try the brute-force approach from the previous section and reduce the 8-dimensional internal metric under the standard circle reduction ansatz \eqref{eq:CircleReduction}. The antisymmetrised metrics in this case are given by
\begingroup
\renewcommand*{\arraystretch}{1.2}
\begin{align}
{\hat{g}}_{{\hat{m}}_1 {\hat{m}}_2, {\hat{n}}_1 {\hat{n}}_2} & = \begin{pmatrix}
e^{4\alpha \phi} g_{m_1 m_2, n_1 n_2} - 2 e^{2(\alpha + \beta)\phi}A_{[m_1} g_{m_2],[n_1} A_{n_2]} & - e^{2(\alpha + \beta)\phi} A_{[m_1} g_{m_2]n_1}\\
e^{2(\alpha + \beta) \phi} g_{m_1[n_1} A_{n_2]} & \frac{1}{2} e^{2(\alpha + \beta)\phi} g_{m_1 n_1}
\end{pmatrix}\,,\\
{\hat{g}}^{{\hat{m}}_1 {\hat{m}}_2, {\hat{n}}_1 {\hat{n}}_2} & = \begin{pmatrix}
e^{-4\alpha \phi} g^{m_1 m_2, n_1 n_2} & e^{-4 \alpha \phi} A^{[m_1} g^{m_2]n_1}\\
- e^{-4\alpha \phi} g^{m_1[n_1} A^{n_2]} & \frac{1}{2} e^{-2(\alpha + \beta)} g^{m_1 n_1} + \frac{1}{2} e^{-4\alpha \phi} ( g^{m_1 n_1} A \cdot A - A^{m_1} A^{n_1})
\end{pmatrix}\,,\\
{\hat{g}}_{{\hat{m}}_1 {\hat{m}}_2 {\hat{m}}_3, {\hat{n}}_1 {\hat{n}}_2 {\hat{n}}_3} & = \begin{pmatrix}
\begin{array}{c}
e^{6\alpha \phi} g_{m_1 m_2 m_3, n_1 n_2 n_3}\\
+ 3 e^{2(2\alpha + \beta)\phi}A_{[m_1} g_{m_2 m_3],[n_1 n_2} A_{n_3]}
\end{array} & e^{2(2\alpha + \beta)\phi} A_{[m_1} g_{m_2 m_3],n_1 n_2}\\
&\\
e^{2(2\alpha + \beta) \phi} g_{m_1 m_2, [n_1 n_2} A_{n_3]} & \frac{1}{3} e^{2(2\alpha + \beta)\phi} g_{m_1 m_2, n_1 n_2}
\end{pmatrix}\,,\\
{\hat{g}}^{{\hat{m}}_1 {\hat{m}}_2 {\hat{m}}_3, {\hat{n}}_1 {\hat{n}}_2 {\hat{n}}_3} & = \begin{pmatrix}
e^{-6\alpha \phi} g^{m_1 m_2 m_3, n_1 n_2 n_3} & - e^{-6\alpha \phi} A^{[m_1} g^{m_2 m_3],n_1 n_2}\\
&\\
- e^{-6\alpha \phi} g^{m_1 m_2,[n_1 n_2} A^{n_3]} &
\begin{array}{c}
\frac{1}{3} e^{-2(2\alpha + \beta)\phi} g^{m_1 m_2, n_1 n_2}\\
+ \frac{1}{3} e^{-6\alpha \phi} \left( g^{m_1 m_2, n_1 n_2} A \cdot A + 2 A^{[m_1} g^{m_2][n_1} A^{n_2]} \right)
\end{array}
\end{pmatrix}\,.
\end{align}
\endgroup
It is simple to check that ${\hat{g}}^{{\hat{m}}_1 {\hat{m}}_2, {\hat{n}}_1 {\hat{n}}_2}$ and ${\hat{g}}^{{\hat{m}}_1 {\hat{m}}_2 {\hat{m}}_3, {\hat{n}}_1 {\hat{n}}_2 {\hat{n}}_3}$ given above are indeed the inverses of ${\hat{g}}_{{\hat{m}}_1 {\hat{m}}_2, {\hat{n}}_1 {\hat{n}}_2}$ and ${\hat{g}}_{{\hat{m}}_1 {\hat{m}}_2 {\hat{m}}_3, {\hat{n}}_1 {\hat{n}}_2 {\hat{n}}_3}$ respectively if we account for the fact that the contraction of the decomposed indices requires the contraction conventions
\begin{align}
{\hat{g}}_{{\hat{m}}_1 {\hat{m}}_2, {\hat{p}}_1 {\hat{p}}_2} {\hat{g}}^{{\hat{p}}_1 {\hat{p}}_2, {\hat{n}}_1 {\hat{n}}_2} & = {\hat{g}}_{{\hat{m}}_1 {\hat{m}}_2, p_1 p_2} {\hat{g}}^{p_1 p_2, {\hat{n}}_1 {\hat{n}}_2}  + {\hat{g}}_{{\hat{m}}_1 {\hat{m}}_2, p_1 z} {\hat{g}}^{p_1 z, {\hat{n}}_1 {\hat{n}}_2} + {\hat{g}}_{{\hat{m}}_1 {\hat{m}}_2, z p_2} {\hat{g}}^{z p_2, {\hat{n}}_1 {\hat{n}}_2}\\ 
& = {\hat{g}}_{{\hat{m}}_1 {\hat{m}}_2, p_1 p_2} {\hat{g}}^{p_1 p_2, {\hat{n}}_1 {\hat{n}}_2} + 2 {\hat{g}}_{{\hat{m}}_1 {\hat{m}}_2, p_1 z} {\hat{g}}^{p_1 z, {\hat{n}}_1 {\hat{n}}_2}\\
{\hat{g}}_{{\hat{m}}_1 {\hat{m}}_2 {\hat{m}}_3, {\hat{p}}_1 {\hat{p}}_2 {\hat{p}}_3} {\hat{g}}^{{\hat{p}}_1 {\hat{p}}_2 {\hat{p}}_3, {\hat{n}}_1 {\hat{n}}_2 {\hat{n}}_3} & = \!\!\!
\begingroup
\renewcommand*{\arraystretch}{1.5}
\begin{array}[t]{l}
{\hat{g}}_{{\hat{m}}_1 {\hat{m}}_2 {\hat{m}}_3, p_1 p_2 p_3} {\hat{g}}^{p_1 p_2 p_3, {\hat{n}}_1 {\hat{n}}_2 {\hat{n}}_3} + {\hat{g}}_{{\hat{m}}_1 {\hat{m}}_2 {\hat{m}}_3, p_1 p_2 z} {\hat{g}}^{p_1 p_2 z, {\hat{n}}_1 {\hat{n}}_2 {\hat{n}}_3}\\
\qquad + {\hat{g}}_{{\hat{m}}_1 {\hat{m}}_2 {\hat{m}}_3, p_1 z p_3} {\hat{g}}^{p_1 z p_3, {\hat{n}}_1 {\hat{n}}_2 {\hat{n}}_3} + {\hat{g}}_{{\hat{m}}_1 {\hat{m}}_2 {\hat{m}}_3, z p_2 p_3} {\hat{g}}^{z p_2 p_3, {\hat{n}}_1 {\hat{n}}_2 {\hat{n}}_3}
\end{array}
\endgroup\\
& = {\hat{g}}_{{\hat{m}}_1 {\hat{m}}_2 {\hat{m}}_3, p_1 p_2 p_3} {\hat{g}}^{p_1 p_2 p_3, {\hat{n}}_1 {\hat{n}}_2 {\hat{n}}_3} + 3 {\hat{g}}_{{\hat{m}}_1 {\hat{m}}_2 {\hat{m}}_3, p_1 p_2 z} {\hat{g}}^{p_1 p_2 z, {\hat{n}}_1 {\hat{n}}_2 {\hat{n}}_3}\,.
\end{align}
Then, under these conventions, one may verify that
\begin{align}
{\hat{g}}_{{\hat{m}}_1 {\hat{m}}_2, {\hat{p}}_1 {\hat{p}}_2} {\hat{g}}^{{\hat{p}}_1 {\hat{p}}_2, {\hat{n}}_1 {\hat{n}}_2} & = \delta^{{\hat{n}}_1 {\hat{n}}_2}_{{\hat{m}}_1 {\hat{m}}_2} = \begin{pmatrix}
\delta^{n_1 n_2}_{m_1 m_2} & 0\\ 0 & \frac{1}{2} \delta^{n_1}_{m_1}
\end{pmatrix}\,,\\
{\hat{g}}_{{\hat{m}}_1 {\hat{m}}_2 {\hat{m}}_3, {\hat{p}}_1 {\hat{p}}_2 {\hat{p}}_3} {\hat{g}}^{{\hat{p}}_1 {\hat{p}}_2 {\hat{p}}_3, {\hat{n}}_1 {\hat{n}}_2 {\hat{n}}_3} & = \delta^{{\hat{n}}_1 {\hat{n}}_2 {\hat{n}}_3}_{{\hat{m}}_1 {\hat{m}}_2 {\hat{m}_3}} = \begin{pmatrix}
\delta^{n_1 n_2 n_3}_{m_1 m_2 m_3} & 0\\
0 & \frac{1}{3} \delta^{n_1 n_2}_{m_1 m_2}
\end{pmatrix}\,.
\end{align}
However, the adjoint block $\operatorname{diag}[ {\hat{g}}^{\hat{m}_1 \hat{n}_1} {\hat{g}}_{\hat{m}_2 \hat{n}_2} - \frac{1}{8} \delta^{\hat{m}_1}_{\hat{m}_2} \delta^{\hat{n}_1}_{\hat{n}_2}, 1]$ becomes troublesome as it contributes more off-diagonal terms than the $\operatorname{SL}(5)$ case, since ${\hat{g}}^{-1} \hat{g}$ decomposes to a $4 \times 4$ block matrix under $\operatorname{GL}(7)$ (rather than the $2 \times 2$ block matrices that the other terms give), and the $\delta \delta$ piece that appears in only some of the components in this decomposition disrupts a KK-type ansatz. As the generalised metric becomes more cumbersome it may be more efficient to consider the reduction of the generators or generalised coordinates for a qualitative picture of the reduction instead. Returning to the $E_{8(8)}$ coordinate representation $\mathcal{R}_1 = \mathbf{248}$, we decompose it under $\operatorname{SL}(9)$ to give 
\begin{align}\label{eq:SL9Decomp}
\mathbf{248} \rightarrow \mathbf{80} \oplus \mathbf{84} \oplus \overbar{\mathbf{84}}\,.
\end{align}
In terms of generators, we have
\begin{align}
\{ {\hat{T}}^{\hat{M}} \} & \xrightarrow{\operatorname{SL}(9)} \{ E^{\hat{m}}{}_{\hat{n}}, Z^{\hat{m}_1 \hat{m_2} \hat{m}_3}, Z_{\hat{m}_1 \hat{m}_2 \hat{m}_3} \}\,,
\end{align}
where $\hat{m}, \hat{n} = 1, \ldots, 9$. These satisfy the algebra\cite{Rosabal:2014rga}
\begin{subequations}
\begin{align}
[E^{\hat{m}_1}{}_{\hat{m}_2}, E^{\hat{n}_1}{}_{\hat{n}_2}] & = \delta^{\hat{n}_1}_{\hat{m}_2} E^{\hat{m}_1}{}_{\hat{n}_2} - \delta^{\hat{m}_1}_{\hat{n}_2} E^{\hat{n}_1}{}_{\hat{m}_2}\,,\\
[E^{\hat{m}_{1}}{}_{\hat{m}_2}, Z^{\hat{n}_1 \hat{n}_2 \hat{n}_3}] & = + \left( 3 \delta_{\hat{m}_2}^{[\hat{n}_1} Z^{\hat{n}_2 \hat{n}_3] \hat{m}_1} - \frac{1}{3} \delta^{\hat{m}_1}_{\hat{m}_2} Z^{\hat{n}_1 \hat{n}_2 \hat{n}_3} \right) \,,\\
[E^{\hat{m}_{1}}{}_{\hat{m}_2}, Z_{\hat{n}_1 \hat{n}_2 \hat{n}_3}] & = - \left( 3 \delta_{[\hat{n}_1}^{\hat{m}_1} Z_{\hat{n}_2 \hat{n}_3] \hat{m}_2} - \frac{1}{3} \delta^{\hat{m}_1}_{\hat{m}_2} Z_{\hat{n}_1 \hat{n}_2 \hat{n}_3} \right)\,,\\
[Z^{\hat{m}_1 \hat{m}_2 \hat{m}_3}, Z^{\hat{n}_1 \hat{n}_2 \hat{n}_3}] & = - \frac{1}{3!} \epsilon^{\hat{m}_1 \hat{m}_2 \hat{m}_3 \hat{n}_1 \hat{n}_2 \hat{n}_3 \hat{p}_1 \hat{p}_2 \hat{p}_3} Z_{\hat{p}_1 \hat{p}_2 \hat{p}_3}\,,\\
[Z^{\hat{m}_1 \hat{m}_2 \hat{m}_3}, Z_{\hat{n}_1 \hat{n}_2 \hat{n}_3}] & = 18 \delta^{[\hat{m}_1 \hat{m}_2}_{[\hat{n}_1 \hat{n}_2} E^{\hat{m}_3]}{}_{\hat{n}_3]}\,,\\
[Z_{\hat{m}_1 \hat{m}_2 \hat{m}_3}, Z_{\hat{n}_1 \hat{n}_2 \hat{n}_3}] & = + \frac{1}{3!} \epsilon_{\hat{m}_1 \hat{m}_2 \hat{m}_3 \hat{n}_1 \hat{n}_2 \hat{n}_3 \hat{p}_1 \hat{p}_2 \hat{p}_3} Z^{\hat{p}_1 \hat{p}_2 \hat{p}_3}\,.
\end{align}
\end{subequations}
Under $\operatorname{GL}(8)$, each of the representations in \eqref{eq:SL9Decomp} decompose as
\begin{subequations}
\begin{align}
\mathbf{80} & \xrightarrow{\operatorname{GL}(8)} \mathbf{63}_0 \oplus \mathbf{8}_{+9} \oplus \overbar{\mathbf{8}}_{-9} \oplus \mathbf{1}_0\,,\\
\mathbf{84} & \xrightarrow{\operatorname{GL}(8)} \mathbf{56}_{+3} \oplus \mathbf{28}_{-6}\,,\\
\overbar{\mathbf{84}} & \xrightarrow{\operatorname{GL}(8)} \overbar{\mathbf{56}}_{-3} \oplus \overbar{\mathbf{28}}_{+6}\,,
\end{align}
\end{subequations}
whilst the generators break down according to
\begin{subequations}
\begin{align}
\{ E^{\hat{m}}{}_{\hat{n}} \} & \xrightarrow{\operatorname{GL}(8)} \{ E^m{}_n, E^m{}_9, E^9{}_m, E^9{}_9 \}\,,\\
\{ Z^{\hat{m}_1 \hat{m}_2 \hat{m}_3} \} & \xrightarrow{\operatorname{GL}(8)} \{ Z^{m_1 m_2 m_3}, Z^{m_1 m_2 9} \}\,,\\
\{ Z_{\hat{m}_1 \hat{m}_2 \hat{m}_3} \} & \xrightarrow{\operatorname{GL}(8)} \{Z_{m_1 m_2 m_3}, Z_{m_1 m_2 9} \}\,.
\end{align}
\end{subequations}
Associating the index structures above to each representation , we see that the $E_{8(8)}$ coordinates in this notation are
\begin{align}\label{eq:E8SL9Coords}
{\hat{Y}}^{\hat{M}} = (Y^m{}_9, Y_{m_1 m_2 9}, Y^{m_1 m_2 m_3}, Y^m{}_n, Y^9{}_9, Y_{m_1 m_2 m_3}, Y^{m_1 m_2 9}, Y^9{}_m)\,,
\end{align}
which is the familiar decomposition of $E_{8(8)}$ where each set of coordinates correspond to the usual coordinates and the wrappings modes\footnote{Actually, there is an additional subtlety; the $\mathbf{63} \oplus \mathbf{1}$ contains an additional 8 coordinates over the KK6 wrapping modes which are thought to corresponds to the wrapping modes of non-supersymmetric branes. The string theory interpretation of this is given in \cite{deBoer:2012ma} whilst the $E_{11}$ picture was given in \cite{Kleinschmidt:2011vu}.} of the M2, M5, KK6, $5^3$, $2^6$ and $0^{(1,7)}$ branes. Note that the last few objects, which may be less familiar to the reader, are \emph{exotic branes}---highly non-perturbative objects that include concrete realisations of Hull's T-folds and U-folds (which can still be described geometrically, although only locally since they require duality transformations to glue together local patches) as well as objects that cannot be described geomerically even on local patches. We shall not describe such objects in any more detail but refer the reader to \cite{deBoer:2010ud,deBoer:2012ma,Plauschinn:2018wbo} for a description of such objects within string theory. Additionally, there is a growing body of work describing such objects including explicit constructions of such solutions in ExFT, worldvolume actions for exotic 5-branes and the classification of the mixed-symmetry potentials that they couple to\cite{Musaev:2016yon,Bakhmatov:2017les,Berman:2018okd,Otsuki:2019owg,Fernandez-Melgarejo:2018yxq,Kimura:2013khz,Kimura:2013fda,Kimura:2014upa,Kimura:2016anf,Sakatani:2014hba,Lombardo:2016swq,Cook:2009ri}.\par
We shall determine which components of the $E_{8(8)}$ generalised metric enter into the $E_{7(7)}$ generalised metric by determining how these coordinates fall into into $E_{7(7)}$ representations. In order to do so, we consider the decomposition of the $\operatorname{SL}(9)$ generators under another maximal subgroup $\operatorname{GL}(7)\times \operatorname{SL}(2)$ as a stepping stone to reconstructing full $E_{7(7)}$ representations. The relevant decompositions are
\begin{subequations}
\begin{align}\label{eq:SL9SL7}
\mathbf{80} & \xrightarrow{\operatorname{GL}(7) \times \operatorname{SL}(2)} {\mathbf{(\overbar{7},2)}}_{-9} \oplus {\mathbf{(1,3)}}_{0} \oplus {\mathbf{(48,1)}}_0 \oplus {\mathbf{(1,1)}}_0 \oplus {\mathbf{(7,2)}}_{+9}\,,\\
\mathbf{84} & \xrightarrow{\operatorname{GL}(7) \times \operatorname{SL}(2)} {\mathbf{(7,1)}}_{-12} \oplus {\mathbf{(21,2)}}_{-3} \oplus {\mathbf{(35,1)}}_{+6}\,,\\
\mathbf{\overbar{84}} & \xrightarrow{\operatorname{GL}(7) \times \operatorname{SL}(2)} {\mathbf{(\overbar{35},1)}}_{-6} \oplus {\mathbf{(\overbar{21},2)}}_{+3} \oplus {\mathbf{(\overbar{7},1)}}_{+12}\,,
\end{align}
\end{subequations}
whilst the generators break according to ($\check{m}, \check{n} = 1, \ldots, 7$)
\begin{subequations}
\begin{align}
\{E^{\hat{m}}{}_{\hat{n}} \}& \xrightarrow{\operatorname{GL}(7)\times \operatorname{SL}(2)} \{E^{\check{m}}{}_{\check{n}}, E^{\check{m}}{}_8, E^8{}_{\check{n}}, E^8{}_8, E^{\check{m}}{}_9, E^8{}_9, E^9{}_{\check{n}}, E^9{}_8, E^9{}_9\}\\
\{ Z^{\hat{m}_1 \hat{m}_2 \hat{m}_3} \} & \xrightarrow{\operatorname{GL}(7) \times \operatorname{SL}(2)} \{ Z^{\check{m}_1 \check{m}_2 \check{m}_3}, Z^{\check{m}_1 \check{m}_2 8}, Z^{\check{m}_1 \check{m}_2 9}, Z^{\check{m}_1 89} \}\\
\{ Z_{\hat{m}_1 \hat{m}_2 \hat{m}_3} \} & \xrightarrow{\operatorname{GL}(7) \times \operatorname{SL}(2)} \{ Z_{\check{m}_1 \check{m}_2 \check{m}_3}, Z_{\check{m}_1 \check{m}_2 8}, Z_{\check{m}_1 \check{m}_2 9}, Z_{\check{m}_1 89} \}
\end{align}
\end{subequations}
Being explicit, the exact identification of the $\operatorname{GL}(7)\times \operatorname{SL}(2)$ generators with the representations in \eqref{eq:SL9SL7} are
\begin{align}\label{eq:SL7Gen}
\begin{split}
{\mathbf{(\bar{7},2)}}_{-9}: & \{E^8{}_{\check{n}}, E^9{}_{\check{n}}\}\\
{\mathbf{(48,1)}}_0 \oplus {\mathbf{(1,1)}}_0 \oplus {\mathbf{(1,3)}}_0: & \{E^{\check{m}}{}_{\check{n}}, E^8{}_8, E^8{}_9, E^9{}_8, E^9{}_9\}\\
{\mathbf{(7,2)}}_{+9}: & \{ E^{\check{m}}{}_8, E^{\check{m}}{}_9\}\\
{\mathbf{(7,1)}}_{-12}: & \{ Z^{\check{m}_1 89}\}\\
{\mathbf{(21,2)}}_{-3}: & \{ Z^{\check{m}_1 \check{m}_2 8}, Z^{\check{m}_1 \check{m}_2 9}\}\\
{\mathbf{(35,1)}}_{+6}: & \{ Z^{\check{m}_1 \check{m}_2 \check{m}_3}\}\\
{\mathbf{(\bar{35},1)}}_{-6}: & \{ Z_{\check{m}_1 \check{m}_2 \check{m}_3}\}\\
{\mathbf{(\bar{21},2)}}_{+3}: & \{ Z_{\check{m}_1 \check{m}_2 8}, Z_{\check{m}_1 \check{m}_2 9}\}\\
{\mathbf{(\bar{7},1)}}_{+12}: & \{ Z_{\check{m}_1 89}\}\,.
\end{split}
\end{align}
Note that the $\operatorname{SL}(2)$ factor acts on the $T^2$, spanned by the directions $y^8$ and $y^9$, by exchanging $8 \leftrightarrow 9$ as expected.
Comparing to the decomposition of $\mathbf{248}$ under $E_{7(7)}\times \operatorname{SL}(2)$
\begin{align}
\mathbf{248} = & \mathbf{(133,1)}\oplus \mathbf{(56,2)} \oplus \mathbf{(1,3)},
\end{align}
we may reconstruct full $E_{7(7)}$ representations from those appearing in \eqref{eq:SL9SL7} by the $\operatorname{SL}(2)$ representations that appear here:
\begin{subequations}
\begin{align}
{(\mathbf{133,1})} & \xrightarrow{\operatorname{GL}(7) \times \operatorname{SL}(2)} {(\mathbf{7,1})}_{-12} \oplus {(\bar{\mathbf{35}}, \mathbf{1})}_{-6} \oplus \mathbf{(48,1)}_0 \oplus \mathbf{(1,1)}_0 \oplus \mathbf{(35,1)}_{+6} \oplus {(\bar{\mathbf{7}}, \mathbf{1})}_{+12}\,,\\
\mathbf{(1,3)} & \xrightarrow{\operatorname{GL}(7) \times \operatorname{SL}(2)} \mathbf{(1,3)}_0\,,\\
{(\mathbf{56,2})} & \xrightarrow{\operatorname{GL}(7) \times \operatorname{SL}(2)} {(\bar{\mathbf{7}},\mathbf{2})}_{-9} \oplus {(\mathbf{21,2})}_{-3} \oplus {(\bar{\mathbf{21}}, \mathbf{2})}_{+3} \oplus {(\mathbf{7,2})}_{+9}\,.
\end{align}
\end{subequations}
Note, in particular, the doublet of $\mathbf{56}$ representations; $E_{8(8)}$ is large enough to contain two copies of the fundamental representation of $E_{7(7)}$. If we denote the generators of $E_{7(7)}$, decomposed under $\operatorname{GL}(7) \times \operatorname{SL}(2)$, as
\begin{align}
\{{\hat{T}}^{\hat{M}}\} \xrightarrow{E_{7(7)} \times \operatorname{SL}(2)} \{t^{\alpha}, t^\sharp, t^\natural, t^\flat, t^M, t^{\bar{M}}\},
\end{align}
we are now ready to identify how the generators of $E_{8(8)}$ descend to $E_{7(7)} \times \operatorname{SL}(2)$. The only non-trivial identification is for $(\mathbf{1,1}) \oplus (\mathbf{1,3})$. Noting that the Cartan generators of $E_{8(8)}$ are $\{E^1{}_2, \ldots E^7{}_8, R^{678} \}$, we decompose this under $E_{7(7)} \times \operatorname{SL}(2)$ by deleting the node corresponding to $E^1{}_2$ in the extended Dynkin diagram leaving the Cartan generators of $E_{7(7)}$ and $\operatorname{SL}(2)$ to be $\{E^2{}_3, \ldots, E^7{}_8, R^{678}\}$ and $\{E^8{}_9\}$ respectively. Here, the generator $E^8{}_9$ corresponds to the extra node in the extended Dynkin diagram or, equivalently, the final node of the gravity line under $E_{8(8)} \rightarrow \operatorname{SL}(9)$. Thus, the $\operatorname{SL}(2)$ triplet must then be formed from the generators $\{E^8{}_8, E^8{}_9, E^9{}_9\} \equiv \{ t^\sharp, t^\natural, t^\flat\}$ and the $(\mathbf{1,1})$ factor must be given by the remaining $\{E^9{}_8\}$ generator. Thus, we end up with the identification of the generators
\begin{subequations}
\begin{align}
t^{\alpha} : & \{Z^{\check{m}_1 89}, Z_{\check{m}_1 \check{m}_2 \check{m}_3}, E^{\check{m}}{}_{\check{n}}, E^9{}_8, Z^{\check{m}_1 \check{m}_2 \check{m}_3}, Z_{\check{m}_1 89}\},\\
(t^\sharp, t^\natural, t^\flat): & \{ E^8{}_8, E^8{}_9, E^9{}_9\},\\
(t^M, t^{\bar{M}}) : & \{ E^8{}_{\check{n}}, E^9{}_{\check{n}}, Z^{\check{m}_1 \check{m}_2 8}, Z^{\check{m}_1 \check{m}_2 9}, Z_{\check{m}_1 \check{m}_2 8}, Z_{\check{m}_1 \check{m}_2 9}, E^{\check{m}}{}_8, E^{\check{m}}{}_9\},
\end{align}
\end{subequations}
The ranges of the indices should hopefully be self-evident: $\alpha = 1,\ldots, 133$ indexes the adjoint representation of $E_{7(7)}$, $M$ and $\bar{M}$ index distinct 56-dimensional representations and $(\sharp, \natural, \flat)$ denote an $\operatorname{SL}(2)$ triplet of $E_{7(7)}$ singlets. To each of these generators, we assign coordinates with the same index structure in the usual fashion e.g. $Y^{\check{m}_1}$ is associated to $Z^{\check{m}_1 89}$ etc.\par
The above data is now sufficient to reconstruct all of the $E_{7(7)}$ coordinates from the $E_{8(8)}$ coordinates. Since we can trace the origin of the $E_{7(7)}$ generators back to those of $E_{8(8)}$ e.g. $E^{\check{m}}{}_9$ (associated to $Y^{\check{m}}{}_9$) descends from $E^m{}_9$ (associated to $Y^m{}_9$, or the usual coordinates), we may disentangle the two sets of $E_{7(7)}$ generalised coordinates $Y^M$ and $Y^{\bar{M}}$ by demanding that the geometric wrapping modes of $E_{7(7)}$ descend from the geometric wrapping modes of $E_{8(8)}$. This gives (note the mixing of 8 and 9 indices)\footnote{For the exotic branes, we may identify the branes by dualising in 8 dimensions and/or adding full sets of antisymmetric indices $[ \check{m}_1 \ldots \check{m}_7 8]$ (note that the index 9 may be dropped as it is just a relic of the decomposition we took):
\begin{subequations}
\begin{align}
Y^{\check{m}_1}{}_8 & \equiv Y_{\check{m}_2 \ldots \check{m}_7 8,8} \rightarrow 6^1 \text{=KK6}\\
Y_{\check{m}_1 \check{m}_2 8} & \equiv Y_{\check{m}_1 \ldots \check{m}_7 8, \check{m}_1 \check{m}_2 8} \rightarrow 5^3\\
Y^{\check{m}_1 \check{m}_2 9} & \equiv Y_{\check{m}_1 \ldots \check{m}_7 8, \check{m}_3 \ldots \check{m}_7 8} \rightarrow 2^6\\
Y^9{}_{\check{m}_1} & \equiv Y_{\check{m}_1 \ldots \check{m}_7 8 , \check{m}_1 \ldots \check{m}_7 8, \check{m}_1} \rightarrow 0^{(1,7)}.
\end{align}
\end{subequations}
The identifications of the geometric coordinates should hopefully be self-explanatory.}:
\begin{align}
Y^{M} &: \begin{cases}
Y^{\check{m}}{}_9 & \text{from usual coordinates}\\
Y_{\check{m}_1 \check{m}_29} & \text{from M2}\\
Y^{\check{m}_1 \check{m}_28} & \text{from M5}\\
Y^8{}_{\check{m}} & \text{from KK6}
\end{cases}\\
Y^{\bar{M}} &: \begin{cases}
Y^{\check{m}}{}_8 & \text{from KK6}\\
Y_{\check{m}_1 \check{m}_28} & \text{from } 5^3\\
Y^{\check{m}_1 \check{m}_29} & \text{from } 2^6\\
Y^9{}_{\check{m}} & \text{from } 0^{(1,7)}
\end{cases}
\end{align}
The remaining $E_{7(7)}$ coordinates are formed from the following components on the $E_{8(8)}$ side:
\begin{align}
Y^{\alpha} &: \begin{cases}
Y^{\check{m}_1 8 9} & \text{from } 2^6\\
Y_{\check{m}_1 \check{m}_2 \check{m}_3} & \text{from } 5^3\\
Y^{\check{m}_1}{}_{\check{m}_2} & \text{from KK6}\\
Y^9{}_8 & \text{from } 0^{(1,7)}\\
Y^{\check{m}_1 \check{m}_2 \check{m}_3} & \text{from M5}\\
Y_{\check{m}_1 89} & \text{from M2}
\end{cases}\\
(Y^\sharp, Y^\natural, Y^\flat) &: \begin{cases}
Y^8{}_8 & \text{from KK6}\\
Y^8{}_9 & \text{from usual coordinates}\\
Y^9{}_9 & \text{from KK6}
\end{cases}
\end{align}
Actually, from an earlier footnote on the $E_{8(8)}$ coordinates, $Y^{\check{m}}{}_8$ and $Y^8{}_8$ may need to be identified with the duals of the wrapping modes of non-supersymmetric branes. With this, we see that one of the copies of the $\mathbf{56}$ coordinates descends from the geometric sector of $E_{8(8)}$ whilst the other descends from the non-geometric sector of $E_{8(8)}$. From here, it is simple to reconstruct the decomposition of the simplified generalised metric \eqref{eq:E8GenMetric}, if one so wished. However, this alone will not allow us to reduce the full generalised metric (with non-vanishing internal potentials); the brute force method remains the most direct, though troublesome, method to reduce it. Whilst we have taken a rather indirect way of deriving the correspondence between the extended coordinates of the two theories, we hope that we have illustrated how one might follow the reductions of the generalised coordinates for more complicated EFT reductions.\par
We end with a remark on how such a full reduction may still house new ideas. In the $\operatorname{SL}(5)$ case we saw that different ways of identifying the components that enter into the generalised metric of the reduced theory gave rise to reductions to distinct theories. For each $E_{n(n)}$ EFT, one should be able to reduce to at least the $E_{n-1(n-1)}$ EFT as well as the $\operatorname{O}(n-1, n-1)$ DFT. However, as the size of the generalised metric (as well as the complexity of the reduction ansatz) increases, there is more freedom in how we may pick out the components that enter into the reduced generalised metric. It may then be possible that there exists more choices than the two we have highlighted that lead to reductions to theories that have not yet been studied in the literature, particularly if we do not restrict ourselves to circle reductions as we have done here.
\section{Reduction of the section condition}\label{sec:ReductionSection}
We now consider how the section condition for a given ExFT reduces. We shall consider the reduction of the $E_{8(8)}$ EFT section condition to the $E_{7(7)}$ section condition in detail by an explicit reduction of the $Y$-tensor. We shall be more schematic in the reduction of the $\operatorname{SL}(5)$ section condition but shall reduce it to both the $\operatorname{SL}(3) \times \operatorname{SL}(2)$ and $\operatorname{O}(3,3)$ section conditions.
\subsection{\texorpdfstring{$E_{8(8)}$}{E8(8)} EFT to \texorpdfstring{$E_{7(7)}$}{E8(8)} EFT}
In reducing between EFTs, it quickly becomes clear that we need a consistent set of conventions for both theories that will allow us to reduce one to the other. However, the conventions presented in \cite{Hohm:2014fxa} (whilst, of course, internally consistent) are found to be incompatible with those of \cite{Hohm:2013uia} and so we shall have to modify the conventions of both to conform to a consistent set of rules. In particular, we shall adopt the conventions of \cite{Marrani:2010de,deWit:2002vt} which give compatible reductions of the exceptional structure. We first set up some notation speaking first in generality and then restricting to the cases of interest later. Let the structure constants of an algebra $\mathfrak{g}$ be defined through the commutation relations of the generators in the representation $\mathcal{R}$:
\begin{align}
{[t^\alpha, t^\beta]}_M{}^N = f^{\alpha \beta}{}_{\gamma} {(t^\gamma)}_M{}^N
\end{align}
where $\alpha, \beta, \gamma = 1, \ldots, \operatorname{dim} \mathfrak{g}$ are adjoint indices and $M, N = 1, \ldots, \operatorname{dim}\mathcal{R}$ denote the indices of some representation  $\mathcal{R}$ of $\mathfrak{g}$, which may or may not also be the adjoint representation. The Killing form, with the canonical scaling, is defined as
\begin{align}
{\tilde{\kappa}}^{\alpha \beta} \coloneqq \frac{1}{C_{\mathbf{adj.}}} f^{\alpha \gamma}{}_\delta f^{\beta \delta}{}_\gamma\,,
\end{align}
where $C_{\mathbf{adj.}}$ is the quadratic Casimir in the adjoint representation. The quadratic Casimir of a representation $\mathcal{R}$ is defined through the inverse Killing form and generators $t^\alpha$ in $\mathcal{R}$ as
\begin{align}\label{eq:CasimirR}
C_{\mathbf{R}} \delta^N_M \coloneqq {\tilde{\kappa}}_{\alpha \beta} {(t^\alpha)}_M{}^P {(t^\beta)}_P{}^N\,.
\end{align} 
However, rather than work with the Killing form itself, we shall define the rescaled bilinear invariant
\begin{align}
\kappa^{\alpha \beta} \coloneqq \operatorname{Tr} (t^\alpha t^\beta) = {(t^\alpha)}_M{}^N {(t^\beta)}_N{}^M\,.
\end{align}
Taking the trace of \eqref{eq:CasimirR}, we obtain
\begin{align}
C_{\mathcal{R}} \cdot \operatorname{dim} \mathcal{R} = {\tilde{\kappa}}_{\alpha \beta} {(t^\alpha)}_M{}^N {(t^\beta)}_N{}^M = {\tilde{\kappa}}_{\alpha \beta} \kappa^{\alpha \beta}
\end{align}
and so we see that the rescaled and canonical Killing forms are related by
\begin{align}
\kappa^{\alpha \beta} = \frac{C_{\mathcal{R}} \cdot \operatorname{dim} \mathcal{R}}{\operatorname{dim} \mathfrak{g}} {\tilde{\kappa}}^{\alpha \beta}\,.
\end{align}
It is then easy to check that the inverse rescaled Killing form satisfies
\begin{align}\label{eq:Norm}
\kappa_{\alpha \beta} {(t^\alpha)}_M{}^P {(t^\beta)}_P{}^N = \frac{\operatorname{dim}\mathfrak{g}}{\operatorname{dim}\mathcal{R}} \delta^N_M\,.
\end{align}
We thus end up with
\begin{align}
f_{\alpha \gamma \delta} f_{\beta}{}^{\gamma \delta} & = - \frac{\operatorname{dim}\mathfrak{g}}{\operatorname{dim} \mathcal{R}} \frac{C_{\mathbf{adj.}}}{C_{\mathcal{R}}} \kappa_{\alpha \beta}
\end{align}

and we shall use this rescaled Killing-form (henceforth referred to as just `the Killing form') to raise and lower adjoint indices. Finally, we introduce the Dynkin index of a representation $\mathcal{R}$ as
\begin{align}
I_{\mathcal{R}} \coloneqq \frac{\operatorname{dim} \mathcal{R}}{\operatorname{dim} \mathfrak{g}} C_{\mathcal{R}}\,.
\end{align}
In the case that $\mathcal{R}$ is the adjoint representation, the Dynkin index of the adjoint representation $I_{\mathbf{adj.}}$ coincides with the dual Coxeter number $g^\vee$ and so we obtain
\begin{align}
\kappa_{\alpha \beta} = - \frac{I_{\mathcal{R}}}{g^\vee} f_{\alpha \gamma \delta} f_{\beta}{}^{\gamma \delta}\,.
\end{align}
As a consequence of this normalisation, we end up with the orthogonality of the structure constants
\begin{align}\label{eq:NormalisationGenerators}
f^{\alpha \gamma \delta} f_{\beta \gamma \delta} = - \frac{g^\vee}{I_{\mathcal{R}}} \delta^\alpha_\beta\,.
\end{align}
For our purposes, we are interested in $\mathbf{adj.} = \mathcal{R} = \mathbf{248}$ for $E_{8(8)}$ and $\mathbf{adj.} = \mathbf{133}$, $\mathcal{R} = \mathbf{56}$ for $E_{7(7)}$ which have
\begin{align}
\begin{array}{ll}
g^\vee (E_{8(8)}) = 30, & I_{\mathbf{248}} = 30\\
g^\vee (E_{7(7)}) = 18, & I_{\mathbf{56}} = 6\,.\\
\end{array}
\end{align}
For this section, we shall use $\hat{M}, \hat{N} = 1, \ldots, 248$ and $\alpha, \beta = 1, \ldots, 133$ to index the adjoint representations of $E_{8(8)}$ and $E_{7(7)}$ respectively and $M, N = 1, \ldots, 56$ to index the coordinate representation of $E_{7(7)}$. We define the following Killing forms for $E_{8(8)}$ and $E_{7(7)}$ respectively (adorning hats on $E_{8(8)}$ objects).\par
\begin{align}
{\hat{\kappa}}_{\hat{M} \hat{N}} & = f_{\hat{M} \hat{P}}{}^{\hat{Q}} f_{\hat{N} \hat{Q}}{}^{\hat{P}}\\
\kappa_{\alpha \beta} & = \frac{1}{3} f_{\alpha \gamma}{}^\delta f_{\beta \delta}{}^\gamma
\end{align}
Before we continue, we make a note on differing conventions in the literature. In the original $E_{8(8)}$ EFT paper \cite{Hohm:2014fxa}, the authors define
\begin{align}
{\left( {\hat{\mathbb{P}}}_{\mathbf{248}} \right)}^{\hat{M}}{}_{\hat{N}}{}^{\hat{K}}{}_{\hat{L}} = +\frac{1}{60} {\hat{f}}^{\hat{M}}{}_{\hat{N} \hat{P}} {\hat{f}}^{\hat{P} \hat{K}}{}_{\hat{L}}
\end{align}
with normalisation ${\hat{f}}^{\hat{M} \hat{K} \hat{L}} {\hat{f}}_{\hat{N} \hat{K} \hat{L}} = - 60 \delta^{\hat{M}}_{\hat{N}}$. By contrast, our normalisation is dictated by \eqref{eq:NormalisationGenerators} as
\begin{align}
{\hat{f}}^{\hat{M} \hat{P} \hat{Q}} {\hat{f}}_{\hat{N} \hat{P} \hat{Q}} = - \delta^{\hat{M}}_{\hat{N}}\,,
\end{align}
which is the same convention as that used in \cite{Marrani:2010de}. Since both $E_{8(8)}$ and $E_{7(7)}$ possess invariants with which to identify $\mathbf{R}$ and $\overbar{\mathbf{R}}$, we shall define our projectors to act on $\mathbf{R} \otimes \mathbf{R}$ with the following conventions:
\begin{align}
{\left( {\mathbb{P}}_{\mathbf{R}} \right)}_{MN}{}^{KL} {\left( {\mathbb{P}}_{\mathbf{R}} \right)}_{KL}{}^{PQ} & = {\left( {\mathbb{P}}_{\mathbf{R}} \right)}_{MN}{}^{PQ}\\
{\left( {\mathbb{P}}_{\mathbf{R}} \right)}_{MN}{}^{MN} & = \operatorname{rnk} \mathbb{P} = \operatorname{dim} \mathbf{R}\,.
\end{align}
Then, we require 
\begin{align}
{\left({\hat{\mathbb{P}}}_{\mathbf{248}} \right)}_{\hat{K} \hat{L}}{}^{\hat{M} \hat{N}} & = - {\hat{f}}_{\hat{K} \hat{L} \hat{P}} {\hat{f}}^{\hat{P} \hat{M} \hat{N}}\,.
\end{align}
Since the normalisation of the generators have the same sign but the projector differs from \cite{Hohm:2014fxa} by a sign (when we identify their ${\left( {\hat{\mathbb{P}}}_{\mathbf{248}} \right)}^{\hat{M}}{}_{\hat{K}}{}^{\hat{N}}{}_{\hat{L}}$ with our ${\hat{\kappa}}^{\hat{M} \hat{S}} {\left({\hat{\mathbb{P}}}_{\mathbf{248}} \right)}_{\hat{S} \hat{K}}{}^{\hat{N} \hat{T}} {\hat{\kappa}}_{\hat{T} \hat{L}}$) we must compensate by introducing a minus sign for every instance of ${\hat{\mathbb{P}}}_{\mathbf{248}}$ such as in the generalised Lie derivative \eqref{eq:OurGenLie}. Note, however, that we do not need to introduce a minus sign for ${\hat{\mathbb{P}}}_{\mathbf{3875}}$. We also note the conventions of \cite{Cederwall:2015ica} which is closer to ours and differs only by the scaling of the generators.\par
In our conventions, the projectors onto various irreps of $E_{8(8)}$ within $\mathbf{248} \otimes \mathbf{248}$ are given by
\begin{align}
{\left( {\hat{\mathbb{P}}}_{\mathbf{1}} \right)}_{\hat{K} \hat{L}}{}^{\hat{M} \hat{N}}  & = \frac{1}{248} {\hat{\kappa}}_{\hat{K} \hat{L}} {\hat{\kappa}}^{\hat{M} \hat{N}}\,\\
{\left({\hat{\mathbb{P}}}_{\mathbf{248}} \right)}_{\hat{K} \hat{L}}{}^{\hat{M} \hat{N}} & = - {\hat{f}}_{\hat{K} \hat{L} \hat{P}} {\hat{f}}^{\hat{P} \hat{M} \hat{N}}\,\\
{\left( {\hat{\mathbb{P}}}_{\mathbf{3875}} \right)}_{\hat{K} \hat{L}}{}^{\hat{M} \hat{N}} & = \frac{1}{7} \delta^{(\hat{M}}_{\hat{K}} \delta^{\hat{N})}_{\hat{L}} - \frac{1}{56} {\hat{\kappa}}_{\hat{K} \hat{L}} {\hat{\kappa}}^{\hat{M} \hat{N}} - \frac{30}{7} {\hat{f}}^{\hat{P} (\hat{M}}{}_{\hat{K}} {\hat{f}}_{\hat{P}}{}^{\hat{N})}{}_{\hat{L}}\label{eq:3875}\,\\
{\left( {\hat{\mathbb{P}}}_{\mathbf{27000}} \right)}_{\hat{K} \hat{L}}{}^{\hat{M} \hat{N}} & = \frac{6}{7} \delta^{(\hat{M}}_{\hat{K}} \delta^{\hat{N})}_{\hat{L}} + \frac{3}{217} {\hat{\kappa}}_{\hat{K} \hat{L}} {\hat{\kappa}}^{\hat{M} \hat{N}} + \frac{30}{7} {\hat{f}}^{\hat{P} (\hat{M}}{}_{\hat{K}} {\hat{f}}_{\hat{P}}{}^{\hat{N})}{}_{\hat{L}}\,\\
{\left({\hat{\mathbb{P}}}_{\mathbf{30380}} \right)}_{\hat{K} \hat{L}}{}^{\hat{M} \hat{N}} & = \delta^{[\hat{M}}_{\hat{K}} \delta^{\hat{N}]}_{\hat{L}} + {\hat{f}}_{\hat{K} \hat{L} \hat{P}} {\hat{f}}^{\hat{P} \hat{M} \hat{N}}\,.
\end{align}
Simple computations will verify that each of these square to themselves and project onto spaces of the correct dimensions. However, showing that ${\hat{\mathbb{P}}}_{\mathbf{3875}}$ and ${\hat{\mathbb{P}}}_{\mathbf{27000}}$ square to themselves will require the use of the identity
\begin{align}
{\hat{f}}_{\hat{K} \hat{R} \hat{M}} {\hat{f}}_{\hat{L}}{}^{\hat{R} \hat{N}} {\hat{f}}_{\hat{P}}{}^{\hat{M} \hat{S}} {\hat{f}}_{\hat{Q} \hat{N} \hat{S}} & = \frac{1}{300} ( {\hat{\kappa}}_{\hat{K} \hat{Q}} {\hat{\kappa}}_{\hat{L} \hat{P}} + 2 {\hat{\kappa}}_{\hat{K}(\hat{L}} {\hat{\kappa}}_{\hat{P)} \hat{Q}}) - \frac{1}{6} ( 2 {\hat{f}}^{\hat{M}}{}_{\hat{K} \hat{P}} {\hat{f}}_{\hat{M} \hat{L} \hat{Q}} - {\hat{f}}^{\hat{M}}{}_{\hat{K} \hat{Q}} {\hat{f}}_{\hat{M} \hat{L} \hat{P}} )\,.
\end{align}
Noting that
\begin{align}
\operatorname{Sym} \left( \mathbf{248} \otimes \mathbf{248} \right) & = \mathbf{1} \oplus \mathbf{3875} \oplus \mathbf{27000}\,,\\
\operatorname{Asym} \left( \mathbf{248} \otimes \mathbf{248} \right) & = \mathbf{248} \oplus \mathbf{30380}\,,
\end{align}
we see that the projectors further obey the completeness relations
\begin{align}
{\left( {\hat{\mathbb{P}}}_{\mathbf{1}} \right)}_{\hat{K} \hat{L}}{}^{\hat{M} \hat{N}} + {\left( {\hat{\mathbb{P}}}_{\mathbf{3875}} \right)}_{\hat{K} \hat{L}}{}^{\hat{M} \hat{N}} + {\left( {\hat{\mathbb{P}}}_{\mathbf{27000}} \right)}_{\hat{K} \hat{L}}{}^{\hat{M} \hat{N}} & = \delta^{(\hat{M}}_{\hat{K}} \delta^{\hat{N})}_{\hat{L}}\,,\\
{\left( {\hat{\mathbb{P}}}_{\mathbf{248}} \right)}_{\hat{K} \hat{L}}{}^{\hat{M} \hat{N}} + {\left( {\hat{\mathbb{P}}}_{\mathbf{30380}} \right)}_{\hat{K} \hat{L}}{}^{\hat{M} \hat{N}} & = \delta^{[\hat{M}}_{\hat{K}} \delta^{\hat{N}]}_{\hat{L}}\,,
\end{align}
and that the sum of all these projectors gives the identity on $\mathbf{248} \otimes \mathbf{248}$, namely $\delta^{\hat{M}}_{\hat{K}} \delta^{\hat{N}}_{\hat{L}}$. Finally, the relation between ${\hat{\mathbb{P}}}_{\mathbf{248}}$ and ${\hat{\mathbb{P}}}_{\mathbf{3875}}$ is modified to
\begin{align}
{\left( {\hat{\mathbb{P}}}_{\mathbf{3875}} \right) }_{\hat{K} \hat{L}}{}^{\hat{M} \hat{N}} & = \frac{1}{7} \delta^{(\hat{M}}_{\hat{K}} \delta^{\hat{N})}_{\hat{L}} - \frac{1}{56} {\hat{\kappa}}_{\hat{K} \hat{L}} {\hat{\kappa}}^{\hat{M} \hat{N}} - \frac{30}{7} {\hat{f}}^{\hat{P} \hat{M}}{}_{\hat{K}} {\hat{f}}_{\hat{P}}{}^{\hat{N}}{}_{\hat{L}} - \frac{15}{7} {\left( {\hat{\mathbb{P}}}_{\mathbf{248}} \right)}_{\hat{K} \hat{L}}{}^{\hat{M} \hat{N}}\,,
\end{align}
as may be verified by using the Jacobi identity
\begin{align}\label{eq:Jacobi}
{\hat{f}}^{\hat{P} \hat{M} \hat{N}} {\hat{f}}_{\hat{P} \hat{K} \hat{L}} = 2 {\hat{f}}^{\hat{P} [ \hat{M}}{}_{\hat{K}} {\hat{f}}_{\hat{P}}{}^{\hat{N}]}{}_{\hat{L}}\,.
\end{align}
For $E_{7(7)}$, our normalisation agrees with \cite{Hohm:2013uia} but our choice of adjoint projector again differs by a sign:
\begin{align}
{\left( {\mathbb{P}}_{\mathbf{133}} \right)}_{KL}{}^{MN} & = - {(t_\alpha)}_{KL} {(t^\alpha)}^{MN}\,,
\end{align}
where ${\left( t^{\alpha} \right)}_{MN} = {\left( t^\alpha \right)}_M{}^P \Omega_{PN} = {\left(t^\alpha \right)}_{(MN)}$ are the generators of $E_{7(7)}$ and $\Omega_{MN}$ is the invariant symplectic form\footnote{Our conventions for contractions with the symplectic form are the standard in the literature:
\begin{align}
V^M = \Omega^{MN} V_N\,, \qquad V_M = V^N \Omega_{NM}\,,
\end{align}
with normalisation $\Omega^{MK} \Omega_{NK} = \delta^M_N$
}. Thus, as in the $E_{8(8)}$ case, we need to introduce a sign into every instance of ${\mathbb{P}}_{\mathbf{133}}$. Note, however, that the normalisation of the generators is still the same and follows from \eqref{eq:Norm}:
\begin{align}
{(t_\alpha)}_{MP} {(t^\alpha)}^{PN} = - {(t_\alpha)}_M{}^P {(t^\alpha)}_P{}^N =  - \frac{19}{8} \delta^N_M\,.
\end{align}
In our conventions, the projectors of $E_{7(7)}$ onto irreps in $\mathbf{56} \otimes \mathbf{56}$ are
\begin{align}
{\left( {\mathbb{P}}_{\mathbf{1}} \right)}_{KL}{}^{MN} & = \frac{1}{56} \Omega_{KL} \Omega^{MN}\\
{\left( {\mathbb{P}}_{\mathbf{133}} \right)}_{KL}{}^{MN} & = - {(t_\alpha)}_{KL} {(t^\alpha)}^{MN}\\
{\left( {\mathbb{P}}_{\mathbf{1463}} \right)}_{KL}{}^{MN} & = \delta^{(M}_K \delta^{N)}_L + {(t_\alpha)}_{KL} {(t^\alpha)}^{MN}\\
{\left( {\mathbb{P}}_{\mathbf{1539}} \right)}_{KL}{}^{MN} & = \delta^{[M}_K \delta^{N]}_L - \frac{1}{56} \Omega_{KL} \Omega^{MN}\,.
\end{align}
As before, these may be verified to square to themselves and project onto spaces of the correct dimension. Analogous to the $E_{8(8)}$ relations, we have
\begin{align}
\operatorname{Sym} \left( \mathbf{56} \otimes \mathbf{56} \right) & = \mathbf{133} \oplus \mathbf{1463}\,,\\
\operatorname{Asym} \left( \mathbf{56} \otimes \mathbf{56} \right) & = \mathbf{1} \oplus \mathbf{1539}\,,
\end{align}
which requires that the projectors satisfy the completeness relations
\begin{align}
{\left( {\mathbb{P}}_{\mathbf{133}} \right)}_{\hat{K} \hat{L}}{}^{\hat{M} \hat{N}} + {\left( {\mathbb{P}}_{\mathbf{1463}} \right)}_{\hat{K} \hat{L}}{}^{\hat{M} \hat{N}} & = \delta^{(M}_K \delta^{N)}_L\,,\\
{\left( {\mathbb{P}}_{\mathbf{1}} \right)}_{\hat{K} \hat{L}}{}^{\hat{M} \hat{N}} + {\left( {\mathbb{P}}_{\mathbf{1539}} \right)}_{\hat{K} \hat{L}}{}^{\hat{M} \hat{N}} & = \delta^{[M}_K \delta^{N]}_L\,,
\end{align}
with their sum giving the identity $\delta^M_K \delta^N_L$ on $\mathbf{56} \otimes \mathbf{56}$. Finally, since our normalisation of the generators is the same as \cite{Hohm:2013uia}, we still have the relation
\begin{align}\label{eq:E7Gen}
{(t_\alpha)}_M{}^K {(t^\alpha)}_N{}^L & = - {\left( \mathbb{P}_{\mathbf{133}} \right)}^K{}_M{}^L{}_N =  \frac{1}{24} \delta^K_M \delta^L_N + \frac{1}{12} \delta^K_N \delta^L_M + {(t_\alpha)}_{MN} {(t^\alpha)}^{KL} - \frac{1}{24} \Omega_{MN} \Omega^{KL} \,.
\end{align}
In the conventions that we employ, the generalised Lie derivative of the $E_{8(8)}$ and $E_{7(7)}$ EFT are
\begin{align}\label{eq:OurGenLie}
{\hat{\mathbb{L}}}_{(\hat{\Lambda}, \hat{\Sigma})} {\hat{V}}^{\hat{M}} & = {\hat{\Lambda}}^{\hat{N}} {\hat{\partial}}_{\hat{N}} {\hat{V}}^{\hat{M}} + 60 {\left({\mathbb{P}}_{\mathbf{248}} \right)}^{\hat{M}}{}_{\hat{K}}{}^{\hat{N}}{}_{\hat{L}} \left( {\hat{\partial}}_{\hat{N}} {\hat{\Lambda}}^{\hat{L}} + \frac{1}{60} {\hat{\Sigma}}_{\hat{R}} {\hat{f}}^{\hat{R} \hat{L}}{}_{\hat{N}} \right) {\hat{V}}^{\hat{K}} + \hat{\lambda} {\hat{\partial}}_{\hat{N}} {\hat{\Lambda}}^{\hat{N}} {\hat{V}}^{\hat{M}}\,\\
{\mathbb{L}}_{\Lambda} V^M & = \Lambda^N \partial_N V^M + 12 {\left( {\mathbb{P}}_{\mathbf{133}} \right)}^M{}_K{}^N{}_L \partial_N \Lambda^K V^L + \lambda \partial_N \Lambda^N V^M\,,
\end{align}
where we have further modified the definition of $\hat{\Sigma}$ to take into account the difference in normalisation of the structure constants in $E_{8(8)}$. Here, ${\hat{\Sigma}}_M$ is a parameter for an extra gauge transformation, not present in other ExFTs, that is required for the generalised Lie derivative to close appropriately. It is a constrained parameter (in the sense that it is treated in the same way as a derivative with respect to the section condition) and appears to be a common feature of 3-dimensional ExFTs, appearing in \cite{Hohm:2017wtr,Hohm:2014fxa,Hohm:2013jma}. In this form, we may read off the $Y$-tensor in these conventions as
\begin{align}
{\hat{Y}}^{\hat{M} \hat{N}}{}_{\hat{K} \hat{L}} & = 60 {\left({\hat{\mathbb{P}}}_{\mathbf{248}} \right)}^{\hat{M}}{}_{\hat{L}}{}^{\hat{N}}{}_{\hat{K}} + 2 \delta^{(\hat{M}}_{\hat{K}} \delta^{\hat{N})}_{\hat{L}}\,,\\
Y^{MN}{}_{KL} & = 12 {\left( {\mathbb{P}}_{\mathbf{133}} \right)}^M{}_L{}^N{}_K + \frac{1}{2} \delta^M_L \delta^N_K + \delta^M_K \delta^N_L\label{eq:E7Y}\\
	& = - 12 {(t_\alpha)}_{KL} {(t^\alpha)}^{MN} - \frac{1}{2} \Omega_{KL} \Omega^{MN}\,.
\end{align}
We choose to break the $E_{8(8)}$ generalised coordinates under $E_{7(7)} \times \operatorname{SL}(2)$ according to\footnote{We have condensed the notation from Section \ref{sec:ReductionLargerGenMetric} for convenience. Here, $Y^{Ma} = (Y^{M1}, Y^{M2}) \equiv (Y^M, Y^{\bar{M}})$ and $Y^i \equiv (Y^\sharp, Y^\natural, Y^\flat)$ of that section.}
\begin{align}\label{eq:E8E7Decomp}
{\hat{Y}}^{\hat{M}} = (Y^\alpha, Y^{Ma}, Y^i)\,.
\end{align} 
Then, with the normalisation that we employ, the $E_{8(8)}$ Killing form breaks under $E_{7(7)}$ according to
\begin{align}
{\hat{\kappa}}_{\hat{M} \hat{N}} & = \operatorname{diag} \left( \kappa_{\alpha \beta}, \Omega_{MN} \varepsilon_{ab}, g_{ij} \right)\\
{\hat{\kappa}}^{\hat{M} \hat{N}} & = \operatorname{diag} \left( \kappa^{\alpha \beta}, \Omega^{MN} \varepsilon^{ab}, g^{ij} \right)\label{eq:KillingDecomp}
\end{align}
where $g_{ij} = -\delta_{ij}$ is the $\operatorname{SL}(2)$ Killing form (taken to be negative-definite\footnote{Note that the Levi-Civita symbol consequently picks up an extra sign in its contractions:
\begin{align}
\varepsilon_{ikl} \varepsilon_j{}^{kl} = -2 g_{ij}
\end{align}}) that raises and lowers $\operatorname{SL}(2)$ adjoint indices and $\varepsilon_{ab}$ is the $\operatorname{SL}(2)$ invariant (i.e. $i,j = 1, 2,3$ and $a,b, = 1,2$). Here, $\kappa^{\alpha \beta}$ is the Killing form on $E_{7(7)}$ with the scaling defined above. The $E_{8(8)}$ structure constants consistent with this Killing form are
\begin{align}\label{eq:E8StructureConstants}
{\hat{f}}_{\hat{M} \hat{N} \hat{K}} & = \begin{cases}
{\hat{f}}_{\alpha \beta \gamma} & = \frac{1}{\sqrt{5}} f_{\alpha \beta \gamma}\,,\\
{\hat{f}}_{Ma Nb i} & = -\frac{1}{\sqrt{30}} \Omega_{MN} {\left( D_i \right)}_{ab}\,,\\
{\hat{f}}_{Ma Nb \alpha} & = -\frac{1}{\sqrt{5}} {(t_\alpha)}_{MN} \varepsilon_{ab}\,,\\
{\hat{f}}_{ijk} & = \frac{1}{\sqrt{30}} \varepsilon_{ijk}\,.
\end{cases}
\end{align}
Here, the $D_i$ are the representation matrices of the $\mathbf{3}$ of $\operatorname{SL}(2)$ which we take to be anti-Hermitian Pauli matrices
\begin{align}
{(D_i)}_a{}^b = \frac{i}{2} {(\sigma_i)}_a{}^b\,.
\end{align}
We may also identify the adjoint representation as the symmetric representation via the relation
\begin{align}
{(D_i)}^{ab} {(D^i)}^{cd} = - \frac{1}{4} \left( \varepsilon^{ac} \varepsilon^{bd} + \varepsilon^{ad} \varepsilon^{bc} \right)\,.
\end{align}
We are now ready to reduce the generalised Lie derivative \eqref{eq:OurGenLie}. Our ansatz for the reduction is to break the $\operatorname{SL}(2)$ covariance by selecting ${\hat{Y}}^{M1} \equiv Y^M$ to be our $E_{7(7)}$ extended coordinates, and applying the following:
\begin{align}\label{eq:E8E7Ansatz}
{\hat{\partial}}_{M1} & = \partial_{M} \qquad {\hat{\partial}}_{M2} = {\hat{\partial}}_{\alpha} = {\hat{\partial}}_i = 0\,,\\
{\hat{V}}^{\hat{M}} & = ({\hat{V}}^{M1} = V^M , 0, 0, 0)\,,\\
{\hat{\Lambda}}^{\hat{M}} & = ({\hat{\Lambda}}^{M1} = \Lambda^M , 0, 0, 0)\,,\\
{\hat{\Sigma}}^{\hat{M}} & = ({\hat{\Sigma}}^{M1} = \Sigma^M , 0, 0, 0)
\end{align}
i.e. drop all coordinate dependence on the extra extended coordinate $\{ {\hat{Y}}^{M2}, {\hat{Y}}^\alpha, {\hat{Y}}^i \}$ and set all components of the fields to zero, apart from those in the $E_{7(7)}$ section. As such, we are only interested in the $M1$ components of all our objects. Consider the restriction of the $E_{8(8)}$ adjoint projector onto $(\mathbf{56},\mathbf{2})$ indices $Ma$:
\begin{align}\label{eq:248Restriction}
{\left( {\hat{\mathbb{P}}}_{\mathbf{248}} \right)}^{Ma}{}_{Kc}{}^{Nb}{}_{Ld} & = - \frac{1}{5} {(t_{\alpha})}_K{}^M {(t^\alpha)}_L{}^N \delta^a_c \delta^b_d - \frac{1}{120} \delta^M_K \delta^N_L ( \delta^a_d \delta^b_c - \varepsilon^{ab} \varepsilon_{cd})\\ 
& = \frac{1}{60} {\left( 12 {\left( {\mathbb{P}}_{\mathbf{133}}\right)}^M{}_K{}^N{}_L \delta^a_c \delta^b_d  - \frac{1}{2} \delta^M_K \delta^N_L \delta^a_d \delta^b_c \right)} + \frac{1}{120} \delta^M_K \delta^N_L \varepsilon^{ab} \varepsilon_{cd}\,.
\end{align}
Upon setting $a=b=c=d=1$ (i.e.\ restricting entirely to the $E_{7(7)}$ section), the last term drops out and we see that the reduction of the adjoint projector of $E_{8(8)}$ yields the adjoint projector in $E_{7(7)}$ as well as an extra $\delta \delta$ term:
\begin{align}\label{eq:ProjectorReduction}
{\left( {\hat{\mathbb{P}}}_{\mathbf{248}} \right)}^{M1}{}_{K1}{}^{N1}{}_{L1} & = \frac{1}{60} {\left( 12 {\left( {\mathbb{P}}_{\mathbf{133}}\right)}^M{}_K{}^N{}_L - \frac{1}{2} \delta^M_K \delta^N_L \right)}\,.
\end{align}
The generalised Lie derivative then reduces as
\begin{align}
{\hat{\mathbb{L}}}_{(\hat{\Sigma}, \hat{\Lambda})} {\hat{V}}^{\hat{M}} & \longrightarrow \Lambda^{N} \partial_{N} V^M + \left( 12 {\left( {\mathbb{P}}_{\mathbf{133}} \right)}^M{}_K{}^N{}_L - \frac{1}{2} \delta^M_K \delta_L^N \right) \left( \partial_N \Lambda^L + \frac{1}{60} \Sigma_R {\hat{f}}^{R L}{}_N \right) V^K + \hat{\lambda} \partial_N \Lambda^N V^M\,,\\
& \longrightarrow \Lambda^{N} \partial_{N} V^M + 12 {\left( {\mathbb{P}}_{\mathbf{133}} \right)}^M{}_K{}^N{}_L \partial_N \Lambda^L + \left(\hat{\lambda} -\frac{1}{2}\right) \partial_N \Lambda^N V^M \,.
\end{align}
In particular, all terms involving the extra gauge parameter $\hat{\Sigma}$ automatically drop out under our ansatz. We thus recover the $E_{7(7)}$ generalised Lie derivative, where the weight under the $E_{7(7)}$ generalised Lie derivative is identified with a shift of the weight under the $E_{8(8)}$ generalised diffeomorphisms:
\begin{align}
\hat{\lambda} \rightarrow \hat{\lambda} - \frac{1}{2} \coloneqq \lambda\,.
\end{align}
There is a naturally interpretation of this if we employ the $Y$-tensor. The reduction of the $Y$-tensor induced by \eqref{eq:ProjectorReduction} is
\begin{align}\label{eq:E8ToE7YTensor}
{\hat{Y}}^{M1 N1}{}_{K1 L1} & = Y^{MN}{}_{KL}\,.
\end{align}
The effective weight ($\lambda_{\text{eff.}} = \lambda_V + \omega$) term reduces exactly and we are left with
\begin{align}
{\hat{\mathbb{L}}}_{(\hat{\Sigma}, \hat{\Lambda})} {\hat{V}}^{\hat{M}} & = {\hat{\Lambda}}^{\hat{N}} {\hat{\partial}}_{\hat{N}} {\hat{V}}^{\hat{M}}  - {\hat{V}}^{\hat{N}} {\hat{\partial}}_{\hat{N}} {\hat{\Lambda}}^{\hat{M}} + {\hat{Y}}^{\hat{M} \hat{N}}{}_{\hat{K} \hat{L}} {\hat{\partial}}_{\hat{N}} {\hat{\Lambda}}^{\hat{K}}  {\hat{V}}^{\hat{L}} + \lambda_{\text{eff.}} {\hat{\partial}}_{\hat{N}} {\hat{\Lambda}}^{\hat{N}} {\hat{V}}^{\hat{M}} - {\hat{\Sigma}}_{\hat{K}} {\hat{f}}^{\hat{K} \hat{M}}{}_{\hat{N}} {\hat{V}}^{\hat{N}}\,\\
& \longrightarrow \Lambda^N \partial_N V^M - V^N \partial_N \Lambda^M + Y^{MN}{}_{KL} \partial_N \Lambda^K V^L + \lambda_{\text{eff.}} \partial_N \Lambda^N V^M\,,\\
& \coloneqq\Lambda^N \partial_N V^M - V^N \partial_N \Lambda^M + Y^{MN}{}_{KL} \partial_N \Lambda^K V^L + \lambda_{\text{eff.}} \partial_N \Lambda^N V^M\,.
\end{align}
In this picture, we thus obtain a transfer of weight from the universal weight to the weight of $V$ such that the effective weight in the two theories remains the same:
\begin{align}
\hat{\lambda} - 1= \lambda - \frac{1}{2} = \lambda_{\text{eff.}}
\end{align}
Recall that the generalised gauge field that forms the starting point of the rather intricate tensor hierarchy has an effective weight of 0 in both theories. The fact that this is not disturbed in this reduction is perhaps to be expected.\par
For completeness, we note the following restrictions to the $(\mathbf{56,2})$ piece of the $E_{8(8)}$ projectors:
\begin{align}\label{eq:ProjRestrictions}
{\left( {\hat{\mathbb{P}}}_{\mathbf{1}} \right)}_{Kc Ld}{}^{Ma Nb} & = \frac{1}{248} \Omega_{KL} \Omega^{MN} \varepsilon^{ab} \varepsilon_{cd}\\
{\left( {\hat{\mathbb{P}}}_{\mathbf{248}} \right)}_{Kc Ld}{}^{Ma Nb} & = - \frac{1}{5} {(t_\alpha)}_{KL} {(t^\alpha)}^{MN} \varepsilon^{ab} \varepsilon_{cd} + \frac{1}{60} \Omega_{KL} \Omega^{MN} \delta^{(a}_d \delta^{b)}_c\\
\begin{split}
{\left( {\hat{\mathbb{P}}}_{\mathbf{3875}} \right)}_{Kc Ld}{}^{Ma Nb} & = \left( \frac{1}{14} \delta^M_K \delta^N_L - \frac{3}{7} {(t_\alpha)}_K{}^M {(t^\alpha)}_L{}^N - \frac{1}{56} \delta^N_K \delta^M_L \right) \delta^a_c \delta^b_d\\
& \qquad + \left( \frac{1}{14} \delta^N_K \delta^M_L - \frac{3}{7} {(t_\alpha)}_K{}^N {(t^\alpha)}_L{}^M - \frac{1}{56} \delta^M_K \delta^N_L \right) \delta^b_c \delta^a_d\\
& \qquad + \frac{1}{56} \left( 2 \delta^{(M}_K \delta^{N)}_L - \Omega_{KL} \Omega^{NM} \right) \varepsilon^{ab} \varepsilon_{cd}
\end{split}\\
\begin{split}
{\left( {\hat{\mathbb{P}}}_{\mathbf{27000}} \right)}_{Kc Ld}{}^{Ma Nb} & = \left( \frac{6}{14} \delta^M_K \delta^N_L + \frac{3}{7} {(t_\alpha)}_K{}^M {(t^\alpha)}_L{}^N + \frac{1}{56} \delta^N_K \delta^M_L \right) \delta^a_c \delta^b_d\\
& \qquad + \left( \frac{6}{14} \delta^N_K \delta^M_L  + \frac{3}{7} {(t_\alpha)}_K{}^N {(t^\alpha)}_L{}^M + \frac{1}{56} \delta^M_K \delta^N_L \right) \delta^b_c \delta^a_d\\
& \qquad - \frac{1}{56} \left( 2 \delta^{(M}_K \delta^{N)}_L - \frac{24}{31} \Omega_{KL} \Omega^{NM} \right) \varepsilon^{ab} \varepsilon_{cd}
\end{split}\\
{\left( {\hat{\mathbb{P}}}_{\mathbf{30380}} \right)}_{Kc Ld}{}^{Ma Nb} & = \frac{1}{5} {(t_\alpha)}_{KL} {(t^\alpha)}^{MN} \varepsilon^{ab} \varepsilon_{cd} - \frac{1}{60} \Omega_{KL} \Omega^{MN} \delta^{(a}_d \delta^{b)}_c + \delta^{[Ma}_{Kc} \delta^{Nb]}_{Ld}\,.
\end{align}
Note that the (anti-)symmetrisation of the composite indices $Ma$ is given by
\begin{align}
\delta^{(Ma}_{Kc} \delta^{Nb)}_{Ld} & = \frac{1}{2} \left( \delta^M_K \delta^N_L \delta^a_c \delta^b_d + \delta^N_K \delta^M_L \delta^b_c \delta^a_d \right)\,,\\
\delta^{[Ma}_{Kc} \delta^{Nb]}_{Ld} & = \frac{1}{2} \left( \delta^M_K \delta^N_L \delta^a_c \delta^b_d - \delta^N_K \delta^M_L \delta^b_c \delta^a_d \right)\,.
\end{align}
The sum of the relevant projectors then still satisfy
\begin{align}
{\left( {\hat{\mathbb{P}}}_{\mathbf{1}} \right)}_{Kc Ld}{}^{Ma Nb} + {\left( {\hat{\mathbb{P}}}_{\mathbf{3875}} \right)}_{Kc Ld}{}^{Ma Nb} + {\left( {\hat{\mathbb{P}}}_{\mathbf{27000}} \right)}_{Kc Ld}{}^{Ma Nb} & = \delta^{(Ma}_{Kc} \delta^{Nb)}_{Ld}\,,\\
{\left( {\hat{\mathbb{P}}}_{\mathbf{248}} \right)}_{Kc Ld}{}^{Ma Nb} + {\left( {\hat{\mathbb{P}}}_{\mathbf{30380}} \right)}_{Kc Ld}{}^{Ma Nb} & = \delta^{[Ma}_{Kc} \delta^{Nb]}_{Ld}\,.
\end{align}
We now look in more detail at the $E_{8(8)}$ section constraints (which is typically viewed as the vanishing of some other representation $\mathcal{R}_2 \subset \mathcal{R}_1 \otimes \mathcal{R}_1$)
\begin{align}\label{eq:E8Section}
{\left( {\hat{\mathbb{P}}}_{\mathbf{1} \oplus \mathbf{248} \oplus \mathbf{3875}} \right)}_{\hat{K} \hat{L}}{}^{\hat{M} \hat{N}} {\hat{C}}_{\hat{M}} \otimes {{\hat{C}^\prime}}_{\hat{N}} & =  0\,,
\end{align}
where ${\hat{C}}_{\hat{M}}, {\hat{C}}^\prime_{\hat{M}} \in \{ {\hat{\partial}}_{\hat{M}}, {\hat{B}}_{\mu \hat{M}}, {\hat{\Sigma}}_{\hat{M}} , {\hat{C}}_{\hat{\mu} \hat{\nu} \hat{M}}{}^{\hat{N}}, \ldots\}$ are covariantly constrained objects. However, since the projectors do not reduce exactly (e.g.\ in \eqref{eq:ProjectorReduction}), it is much more convenient to consider the section constraint in terms of the $Y$-tensor which does reduce exactly \eqref{eq:E8ToE7YTensor}. We thus consider the combination
\begin{align}
{\hat{Y}}^{\hat{M} \hat{N}}{}_{\hat{K} \hat{L}} {\hat{C}}_{\hat{M}} \otimes {\hat{C}^\prime}_{\hat{N}} & = { \left(62 {\hat{\mathbb{P}}}_{\mathbf{1}} + 30 {\hat{\mathbb{P}}}_{\mathbf{248}} +  14 {\hat{\mathbb{P}}}_{\mathbf{3875}} \right)}_{\hat{L} \hat{K}}{}^{\hat{M} \hat{N}} {\hat{C}}_{\hat{M}} \otimes {{\hat{C}}^\prime}_{\hat{N}} = 0\,.
\end{align}
Substituting in the explicit forms of the projectors, the section constraints read
\begin{align}
\hat{\kappa}^{\hat{K} \hat{L}} {\hat{C}}_{\hat{K}} \otimes {\hat{C}}^\prime_{\hat{L}} & = 0\,,\\
\hat{f}^{\hat{P} \hat{K} \hat{L}} {\hat{C}}_{\hat{K}} \otimes {\hat{C}}^\prime_{\hat{L}} & = 0\,,\\
\left( \delta^{(\hat{K}}_{\hat{M}} \delta^{\hat{L})}_{\hat{N}} - 30 {\hat{f}}^{\hat{P} (\hat{K}}{}_{\hat{M}} {\hat{f}}_{\hat{P}}{}^{\hat{L})}{}_{\hat{N}} \right)  {\hat{C}}_{\hat{K}} \otimes {\hat{C}}^\prime_{\hat{L}} & = 0,
\end{align}
The $E_{7(7)}$ analogue of \eqref{eq:E8Section} is
\begin{align}
{\left( {\mathbb{P}}_{\mathbf{1} \oplus \mathbf{133}} \right)}_{MN}{}^{KL} \partial_K \otimes \partial_L & = 0
\end{align}
which can equivalently be obtained from the $Y$-tensor:
\begin{align}
Y^{KL}{}_{MN} \partial_K \otimes \partial_L = {\left( 12 {\mathbb{P}}_{\mathbf{133}} + 28 {\mathbb{P}}_{\mathbf{1}} \right)}_{NM}{}^{KL} \partial_K \otimes \partial_L = 0\,.
\end{align}
Expanding each of these in terms of invariants, the section condition for $E_{7(7)}$ EFT reads
\begin{align}
\Omega^{KL} \partial_K \otimes \partial_L & = 0\,,\\
{(t^{\alpha})}^{KL} \partial_K \otimes \partial_L & = 0\,.
\end{align}
We now expand each of the $E_{8(8)}$ constraints to explicitly verify that they reduce to the $E_{7(7)}$ section constraints. The only non-trivial constraint obtainable from the first of the $E_{8(8)}$ constraints, under the ansatz that only ${\hat{\partial}}_{M1} \neq 0$, is
\begin{align}
\Omega^{KL} \varepsilon^{ab} \partial_{Ka} \otimes \partial_{Lb} \biggr\vert_{a=b=1}=0 
\end{align}
which always vanishes and is thus vacuous on the $E_{7(7)}$ section. Looking at \eqref{eq:E8StructureConstants}, the only two non-vanishing structure constants give
\begin{align}
\hat{f}^{\alpha Ka Lb} {\partial}_{Ka} \otimes {\partial}_{Lb} \biggr\vert_{a=b=1} & = 0 \qquad \longrightarrow \qquad {(t^\alpha)}^{KL} \varepsilon^{ab} {\partial}_{Ka} \otimes {\partial}_{Lb} \biggr\vert_{a=b=1} = 0\,,\\
\hat{f}^{i Ka Lb} {\partial}_{Ka} \otimes {\partial}_{Lb} \biggr\vert_{a=b=1} & = 0 \qquad \longrightarrow \qquad \Omega^{KL} {(D^i)}^{ab} {\partial}_{Ka} \otimes {\partial}_{Lb} \biggr\vert_{a=b=1} = 0\,.
\end{align}
The first is, again, vacuous but the second gives one of the $E_{7(7)}$ section constraints
\begin{align}
\Omega^{KL} \partial_K \otimes \partial_L = 0\,.
\end{align}
The final constraint is symmetric under $\hat{M} \leftrightarrow \hat{N}$ and so there are 6 possible independent constraints. If [$\hat{M} = \alpha$ and $\hat{N} = \beta$] or [$\hat{M} = i$ and $\hat{N} = j$], then it is vacuous since the result is proportional to $\varepsilon^{ab} {\hat{\partial}}_{Ka} \otimes {\hat{\partial}}_{Lb}$. If [$\hat{M} = M$ and $\hat{N} = N$], then we obtain
\begin{align}
\left( \frac{3}{4} \delta^{(K}_M \delta^{L)}_N - 6 {(t_\alpha)}_M{}^{(K} {(t^\alpha)}_N{}^{L)} \right) \partial_{K} \otimes \partial_{L} = 0\,,
\end{align}
which can be brought to the form
\begin{align}
-6 {(t_\alpha)}_{MN} {(t^\alpha)}^{KL} \partial_K \otimes \partial_L = 0
\end{align}
upon substituting in the relation \eqref{eq:E7Gen}, thereby recovering the second $E_{7(7)}$ section constraint. In fact, choosing $[\hat{M} = \alpha \text{ and }  \hat{N} = j]$ gives
\begin{align}
-\frac{\sqrt{6}}{2} {(t_\alpha)}^{KL} {(D_j)}^{(cd)} {\hat{\partial}}_{Kc} \otimes {\hat{\partial}}_{Ld} = 0
\end{align}
and so also recovers the second $E_{7(7)}$ section condition. The remaining two choices, [$\hat{M} = \alpha$ and $\hat{N} = Nc$] or [$\hat{M} = j$ and $\hat{N} = Nc$], both vanish trivially as there are no structure constants of the required index structures. Thus the set of all $E_{8(8)}$ constraints, subject to \eqref{eq:E8E7Ansatz}, recovers both of the $E_{7(7)}$ constraints and nothing else.\par
Of particular interest here is that the maximal subgroup of $E_{8(8)}$ has an additional $\operatorname{SL}(2)$ symmetry over the $E_{7(7)}$ symmetry that is made manifest in the usual EFT. In particular, we have an $\operatorname{SL}(2)$ doublet worth of $E_{7(7)}$ coordinate representations inside the $\mathbf{248}$ of $E_{8(8)}$. In reducing the section constraints above, we explicitly broke the $\operatorname{SL}(2)$ covariance and chose ${\hat{Y}}^{M1} \equiv Y^M$ to be the $E_{7(7)}$ extended coordinates. However, this choice was entirely arbitrary; we could equally have chosen ${\hat{Y}}^{M2} \equiv Y^{\bar{M}}$ (in the notation of Section \ref{sec:ReductionLargerGenMetric}) to be the $E_{7(7)}$ extended coordinates instead. More generally, we could have chosen $a=1$ to be the extended coordinates on one local patch and $a=2$ to be the extended coordinates on another local patch such that we require an $\operatorname{SL}(2)$ transformation to patch the two together in a manner reminiscent of the T- and U-folds of Hull. However, the interpretation of such configurations from the $E_{7(7)}$ perspective is not clear. It is not a non-geometric configuration of the sort that has been previously studied in the literature since the patching $\operatorname{SL}(2)$ transformation is not even a $G$-transformation and is thus not generated by the local symmetries of the supergravity fields. These may be considered as examples of `truly' non-geometric backgrounds, of the sort hypothesised in \cite{Berman:2018okd,Otsuki:2019owg} and in much earlier works such as \cite{Dabholkar:2005ve}, in the sense that they cannot be related to any geometric configurations by duality transformations.\par
\subsection{\texorpdfstring{$\operatorname{SL}(5)$}{SL(5)} EFT to \texorpdfstring{$\operatorname{SL}(3) \times \operatorname{SL}(2)$}{SL(3)xSL(2)} EFT and \texorpdfstring{$\operatorname{O}(3,3)$}{O(3,3)} DFT}
We begin with the $\operatorname{SL(5)}$ section constraint
\begin{align}\label{eq:SL5Section}
{\hat{Y}}^{\hat{M}\hat{N}}{}_{\hat{K}\hat{L}} \partial_{\hat{M}} \otimes \partial_{\hat{N}} & = 3! \delta^{{\hat{\underbar{m}}}_1 {\hat{\underbar{m}}}_2 {\hat{\underbar{n}}}_1 {\hat{\underbar{n}}}_2}_{{\hat{\underbar{k}}}_1 {\hat{\underbar{k}}}_2 {\hat{\underbar{l}}}_1 {\hat{\underbar{l}}}_2}\partial_{{\hat{\underbar{m}}}_1 {\hat{\underbar{m}}}_2} \otimes \partial_{{\hat{\underbar{n}}}_1 {\hat{\underbar{n}}}_2} = 0\,,
\end{align}
where we have used the fact that the $\mathcal{R}_1 = \mathbf{10}$ indices can be written in terms of the 5-dimensional indices ${{\hat{\underbar{m}}}}_1 = 1, \ldots, 5$ as an antisymmetric pair $\hat{M} = [{\hat{\underbar{m}}}_1 {\hat{\underbar{m}}}_2]$. It is well-known that the section condition has only two inequivalent solutions; the so-called M-theory section and the Type IIB section. These are solutions in the sense that they give rise to theories with no further constraints. Here, we consider the $\operatorname{SL}(3) \times \operatorname{SL}(2)$ and $\operatorname{O}(3,3)$ ExFTs as arising from \emph{partial} solutions to the section conditions; choices of dropped coordinate dependences that give rise to theories with residual constraints.\par
These come in two classes. The first are obtained from imposing only a subset of the constraints that solve the section condition in the usual manner and these are expected to give the section conditions of the lower-dimensional EFTs. The $\operatorname{SL}(5) \rightarrow \operatorname{SL}(3) \times \operatorname{SL}(2)$ reduction falls under this category. Further imposing the M-theory or Type IIB sections of the child ExFT should yield the M-theory or the Type IIB solutions of the child theory; this partial solution should be extensible to a full solution of the parent ExFT's section constraint. The second is a partial solution that cannot be extended to a full M-theory or Type IIB solution in the manner described above. Whilst it may sound like this in conflict with the usual narrative that there are only two inequivalent solutions, we emphasise that this is not the case since it is not a full solution of the section condition. The $\operatorname{SL}(5) \rightarrow \operatorname{O}(3,3)$ reduction is an example of such a reduction.\par
We begin with a brief description of the usual story. For the M-theory section, we decompose $\operatorname{SL}(5)$ under $\operatorname{GL}(4)$ according to
\begin{align}
\mathbf{10} \rightarrow \mathbf{4}_{-3} \oplus \mathbf{6}_2
\end{align}
which corresponds to the decomposition of the generalised coordinates
\begin{align}
{\hat{Y}}^{[\hat{\underbar{m}}_1 \hat{\underbar{m}}_2]} = ({\hat{Y}}^{\hat{m}_1 5}, {\hat{Y}}^{\hat{m}_1 \hat{m}_2}) = \left( Y^{\hat{m}_1 \hat{m}_2}, \frac{1}{2} \varepsilon^{\hat{m}_1 \hat{m}_2 \hat{n}_1 \hat{n}_2} Y_{\hat{n}_1 \hat{n}_2} \right)\,.
\end{align}
The section condition decomposes under this branching as (here, we have rewritten the generalised Kronecker delta in terms of $\varepsilon$-symbols)
\begin{align}\label{eq:SL5GL4Section}
\begin{aligned}
\varepsilon^{\hat{k} \hat{m}_1 \hat{m}_2 \hat{n}_1 5} \left( \partial_{\hat{m}_1 \hat{m}_2} \otimes \partial_{\hat{n}_1 5} + \partial_{\hat{n}_1 5} \otimes \partial_{\hat{m}_1 \hat{m}_2} \right) & = 0\,,\\
\varepsilon^{5 \hat{m}_1 \hat{m}_2 \hat{n}_1 \hat{n}_2} \partial_{\hat{m}_1 \hat{m}_2} \otimes \partial_{\hat{n}_1 \hat{n}_2} & = 0\,.
\end{aligned}
\end{align}
It is then easy to see that these are solved by dropping all coordinate dependences on the membrane wrapping coordinates
\begin{align}\label{eq:MThSoln}
\partial_{\hat{m}_1 \hat{m}_2} \sim \partial^{\hat{n}_1 \hat{n}_2} = 0 \rightarrow \text{M-theory}\,.
\end{align}
For the Type IIB section, we instead decompose $\operatorname{SL}(5)$ under $\operatorname{SL}(3) \times \operatorname{SL}(2)$ as
\begin{align}
\mathbf{10} \rightarrow {(\bar{\mathbf{3}}, \mathbf{1})}_{4} \oplus {(\mathbf{3,2})}_{-1} \oplus {(\mathbf{1,1})}_{-6}\,.
\end{align}
The index splitting $\hat{\underbar{m}} = (m, \alpha)$ induces the decomposition of the generalised coordinates
\begin{align}
{\hat{Y}}^{[{\hat{\underbar{m}}}_1 {\hat{\underbar{m}}}_2]} \rightarrow \left( Y^{[m_1 m_2]}, Y^{m_1 \alpha}, Y^{[\alpha \beta]} \right)\,,
\end{align}
where $m =1,2,3$ and $\alpha = 4,5$, and the decomposition of the section condition 
\begin{align}\label{eq:IIBSplitting}
\begin{aligned}
\delta^{m_1 m_2 n_1}_{k_1 l_1 l_2} \delta^\alpha_\gamma \left( \partial_{m_1 m_2} \otimes \partial_{n_1 \alpha} + \partial_{n_1 \alpha} \otimes \partial_{m_1 m_2} \right) & = 0\,,\\
\delta^{m_1 n_1}_{k_1 l_1} \delta^{\alpha \beta}_{\gamma \delta} \left( \partial_{m_1 \alpha} \otimes \partial_{n_1 \beta} - \partial_{m_1 n_1} \otimes \partial_{\alpha \beta} - \partial_{\alpha \beta} \otimes \partial_{m_1 n_1} \right) & = 0\,.
\end{aligned}
\end{align}
The IIB section is then given by the choice 
\begin{align}\label{eq:IIBSoln}
\partial_{m \alpha} = \partial_{\alpha \beta} = 0 \rightarrow \text{IIB}\,.
\end{align}
We now turn to how we can recover the $\operatorname{SL}(3) \times \operatorname{SL}(2)$ EFT and $\operatorname{O}(3,3)$ DFT section conditions from the above. Rather than follow the M-theory reduction, we consider what happens if we choose to set $\partial_{\hat{m}_1 5} = 0$ instead of the M-theory solution \eqref{eq:MThSoln}. This does \emph{not} fully solve the section condition and \eqref{eq:SL5GL4Section} instead reduces to
\begin{align}
\varepsilon^{\hat{m}_1 \hat{m}_2 \hat{n}_1 \hat{n}_2}  \partial_{\hat{m}_1 \hat{m}_2} \otimes \partial_{\hat{n}_1 \hat{n}_2}= 0\,.
\end{align}
To find solutions for this we further split $\hat{m} = ( m, 4)$, similar to the Type IIB solution, for which 
\begin{align}
\varepsilon^{m_1 m_2 n_1 4} \left( \partial_{m_1 m_2} \otimes \partial_{n_1 4} + \partial_{n_1 4} \otimes \partial_{m_1 m_2} \right) = 0\,.
\end{align}
Owing to the notation $\otimes$, this can be compressed down to a single term
\begin{align}
\varepsilon^{m_1 m_2 n_1} \partial_{m_1 m_2} \otimes \partial_{n_1 4} = 0\,.
\end{align}
 If we identify $\varepsilon^{m_1 m_2 n_1} \partial_{m_1 m_2} \sim \partial^{n_1}$ and $\partial_{n_1 4} \sim \partial_{n_1}$ we obtain the $\operatorname{O}(3,3)$ DFT section condition
\begin{align}
\partial^m \otimes \partial_m = 0\,.
\end{align}
We stress that, despite yielding a 3-dimensional section, this is \emph{not} obtainable from the Type IIB solution since the usual coordinates of the DFT descend wholly from the usual coordinates of the M-theory section (recall that a Type IIB section shares two coordinates with the M-theory section but takes a third coordinate from the membrane wrapping directions). However, nor should it be thought of as a Type IIA `section', since it was constructed as an independent (partial) solution to the M-theory section. It is perhaps better thought of an independent path through which one can obtain the Type IIA theory from the $\operatorname{SL}(5)$ theory.\par
The $\operatorname{SL}(3) \times \operatorname{SL}(2)$ section condition can be recovered from a similar analysis applied to the alternate splitting \eqref{eq:IIBSplitting}. Rather than taking the solution that gives the Type IIB section \eqref{eq:IIBSoln}, if we instead choose to set
\begin{align}\label{eq:SL3SL2SectionSolution}
\partial_{m_1 m_2} = 0
\end{align}
then the only non-trivial remaining constraint is
\begin{align}
\delta^{m_1 n_1}_{k_1 l_1} \delta^{\alpha \beta}_{\gamma \delta} \partial_{m_1 \alpha} \otimes \partial_{n_1 \beta} = 0
\end{align}
which is the $\operatorname{SL}(3) \times \operatorname{SL}(2)$ section condition. We have summarised the ways that one obtains the various theories that we discussed above in Table \ref{tab:MTheorySolutions}. We mention a couple of things to note. Firstly, since we obtained the $\operatorname{SL}(3) \times \operatorname{SL}(2)$ generalised metric from a KK reduction, we should further impose the KK isometry $\partial_{\alpha \beta} = 0$. However, since this is independent of the section constraint, we have not listed this in the table. Additionally, the table is not exhaustive; a circle reduction of the generalised metric in the Type IIB parametrisation is also likely to give further partial solutions that we have not discussed here but we expect that their respective section conditions can be recovered in an analogous fashion.\par
\begin{table}
\centering
\begin{tabulary}{0.7\textwidth}{LCL}
\toprule
Constraint & Residual Section Constraint & Resulting Theory\\
\midrule
$\partial_{\hat{m}\hat{n}} = 0$ & $-$ & M-theory\\
$\partial_{m \alpha} = \partial_{\alpha \beta} = 0$ & $-$ & IIB\\
$\partial_{\hat{m}5} = 0$ & $\partial^m \otimes \partial_m = 0$ & $\operatorname{O}(3,3) \text{ DFT}$\\
$\partial_{mn} = 0$ & $\delta^{mn}_{pq} \delta^{\alpha \beta}_{\gamma \delta} \partial_{m\alpha} \otimes \partial_{n \beta} = 0$ & $\operatorname{SL}(3) \times \operatorname{SL}(2) \text{ EFT}$\\
\bottomrule
\end{tabulary}
\caption{The possible solutions and partial solutions that can be obtained from the $\operatorname{SL}(5)$ section constraint.}
\label{tab:MTheorySolutions}
\end{table}
However, in the absence of such explicit calculations, it is not clear which choices of partial solutions admit such interpretations. For example, rather than taking the full IIB section, one might be tempted to consider setting only $\partial_{m \alpha} = 0$ to give rise to a constrained theory with some residual section condition. However, there is no reason to expect that such arbitrary choices will lead to recognisable ExFTs. A full classification would likely be based on considering subgroups of $\hat{G}$ which have representations that are consistent with the branching of the parent ExFT's representations under that reduction (whether it be a circle/torus reduction or more general ansatzes). However, it is hopefully clear that the much more involved section conditions of larger EFTs may house more derivative EFTs than one might initially expect.

\section{Rewriting the \texorpdfstring{$\mathcal{B}$-$\mathcal{F}$}{BF} Term in \texorpdfstring{$E_{8(8)}$}{E8(8)} EFT}\label{sec:BF}
Despite the similarity in the way that the EFTs are constructed, even a cursory look reveals that the details of the theories differ wildly by dimension. Focusing on the $d=3$ and $d=4$ EFTs, we have already outlined the novel features of the $E_{8(8)}$ coordinates, generalised Lie derivative and generalised metric but its differences from $E_{7(7)}$ EFT extends to the action as well. The former possesses the signature vector-scalar duality of $d=3$ theories and a full Lagrangian description whilst the vector-vector duality of the latter is demoted to an extra condition on a pseudo-action as a twisted self-duality constraint on the generalised fieldstrength $\mathcal{F}_{\mu \nu}{}^M$. Additionally, the Chern-Simons term of $E_{8(8)}$ EFT must somehow source part of the topological and Yang-Mills terms of $E_{7(7)}$ EFT (since these are where the gauge sectors are encoded in the respective theories), but the identification is far from obvious.\par
In this section, we study how a subset of the terms in the $E_{8(8)}$ EFT Lagrangian density rearrange themselves into the terms found in the $E_{7(7)}$ EFT Lagrangian density. The full identification of the reduction will also involve a careful study of the topological terms in the two theories but this is complicated by the fact that one can add and remove terms that vanish under section condition (including terms that involve covariantly constrained objects) to both theories. Such topological terms may also involve boundary terms of the type described in \cite{Berman:2011kg} and are well beyond the scope of this paper.\par
Before we proceed, we first make a list of the pieces of the $E_{7(7)}$ fields that need to be identified from amongst the $E_{8(8)}$ fields. The external coordinates of $E_{7(7)}$ EFT are given by $\mu = (\hat{\mu}, 4)$ where $\hat{\mu} = 1,2,3$ are the external coordinates of the $E_{8(8)}$ EFT and the direction $4$ is to be identified from the internal coordinates in $E_{8(8)}$. The $E_{7(7)}$ EFT gauge fields are $\{\mathcal{A}_\mu{}^M, \mathcal{B}_{\mu \nu M}, \mathcal{B}_{\mu \nu\alpha}\}$ where $M$ and $\alpha$ index the $\mathbf{56}$ coordinate- and $\mathbf{133}$ adjoint-representations of $E_{7(7)}$ EFT respectively. They decompose under this $3+1$ splitting to components of the form
\begin{align}
{\mathcal{A}}_{\mu}{}^M = \begin{pmatrix} {\mathcal{A}}_{\hat{\mu}}{}^M\\ {\mathcal{A}}_4{}^M \end{pmatrix}\,, \qquad {\mathcal{B}}_{\mu \nu M} = \begin{pmatrix} {\mathcal{B}}_{\hat{\mu} \hat{\nu} M}\\ {\mathcal{B}}_{\hat{\mu} 4M} \end{pmatrix}\,, \qquad {\mathcal{B}}_{\mu \nu \alpha} = \begin{pmatrix} {\mathcal{B}}_{\hat{\mu} \hat{\nu} \alpha}\\ {\mathcal{B}}_{\hat{\mu} 4 \alpha} \end{pmatrix}\,.
\end{align}
Some of these have an obvious $E_{8(8)}$ origin as follows:
\begin{align}\label{eq:E7E8GaugeFields}
{\mathcal{A}}_{\hat{\mu}}{}^M = {\hat{\mathcal{A}}}_{{\hat{\mu}}}{}^M\,, \qquad {\mathcal{B}}_{\hat{\mu} 4M} \sim {\hat{\mathcal{B}}}_{\hat{\mu} M} + \ldots \,, \qquad {\mathcal{B}}_{\hat{\mu} 4\alpha} \sim {\hat{\mathcal{B}}}_{\hat{\mu} \alpha} + \ldots\,,
\end{align}
which are sourced from the $E_{8(8)}$ generalised gauge fields, decomposed under $E_{7(7)}$, as ${\hat{\mathcal{A}}}_{{\hat{\mu}}}{}^{\hat{M}} = ( {\hat{\mathcal{A}}}_{{\hat{\mu}}}{}^{M}, \ldots )$ and ${\hat{\mathcal{B}}}_{\hat{\mu} \hat{M}} = ({\hat{\mathcal{B}}}_{\hat{\mu} M}, {\hat{\mathcal{B}}}_{\hat{\mu} \alpha} , \ldots )$.  We are thus left with three components which we have yet to determine the $E_{8(8)}$ origins of and we denote them as $\hat{\Phi}$ to differentiate them from the other fields:
\begin{align}
{\mathcal{A}}_{\mu}{}^M = \begin{pmatrix} {\hat{\mathcal{A}}}_{\hat{\mu}}{}^M\\ {\hat{\Phi}}^M \end{pmatrix}\,, \qquad {\mathcal{B}}_{\mu \nu M} = \begin{pmatrix} {\hat{\Phi}}_{\hat{\mu} \hat{\nu} M}\\ {\hat{\mathcal{B}}}_{\hat{\mu}M} + \ldots \end{pmatrix}\,, \qquad {\mathcal{B}}_{\mu \nu \alpha} = \begin{pmatrix} {\hat{\Phi}}_{\hat{\mu} \hat{\nu} \alpha}\\ {\hat{\mathcal{B}}}_{\hat{\mu} \alpha} + \ldots \end{pmatrix}\,.
\end{align}
We have used ellipses to denote corrections to the na\"{i}ve identification. Similarly, the $E_{7(7)}$ generalised field strength of $\mathcal{A}$ decomposes under the KK splitting into the components
\begin{align}
\mathcal{F}_{\mu \nu}{}^M = \begin{pmatrix} \mathcal{F}_{\hat{\mu} \hat{\nu}}{}^M\\ \mathcal{F}_{\hat{\mu}4}{}^M \end{pmatrix}\,.
\end{align}
The upper components have an obvious origin in the $E_{8(8)}$ analogue ${\hat{\mathcal{F}}}_{\hat{\mu} \hat{\nu}}{}^{\hat{M}} = ({\hat{\mathcal{F}}}_{\hat{\mu} \hat{\nu}}{}^M , \ldots )$ such that (note we have already shown the generalised Lie derivative already reduces correctly)
\begin{align}
\mathcal{F}_{\hat{\mu} \hat{\nu}}{}^M = {\hat{\mathcal{F}}}_{\hat{\mu} \hat{\nu}}{}^M\,.
\end{align}
The unidentified piece $\mathcal{F}_{\hat{\mu} 4}{}^M$ is given in terms of $E_{7(7)}$ variables as
\begin{align}
F_{\hat{\mu} 4}{}^M & = \partial_{\hat{\mu}} {\mathcal{A}}_4{}^M - \cancel{\partial_4 {\mathcal{A}}_{\hat{\mu}}{}^M} - \frac{1}{2} \left( \mathbb{L}_{\mathcal{A}_{\hat{\mu}}} {\mathcal{A}}_4{}^M - \mathbb{L}_{{\mathcal{A}}_4}  {\mathcal{A}}_{\hat{\mu}}{}^M \right)\\
	& = \mathcal{D}_{\hat{\mu}} {\mathcal{A}}_4{}^M + {(\! ( \mathcal{A}_{\hat{\mu}},  {\mathcal{A}}_4{})\!)}^M\,,\\
{\mathcal{F}}_{\hat{\mu} 4}{}^M & = F_{\hat{\mu} 4}{}^M  - 12 {(t^\alpha)}^{MN} \partial_N {\mathcal{B}}_{\hat{\mu} 4 \alpha} - \frac{1}{2} \Omega^{MN} {\mathcal{B}}_{\hat{\mu} 4 N}\,,
\end{align}
where $(\!( \cdot, \cdot )\!)$ is the symmetric part of the generalised Lie derivative and $\mathcal{D}_\mu \coloneqq \partial_\mu - \mathbb{L}_{\mathcal{A}_\mu}$ is the Lie-covariantised derivative. We rewrite this in terms of the $E_{8(8)}$ variables that we identified above, giving
\begin{align}
{F}_{\hat{\mu} 4}{}^M & = {\hat{\mathcal{D}}}_{\hat{\mu}} {\hat{\Phi}}^M + {( \! ( \hat{\mathcal{A}}_{\hat{\mu}}, \hat{\Phi} )\!)}^M\\
	& = {\hat{\mathcal{D}}}_{\hat{\mu}} {\hat{\Phi}}^M - 12 {(t^\alpha)}^{MN} \partial_N \left( {(t_{\alpha})}_{PQ} {\hat{\mathcal{A}}}_{\hat{\mu}}{}^P {\hat{\Phi}}^Q \right) + \frac{1}{2} \Omega^{MN} (\partial_N {\hat{\mathcal{A}}}_{\hat{\mu}}{}^P {\hat{\Phi}}_P + \partial_N {\hat{\Phi}}^P {\hat{\mathcal{A}}}_{\hat{\mu} P})\\
{\mathcal{F}}_{\hat{\mu} 4}{}^M	& = {\hat{\mathcal{D}}}_{\hat{\mu}} {\hat{\Phi}}^M - 12 {(t^\alpha)}^{MN} \partial_N \left( {\mathcal{B}}_{\hat{\mu} 4 \alpha} + {(t_{\alpha})}_{PQ} {\hat{\mathcal{A}}}_{\hat{\mu}}{}^P {\hat{\Phi}}^Q \right) - \frac{1}{2} \Omega^{MN} ({\mathcal{B}}_{\hat{\mu} 4 N} - \partial_N {\hat{\mathcal{A}}}_{\hat{\mu}}{}^P {\hat{\Phi}}_P - \partial_N {\hat{\Phi}}^P {\hat{\mathcal{A}}}_{\hat{\mu} P})
\end{align}
and so the components of the $E_{7(7)}$ field strength are given in terms of $E_{8(8)}$ fields as
\begin{align}
\mathcal{F}_{\mu \nu}{}^M & = \begin{pmatrix} {\hat{\mathcal{F}}}_{\hat{\mu} \hat{\nu}}{}^M\\ {\hat{\mathcal{D}}}_{\hat{\mu}} {\hat{\Phi}}^M - 12 {(t^\alpha)}^{MN} \partial_N {\hat{b}}_{\hat{\mu} \alpha} - \frac{1}{2} \Omega^{MN} {\hat{b}}_{\hat{\mu} N}
\end{pmatrix}\,,\\
{\hat{b}}_{\hat{\mu} \alpha} & = {\mathcal{B}}_{\hat{\mu} 4 \alpha} + {(t_{\alpha})}_{PQ} {\hat{\mathcal{A}}}_{\hat{\mu}}{}^P {\hat{\Phi}}^Q\,,\\
{\hat{b}}_{\hat{\mu} M} & = {\mathcal{B}}_{\hat{\mu} 4 N} - \partial_N {\hat{\mathcal{A}}}_{\hat{\mu}}{}^P {\hat{\Phi}}_P - \partial_N {\hat{\Phi}}^P {\hat{\mathcal{A}}}_{\hat{\mu} P}\,.
\end{align}
Note, in particular, that $\hat{b}_{\hat{\mu} \bullet}$ is still given in terms of a mixture of $E_{8(8)}$ objects $(\hat{\mathcal{A}}_{\hat{\mu}}{}^{\hat{M}}, \hat{\Phi}^{\hat{M}})$ and $E_{7(7)}$ objects ($\mathcal{B}_{\hat{\mu}\hat{4} \bullet}$). This is due to the fact that are not identifying the $E_{8(8)}$ $\hat{\mathcal{B}}_{\hat{\mu}}$ field as the $\hat{\mu}$-$4$ component of the $E_{7(7)}$ ${\mathcal{B}}_{\mu \nu}$ directly, as we indicated by ellipses previously. We shall clear up the relation between all these fields below.\par
Our starting point is the sum of the kinetic term for the scalar sector and the $\hat{\mathcal{B}}$-$\hat{F}$ contribution to the Chern-Simons term of $E_{8(8)}$ EFT (as before, we adorn all objects and indices in $d=3$ with hats). The kinetic term can be written in terms of the scalar current,
\begin{align}{\hat{\jmath}}_{\hat{\mu}}{}^{\hat{M}} \coloneqq \frac{1}{60} {\hat{f}}^{\hat{M}}{}_{\hat{K}}{}^{\hat{L}} {\hat{\mathcal{M}}}^{\hat{K} \hat{Q}} {\hat{\mathcal{D}}}_{\hat{\mu}}{\hat{\mathcal{M}}}_{\hat{Q} \hat{L}}\,,
\end{align}
and the uncovariantised field strength $\hat{F}$ may be improved to the fully covariantised field strength $\hat{\mathcal{F}}$ since the two differ only by terms that vanish under the section condition. Our starting point is the Lagrangian
\begin{align}\label{eq:E8ActionPart}
{\hat{\mathcal{L}}}_{\text{kin.}} + {\hat{\mathcal{L}}}_{\text{CS}} & = \frac{1}{240} {\hat{\mathcal{D}}}_{\hat{\mu}} {\hat{\mathcal{M}}}_{\hat{M} \hat{N}} {\hat{\mathcal{D}}}^{\hat{\mu}} {\hat{\mathcal{M}}}^{\hat{M} \hat{N}} + \frac{1}{2} \varepsilon^{\hat{\mu} \hat{\nu} \hat{\rho}} {\hat{F}}_{\hat{\mu} \hat{\nu}}{}^{\hat{M}} {\hat{\mathcal{B}}}_{\hat{\rho}\hat{M}} + \ldots\\
	& = - \frac{\hat{e}}{4} {\hat{g}}^{\hat{\mu} \hat{\nu}} {\hat{\jmath}}_{\hat{\mu}}{}^{\hat{M}} {\hat{\jmath}}_{\hat{\nu} \hat{M}} + \frac{1}{2} \varepsilon^{\hat{\mu} \hat{\nu} \hat{\rho}} {\hat{\mathcal{F}}}_{\hat{\mu} \hat{\nu}}{}^{\hat{M}} {\hat{\mathcal{B}}}_{\hat{\rho}\hat{M}} + \ldots\label{eq:KinCS}\,,
\end{align}
where the ellipses represent terms that are independent of $\hat{\mathcal{B}}$-field and so it is sufficient to vary \eqref{eq:KinCS} to obtain the $\hat{\mathcal{B}}$-field equations of motion. In doing so, we use the fact that the scalar current is given explicitly by
\begin{align}
{\hat{\jmath}}_{\hat{\mu}}{}^{\hat{M}} & = \frac{1}{60} {\hat{f}}^{\hat{M}}{}_{\hat{K}}{}^{\hat{L}} {\hat{\mathcal{M}}}^{\hat{K}\hat{Q}} (\partial_{\hat{\mu}} - {\hat{\mathcal{A}}}_{\hat{\mu}}{}^{\hat{R}} \partial_{\hat{R}} ) {\hat{\mathcal{M}}}_{\hat{Q}\hat{L}} + ({\hat{\mathcal{M}}}^{\hat{M}\hat{N}} + {\hat{\kappa}}^{\hat{M}\hat{N}})({\hat{f}}_{\hat{N}}{}^{\hat{S}}{}_{\hat{T}} \partial_{\hat{S}} {\hat{\mathcal{A}}}_{\hat{\mu}}{}^{\hat{T}} + {\hat{\mathcal{B}}}_{\hat{\mu} \hat{N}})\label{eq:JBRelation}\,,
\end{align}
which gives us an expression for $\hat{\jmath}$ in terms of $\hat{\mathcal{B}}$.  The resulting equation of motion for $\hat{\mathcal{B}}$ is then
\begin{align}\label{eq:VSDuality}
{\hat{\epsilon}}^{\hat{\mu} \hat{\rho} \hat{\sigma}} {\hat{\mathcal{F}}}_{\hat{\rho} \hat{\sigma}}{}^{\hat{M}} = 2 {\hat{g}}^{\hat{\mu} \hat{\nu}} {\hat{\jmath}}_{\hat{\nu}}{}^{\hat{M}}\,,
\end{align}
which relates the fields in the scalar coset to the vector fields just as in a vector-scalar duality relation. 

Multiplying both sides of \eqref{eq:JBRelation} by ${\hat{\jmath}}_{\hat{\nu} \hat{M}}$ gives an expression of the form ${\hat{\mathcal{B}}}_{\hat{\mu} \hat{M}} {\hat{\jmath}}_{\hat{\nu}}{}^{\hat{M}} = {\hat{\jmath}}_{\hat{\mu} \hat{M}} {\hat{\jmath}}_{\hat{\nu}}{}^{\hat{M}} + \ldots$,
 from which we may rewrite both terms in \eqref{eq:E8ActionPart} in terms of ${\hat{\jmath}}^2$ terms as
\begin{align}\label{eq:KinCS2}
{\hat{\mathcal{L}}}_{\text{kin.}} + {\hat{\mathcal{L}}}_{\text{CS}} & = \frac{3}{4} \hat{e} {\hat{g}}^{\hat{\mu} \hat{\nu}} {\hat{\jmath}}_{\hat{\mu}}{}^{\hat{M}} {\hat{\jmath}}_{\hat{\nu} \hat{M}} + \ldots\,.
\end{align}
The final ingredient that we need to proceed is to note that, in analogy with the left-invariant currents in $\sigma$-models, we may write\footnote{One way to see this is to start with the Bianchi identity for $\mathcal{F}$:
\begin{align}
0 & = {\hat{\mathcal{D}}}_{\hat{\mu}} \left(  \varepsilon^{\hat{\mu} \hat{\rho} \hat{\sigma}} \hat{\mathcal{F}}_{\hat{\rho} \hat{\sigma}}{}^{\hat{M}} \right) \otimes {\hat{C}}_{\hat{N}}\qquad \Rightarrow \qquad 0 = {\hat{\mathcal{D}}}_{\hat{\mu}} \left(  2 {\hat{e}} \hat{\jmath}^{\hat{\mu} \hat{M}}\right) \otimes {\hat{C}}_{\hat{N}}\,,
\end{align}
where ${\hat{C}}_{\hat{M}}$ is a covariantly constrained object. We implement this by a Lagrange multiplier $+2 {\hat{\mathcal{D}}}_{\hat{\mu}} \left( \hat{e} {\hat{\jmath}}^{\hat{\mu} \hat{M}} \right) {\hat{\chi}}_{\hat{N}}$, where ${\hat{\chi}}_{\hat{N}}$ is covariantly constrained. Varying \eqref{eq:KinCS2} with respect to ${\hat{\jmath}}^{\hat{\mu} \hat{M}}$ yields that it can be written as the total derivative of ${\hat{\chi}}_{\hat{N}}$ (actually this requires more care since the variation of $\hat{\jmath}$ is not unconstrained but rather given in terms of the variations of ${\hat{\mathcal{A}}}$ and $\hat{\mathcal{B}}$ and so, strictly speaking, we may need some projectors to act on an unconstrained $\hat{\chi}$).}
\begin{align}\label{eq:JDChi}
{\hat{\jmath}}_{\hat{\mu} \hat{M}} = {\hat{\mathcal{D}}}_{\hat{\mu}} {\hat{\chi}}_{\hat{M}}\,,
\end{align}
for some covariantly constrained ${\hat{\chi}}_{\hat{M}}$ (a scalar on the external space).\par
We now return to \eqref{eq:KinCS2} and split it into 3 terms so that we can proceed to identify the necessary pieces needed for the $E_{7(7)}$ theory:
\begin{align}
{\hat{\mathcal{L}}}_{\text{kin.}} + {\hat{\mathcal{L}}}_{\text{CS}} & = a \hat{e} {\hat{g}}^{\hat{\mu} \hat{\nu}} {\hat{\jmath}}_{\hat{\mu}}{}^{\hat{M}} {\hat{\jmath}}_{\hat{\nu} \hat{M}} + b \hat{e} {\hat{g}}^{\hat{\mu} \hat{\nu}} {\hat{\jmath}}_{\hat{\mu}}{}^{\hat{M}} {\hat{\jmath}}_{\hat{\nu} \hat{M}} + c \hat{e} {\hat{g}}^{\hat{\mu} \hat{\nu}} {\hat{\jmath}}_{\hat{\mu}}{}^{\hat{M}} {\hat{\jmath}}_{\hat{\nu} \hat{M}} + \ldots\,,
\end{align}
subject to $a + b + c = \frac{3}{4}$. The first term, we rewrite back in terms of the scalars$\mathcal{D} \mathcal{\hat{M}} \mathcal{D} {\hat{\mathcal{M}}}^{-1}$ as
\begin{align}
a \hat{e} {\hat{g}}^{\hat{\mu} \hat{\nu}} {\hat{\jmath}}_{\hat{\mu}}{}^{\hat{M}} {\hat{\jmath}}_{\hat{\nu} \hat{M}} & = - \frac{a}{60} \hat{e} {\hat{g}}^{\hat{\mu} \hat{\nu}} {\hat{\mathcal{D}}}_{\hat{\mu}} {\hat{\mathcal{M}}}_{\hat{M} \hat{N}} {\hat{\mathcal{D}}}_{\hat{\nu}} {\hat{\mathcal{M}}}^{\hat{M} \hat{N}} = - \frac{a}{60} \hat{e} {\hat{g}}^{\hat{\mu} \hat{\nu}} {\hat{\mathcal{D}}}_{\hat{\mu}} \mathcal{M}_{MN} {\hat{\mathcal{D}}}_{\hat{\nu}} \mathcal{M}^{MN} + \ldots\,,
\end{align}
where we have expanded the $E_{8(8)}$ coordinates under the $E_{7(7)}$ decomposition \eqref{eq:E8E7Decomp} and restricted to the $E_{7(7)}$ coordinates (the $\operatorname{SL}(2)$ index fixed to $a=1$). We have also used the fact that the principal contribution to the $M1$-$N1$ component of the generalised metric should be the $E_{7(7)}$ generalised metric $\hat{\mathcal{M}}_{M1 N1} = \mathcal{M}_{MN} + \ldots$.
 If we assume that the fourth direction is a (generalised) isometry of the fields, which we can describe by 
\begin{align}
\mathbb{L}_{\mathcal{A}_4} \bullet = 0 \qquad \Rightarrow \qquad \partial_4 \bullet = 0 \, , \label{isoconst}
\end{align}
 we can simply replace the covariant derivatives with the 4-dimensional completions as ${\hat{\mathcal{D}}}_{\hat{\mu}} \rightarrow \mathcal{D}_\mu$. Finally, using the KK ansatz of the inverse external metric \eqref{eq:KKInverse}, we have $g^{\hat{\mu} \hat{\nu}} = e^{- 2 \alpha \phi} {\hat{g}}^{\hat{\mu}\hat{\nu}}$ for which $e = e^{3 \alpha \phi} \hat{e}$. Thus, we end up with
\begin{align}
a \hat{e} {\hat{g}}^{\hat{\mu} \hat{\nu}} {\hat{\jmath}}_{\hat{\mu}}{}^{\hat{M}} {\hat{\jmath}}_{\hat{\nu} \hat{M}} & = - \frac{a}{60} e e^{\alpha \phi} g^{\mu \nu} \mathcal{D}_\mu \mathcal{M}_{MN} \mathcal{D}_{\nu} \mathcal{M}^{MN} + \ldots\,.
\end{align}
For the second term, we write one of the $\hat{\jmath}$ in terms of $\hat{\mathcal{F}}$ using vector-scalar relation \eqref{eq:VSDuality} and the other using \eqref{eq:JDChi} to give
\begin{align}
b \hat{e} {\hat{g}}^{\hat{\mu} \hat{\nu}} {\hat{\jmath}}_{\hat{\mu}}{}^{\hat{M}} {\hat{\jmath}}_{\hat{\nu} \hat{M}} & = \frac{be}{2} e^{-3 \alpha \phi} \epsilon^{\hat{\nu} \hat{\rho} \hat{\sigma}} {\hat{\mathcal{F}}}_{\hat{\rho} \hat{\sigma}}{}^{\hat{M}} {\hat{\mathcal{D}}}_{\hat{\nu}} {\hat{\chi}}_{\hat{M}} = \frac{be}{2} e^{-3 \alpha \phi} \epsilon^{\hat{\nu} \hat{\rho} \hat{\sigma}} {\hat{\mathcal{F}}}_{\hat{\rho} \hat{\sigma}}{}^{M} {\hat{\mathcal{D}}}_{\hat{\nu}} {\hat{\chi}}^{N} \Omega_{NM}+ \ldots\,.
\end{align}
Finally, for the third term, we use the relations \eqref{eq:JBRelation} and \eqref{eq:VSDuality}, to rewrite it as
\begin{align}
c \hat{e} {\hat{g}}^{\hat{\mu} \hat{\nu}} {\hat{\jmath}}_{\hat{\mu}}{}^{\hat{M}} {\hat{\jmath}}_{\hat{\nu} \hat{M}} & =  \frac{c\hat{e}}{2} \epsilon^{\hat{\mu} \hat{\rho} \hat{\sigma}} {\hat{\mathcal{F}}}_{\hat{\rho} \hat{\sigma}}{}^{\hat{M}} ( {\hat{\mathcal{B}}}_{\hat{\mu} \hat{M}} + {\hat{f}}_{\hat{M}}{}^{\hat{K}}{}_{\hat{L}} \partial_{\hat{K}} {\hat{\mathcal{A}}}_{\hat{\mu}}{}^{\hat{L}} + \ldots)\\
	& = \frac{c \hat{e}}{2} \epsilon^{\hat{\mu} \hat{\rho} \hat{\sigma}} {\hat{\mathcal{F}}}_{\hat{\rho} \hat{\sigma}}{}^{Ma} \Omega_{MN} \left( - \Omega^{NK} {\hat{\mathcal{B}}}_{\hat{\mu}K a} + \sqrt{12} {(t_\alpha)}^{NK} \partial_{Ka} {\hat{\mathcal{A}}}_{\hat{\mu}}{}^{\alpha} + \ldots \right)\,,
\end{align}
where we have expanded the contracted $E_{8(8)}$ indices in terms of $E_{7(7)} \times \operatorname{SL}(2)$ indices and focused on the $\hat{M} = Ma$ pieces (the $(\mathbf{56,2})$ representations) of the contracted indices (the decomposition of the Killing form and structure constants are given in \eqref{eq:KillingDecomp} and \eqref{eq:E8StructureConstants} respectively\footnote{Strictly speaking, we should rescale the structure constants by $\sqrt{60}$, relative to \eqref{eq:E8StructureConstants}, since we have reverted to the conventions of \cite{Hohm:2014fxa} for this section. However, we shall demonstrate later that we do not need to worry about the precise scaling (at least at the classical level).}). Since we only want non-trivial derivatives in the $a=1$ direction---the directions that form the $E_{7(7)}$ generalised coordinates---we pick out only the $M1 \equiv M$ indices:
\begin{align}
c \hat{e} {\hat{g}}^{\hat{\mu} \hat{\nu}} {\hat{\jmath}}_{\hat{\mu}}{}^{\hat{M}} {\hat{\jmath}}_{\hat{\nu} \hat{M}} & = \frac{c e}{2} e^{-3\alpha \phi} \epsilon^{\hat{\mu} \hat{\rho} \hat{\sigma}} {\hat{\mathcal{F}}}_{\hat{\rho} \hat{\sigma}}{}^{M} \Omega_{MN} \left( - \Omega^{NK} {\hat{\mathcal{B}}}_{\hat{\mu}K } + \sqrt{12} {(t_\alpha)}^{NK} \partial_{K} {\hat{\mathcal{A}}}_{\hat{\mu}}{}^{\alpha} + \ldots \right)\,.
\end{align}
Thus, the sum of three terms can be written as
\begingroup
\renewcommand{\arraystretch}{1.5}
\begin{align}
\begin{array}{ll}
{\hat{\mathcal{L}}}_{\text{kin.}} + {\hat{\mathcal{L}}}_{\text{CS}} & = - \frac{a}{60} e e^{\alpha \phi} g^{\mu \nu} \mathcal{D}_\mu \mathcal{M}_{MN} \mathcal{D}_{\nu} \mathcal{M}^{MN} \\
& \qquad + e e^{-3 \alpha \phi}  \epsilon^{\hat{\mu} \hat{\rho} \hat{\sigma}} {\hat{\mathcal{F}}}_{\hat{\rho} \hat{\sigma}}{}^{M} \Omega_{MN} \left( - \frac{b}{2} {\hat{\mathcal{D}}}_{\hat{\mu}} {\hat{\chi}}^{N} -  \frac{c}{2} \Omega^{NK} {\hat{\mathcal{B}}}_{\hat{\mu} K} + c\sqrt{3} {(t^\alpha)}^{NK} \partial_K {\hat{\mathcal{A}}}_{\hat{\mu} \alpha} \right) + \ldots\\
& = - \frac{a}{60} e e^{\alpha \phi} g^{\mu \nu} \mathcal{D}_\mu \mathcal{M}_{MN} \mathcal{D}_{\nu} \mathcal{M}^{MN}\\
& \qquad + e^{-3 \alpha \phi}  \varepsilon^{\hat{\mu} \hat{\rho} \hat{\sigma}} {\mathcal{F}}_{\hat{\rho} \hat{\sigma}}{}^{M} \Omega_{MN} \left( - \frac{b}{2} {\hat{\mathcal{D}}}_{\hat{\mu}} {\hat{\chi}}^{N} -  \frac{c}{2} \Omega^{NK} {\hat{\mathcal{B}}}_{\hat{\mu} K} + c\sqrt{3} {(t^\alpha)}^{NK} \partial_K {\hat{\mathcal{A}}}_{\hat{\mu} \alpha} \right) + \ldots \,,
\end{array}
\end{align}
\endgroup
where we have replaced $\hat{\mathcal{F}} \rightarrow \mathcal{F}$ in the second line. Then, for judicious choices of $b$ and $c$, the term in parentheses is of the form of ${\mathcal{F}}_{\hat{\mu} 4}{}^{N}$ once we identify
\begin{align}
- \frac{b}{2} {\hat{\chi}}^{\hat{N}} & = {\hat{\Phi}}^{\hat{N}}\,,\label{eq:ChiScaling}\\
c {\hat{\mathcal{B}}}_{\hat{\mu} K} & = {\hat{b}}_{\hat{\mu} K} = {\mathcal{B}}_{\hat{\mu} 4 K} - \partial_K {\hat{\mathcal{A}}}_{\hat{\mu}}{}^P {\hat{\Phi}}_P - \partial_K {\hat{\Phi}}^P {\hat{\mathcal{A}}}_{\hat{\mu} P}\,,\\
c \sqrt{3} {\hat{\mathcal{A}}}_{\hat{\mu} \alpha} & = -12 {\hat{b}}_{\hat{\mu} \alpha} = -12 ({\mathcal{B}}_{\hat{\mu} 4 \alpha} + {(t_{\alpha})}_{PQ} {\hat{\mathcal{A}}}_{\hat{\mu}}{}^P {\hat{\Phi}}^Q )\,.
\end{align}
The last two equations then give the precise form of the $E_{7(7)}$ components in terms of $E_{8(8)}$ fields that we previously denoted by ellipses in \eqref{eq:E7E8GaugeFields}. After the above manipulations, we are then left with
\begin{align}
{\hat{\mathcal{L}}}_{\text{kin.}} + {\hat{\mathcal{L}}}_{\text{CS}} & = - \frac{a}{60} e e^{\alpha \phi} g^{\mu \nu} \mathcal{D}_\mu \mathcal{M}_{MN} \mathcal{D}_{\nu} \mathcal{M}^{MN} + e^{-3 \alpha \phi}  \varepsilon^{\hat{\mu} \hat{\rho} \hat{\sigma}} {\mathcal{F}}_{\hat{\rho} \hat{\sigma}}{}^M \Omega_{MN} {\mathcal{F}}_{\hat{\mu}4}{}^N + \ldots\\
	&  = - \frac{a}{60} e e^{\alpha \phi} g^{\mu \nu} \mathcal{D}_\mu \mathcal{M}_{MN} \mathcal{D}_{\nu} \mathcal{M}^{MN} + \frac{1}{4} e^{-3 \alpha \phi} \varepsilon^{\mu \nu \rho \sigma} {\mathcal{F}}_{\mu \nu}{}^M \Omega_{MN} {\mathcal{F}}_{\rho \sigma}{}^N + \ldots\,. \label{fwedgef}
\end{align}
We are now very close to the desired result. The final piece of the puzzle is to use the $E_{7(7)}$ twisted self-duality constraint for $\mathcal{F}_{\mu \nu}{}^M$:
\begin{align}
\mathcal{F}_{\mu \nu}{}^M= \frac{1}{2} {\mathcal{M}}^M{}_N \sqrt{-g} \varepsilon_{\mu \nu}{}^{\rho \sigma} \mathcal{F}_{\rho \sigma}{}^N \, , \label{twisted}
\end{align}
where $\mathcal{M}^M{}_N = \mathcal{M}^{MP} \Omega_{PN}$.

With this final ingredient, when the equations of motion from (\ref{fwedgef}) are combined with the twisted self-duality constraint then they give the same equations of motion as derived from the $E_{7(7)}$ pseudo-action\footnote{The careful reader should rightly be worried about the status of this pseudo-action and the insertion of the self duality constraint. The pragmatic way to think about is that it is like the pseudo-action for chiral boson where the chirality constraint is imposed on the resulting equations of motion. For a more rigorous approach it is be possible to construct a PST type action \cite{Pasti:2012wv} or one could follow the recent approach of Sen \cite{Sen:2019qit}}:
Equivalently, the equations of motion of $E_{8(8)}$ EFT reduces (for appropriate choices of the coefficients $a,b$ and $c$) to the on-shell equations of motion of $E_{7(7)}$ EFT. Note that, since we started with (part of) the topological term in $E_{8(8)}$, the ellipsis will include terms that enter into the topological term in $E_{7(7)}$ but do not effect the equations of motion.\par
We make one final remark regarding the relative scaling coefficients $a,b$ and $c$ which were constrained to satisfy $a + b + c = \frac{3}{4}$. We actually have an additional freedom in rescaling ${\hat{\chi}}_{\hat{M}}$, which affects $b$ in \eqref{eq:ChiScaling}, and so we can always find $a, b$ and $c$ such that the reduction of the terms to those in the $E_{7(7)}$ Lagrangian is exact. \par

We stress that the recombination of terms to produce the required form of the $E_{7(7)}$ theory is highly non-trivial. It rests on:
\begin{itemize}
\item
the equations of motion of the $\hat{\mathcal{B}}$ field in the $E_{8(8)}$ EFT giving the $d=3$ vector-scalar relation \eqref{eq:VSDuality}; 
\item
the twisted self-duality relation of the $E_{7(7)}$ vector fields \eqref{twisted};
\item
and the isometry condition \eqref{isoconst}.
\end{itemize}

Finally let us comment that getting from the $\hat{\mathcal{B}}\mathcal{F}$ term to a Yang-Mills term by spontaneously breaking the symmetry and integrating out a field is very similar to the Mukhi-Papageorgakis mechanism in Bagger-Lambert theory \cite{Mukhi:2008ux}. There the Chern-Simons scalar theory was turned into a Yang-Mills theory using the equations of motion from integrating out one of the vector fields after one of the scalars is given a constant vacuum expectation value. The situation is similar here where the $\hat{\mathcal{B}}$-field is integrated out after the $E_{8(8)}$ symmetry is broken to $E_{7(7)}$.

\section{Discussion}\label{sec:Discussion}
In this paper we have considered some of the aspects of reductions between ExFTs, with a particular focus on EFT-to-EFT reductions. We began with explicit examples of the dimensional reduction of the generalised metric in ExFTs. For the $\operatorname{SL}(5)$ generalised metric, we described how both the $\operatorname{SL}(3) \times \operatorname{SL}(2)$ and $\operatorname{O}(3,3)$ generalised metrics could be obtained by a Kaluza-Klein reduction by different identifications of the section. In doing so, we suggested a generalised Kaluza-Klein ansatz that both the DFT and $\operatorname{SL}(5)$ EFT generalised metrics respected and argued that both the conventional Kaluza-Klein ansatz and the ansatz of \cite{Thompson:2011uw} could be understood as a special case of this generalised Kaluza-Klein reduction for which certain components of the $Y$-tensor vanish.\par
We outlined some of the difficulties faced when trying to reduce the generalised metrics of larger EFTs. These include the appearance of more on-shell degrees of freedom (the $C_{(6)}$ entering in $E_{6(6)}$ EFT and above and the dual graviton appearing in $E_{8(8)}$ EFT) as well as the increase in the number of blocks that appear in the generalised metric. These suggest that the generalised KK ansatz may not be the whole story. \par
We then considered the reduction of the section condition for $E_{8(8)}$ and $\operatorname{SL}(5)$ EFTs. The former emphasised the fact that we require a consistent set of conventions between the parent and child theories. We showed explicitly that, amongst other things, the projector onto the adjoint does not reduce exactly (rather acquiring a $\delta \delta$ term) but that the $Y$-tensor does and gave the interpretation that the \emph{effective} weight remains unchanged between theories, even if though universal weight of the two theories differ. The reduction of the $\operatorname{SL}(5)$ section condition showed how the section conditions of smaller ExFTs can be obtained from the parent EFT as \emph{partial} solutions to the section condition. In both instances, we raised the prospect of some rather intriguing phenomena. For the $E_{8(8)} \rightarrow E_{7(7)}$ reduction, we suggested that the $(\mathbf{56},\mathbf{2})$ within the $\mathbf{248}$ could allow for the section conditions to be solved independently on different local patches in such a way that they require the residual $\operatorname{SL}(2)$ symmetry to patch together correctly. For the $\operatorname{SL}(5)$ case, we instead suggested that taking partial solutions into consideration may allow for more ExFT reductions than one may have initially expected.\par
In the final section, we described how one of the topological terms of $E_{8(8)}$ EFT is better understood as a $BF$ term that, taken together with the kinetic term for the scalar sector, reproduces the kinetic and Yang-Mills term of $E_{7(7)}$ EFT (more precisely, the equations of motion that we obtain from this rewriting agree with the on-shell equations of motion of $E_{7(7)}$ EFT) upon employing vector-scalar duality and the twisted self-duality condition.\par
There is still plenty left to explore in terms of reductions between ExFTs. The most conspicuous omission in the present paper is the reduction of the full tensor hierarchy; it is a non-trivial problem to determine how the on-shell degrees of freedom need to be reshuffled into tensors of the lower-dimensional EFT to reconstruct its tensor hierarchy. On a related note, the reduction of the topological terms in each theory remains to be studied.
\par
Obvious extensions to the ideas presented here would be to consider reductions of EFTs on more general spaces \cite{Malek:2016vsh} or even the Scherk-Schwarz reductions of EFTs \cite{Berman:2012uy}. It would also be interesting to construct explicit solutions using the KK type gauge fields in the reduction along the lines of \cite{Berman:2014hna}. The reductions described here might also be useful in relating different non-Riemannian solutions of EFT and DFT following  \cite{Berman:2019izh} and \cite{Morand:2017fnv,Cho:2018alk,Cho:2019ofr}.

\section*{Acknowledgements}

David Berman is supported by STFC grant ST/L000415/1, ``String Theory, Gauge Theory and Duality'' and Ray Otsuki is supported by an STFC studentship. We thank Chris Blair, Gianluca Inverso, Jeong-Hyuck Park and Felix Rudolph for discussions

\setcounter{section}{0}
\renewcommand{\thesection}{\Alph{section}}
\begin{appendices}
\section{Appendix}
We use $\varepsilon$ and $\epsilon$ to distinguish between the alternating symbol and the Levi-Civita tensor. In particular
\begin{align}
\varepsilon_{m_1 m_2 \ldots m_n} = \varepsilon^{m_1 m_2 \ldots m_n} & = \begin{cases}
\begin{array}{rl}
+1\,, & \text{Even permutation of indices.}\\
-1\,, & \text{Odd permutation of indices.}\\
0\,, & \text{Otherwise.}\\
\end{array}
\end{cases}\,,\\
\epsilon_{m_1 m_2 \ldots m_n} & = \sqrt{-g} \varepsilon_{m_1 m_2 \ldots m_n}\,,\\
\epsilon^{m_1 m_2 \ldots m_n} & = \frac{1}{\sqrt{-g}} \varepsilon^{m_1 m_2 \ldots m_n}\,.
\end{align}
\end{appendices}
\bibliographystyle{unsrt}
\bibliography{\jobname}
\end{document}